\pdfoutput=1
\documentclass[10pt,amssymb,aps,longbibliography,
nofootinbib, pra]{revtex4-2}

\usepackage{graphicx} 
\usepackage[leqno]{amsmath}
\usepackage{bbm}
\usepackage{multirow}
\usepackage{float}
\usepackage{xcolor}
\usepackage{bbold}
\usepackage{array}
\usepackage{hyperref}
\usepackage{braket}
\newcommand{\ketbra}[2]{| #1 \rangle\!\langle #2| }
\newcommand{\qroam}{\ensuremath{\text{QROAM}}}
\newcommand{\selxor}{\ensuremath{\text{SELECT-XOR}}}
\newcommand{\selz}{\ensuremath{\text{SELECT-Z}}}
\newcommand{\addpg}{\ensuremath{\text{ADD}}}
\usepackage[capitalise]{cleveref}

\definecolor{rlgreen}{HTML}{006F62}   
\definecolor{rlblue}{HTML}{4B5BFF}    
\hypersetup{
  colorlinks   = true, 
  urlcolor     = rlblue, 
  filecolor    = rlblue, 
  runcolor     = rlblue, 
  linkcolor    = rlblue, 
  citecolor    = rlgreen  
}
\makeatletter
\def\@doi#1{\endgroup\href{\doibase#1}{doi:\discretionary{}{}{}#1}}
\makeatother

\usepackage[compat=0.6]{yquant}
\usetikzlibrary{fit,quotes,decorations.pathreplacing,positioning,calc}

\pgfdeclareshape{yquant-multictrlshape}{
   \inheritsavedanchors[from=yquant-circle]%
   \foreach \anc in {center, north, north east, east, south east, south, south west, west, north west} {%
      \inheritanchor[from=yquant-circle]{\anc}%
   }%
   \inheritanchorborder[from=yquant-circle]%
   \backgroundpath{%
      \pgfpathmoveto{\pgfqpoint{-0.707107\dimexpr\xradius\relax}{-0.707107\dimexpr\yradius\relax}}%
      \pgfpathlineto{\pgfqpoint{0.707107\dimexpr\xradius\relax}{0.707107\dimexpr\yradius\relax}}%
      \pgfpathellipse{\pgfpointorigin}%
                     {\pgfqpoint{\xradius}{0pt}}%
                     {\pgfqpoint{0pt}{\yradius}}%
   }%
   \inheritclippath[from=yquant-circle]%
}

\yquantset{multictrl/.style={every negative control/.style={shape=yquant-multictrlshape,radius=1.05mm,draw,fill=white}}}
\newcommand{\opencontrol}[1]{\node[circle, draw, fill=white, inner sep=0pt, minimum size=1mm, line width=0.4pt] at (#1) {};}

\begin{document}
\title{Quantum Simulation of First-Quantised Chemistry with a Sum-of-Squares Representation}
\author{Aleksei V. Ivanov}
\email[Corresponding author: ]{aleksei.ivanov@riverlane.com}
\author{Christoph S\"underhauf}
\author{Jon Kunjummen}
\author{Nick S. Blunt}
\author{Róbert Izsák}
\affiliation{Riverlane, Cambridge, UK}

\begin{abstract}
    Calculating molecular energies is expected to be an important application of quantum computers, but the most efficient algorithms to achieve this typically have significant space requirements. Scaling quantum processors to thousands of logical qubits is a daunting engineering challenge, so developing algorithms that run with reduced qubit requirements is of particular importance. Moreover, most algorithms are benchmarked on active spaces with about as many orbitals as electrons, missing dynamical correlation which is necessary to describe real chemistry. Capturing this correlation requires several times more orbitals than electrons, even in compact atomic-orbital basis sets. This low-filling regime favours first quantisation, where each electron is represented by a number of qubits that grows only logarithmically with the number of orbitals. Whether this approach translates into a practical advantage in orbital basis sets of realistic size, however, has been unclear. Here, we develop a first-quantised phase estimation algorithm for generic orbital basis sets, based on a non-convex sum-of-squares representation of the Hamiltonian and spectral-gap amplification. We compare it with the state-of-the-art second-quantised algorithm, compiling both with the same techniques at the logical level. For six molecules with 4 to 24 electrons in 5 to 14 times as many orbitals, our first-quantised method needs up to 16 times fewer Toffoli gates than the second-quantised algorithm in the low-qubit regime, where only a few hundred logical qubits are available. It stays cheaper up to between several hundred and about 1\,400 qubits, and its smallest qubit count is 1.4 to 2 times lower than in second quantisation. In the low-qubit and low-filling regime, our algorithm is therefore the method of choice for chemistry in orbital basis sets.
\end{abstract}

\maketitle

{\small\makeatletter\def\l@f@section{\addpenalty{\@secpenalty}\addvspace{0.5em plus\p@}}\makeatother
\tableofcontents}  

\section{Introduction}
Calculating energies of molecules and materials is of crucial importance in predicting chemical properties. Developing efficient algorithms to perform such calculations on a quantum computer~\cite{lloyd_universal_1996,abrams_quantum_1999,cao_quantum_2019,mcardle_quantum_2020,bauer_quantum_2020} is an active field of research that demonstrates consistent progress in reducing computational resources~\cite{guzik_simulated_quantum_computation_2005,reiher_elucidating_2017,babbush_low-depth_2018,babbush_encoding_2018,berry_qubitization_2019,kivlichan_improved_2020,motta_low_2021,lee_even_2021,su_yuan_fault-tolerant_2021,cohn_quantum_2021,goings_reliably_2022,delgado_simulating_2022,blunt_perspective_2022,izsak_quantum_2023,ivanov_quantum_2023,zini_quantum_2023,rubin_fault-tolerant_2023,blunt_statistical_2023,clinton_towards_2024,rocca_reducing_the_runtime_2024,oumarou_accelerating_quantum_2024,berry_quantum_2024,georges_quantum_2025,da_jornada_comprehensive_2025,ivanov_quantum_computation_of_electronic_2025,caesura_faster_quantum_chemistry_2025,bay-smidt_fault-tolerant_2025,low_fast_2025,ollitrault_improving_the_runtime_2025,bhardwaj2026fault,bay-smidt_quantum_2026,bothe_more_2026}. The vast majority of these quantum algorithms are benchmarked on active-space problems where the Hilbert space is truncated to a small subspace, in which the number of electrons is approximately equal to the number of orbitals. This is motivated by the classical methodology, where such an active space calculation is often used to treat static correlation, followed by a perturbative calculation to capture dynamical correlation. However, it is unknown how to efficiently (that is, with small Toffoli, qubit and shot counts) perform perturbation theory to correct active-space quantum phase estimation (QPE) calculations, such as for the widely-benchmarked FeMoco problem~\cite{reiher_elucidating_2017,li_the_electronic_complexity_2019}. It is well known that disregarding this dynamical correlation can lead to both quantitatively and qualitatively incorrect results~\cite{roemelt_2018,blunt_efficient_2020}. Therefore, to accurately describe real chemistry, it will be desirable to perform QPE in the ``low-filling'' regime, where the number of orbitals in the active space is several times the number of electrons. In this case, it might be advantageous to use first-quantised approaches where the number of qubits scales only logarithmically with the size of the basis set~\cite{abrams_simulation_1997}. First-quantised algorithms have been developed in real space~\cite{wiesner_simulations_1996,zalka_simulating_1998,kassal_polynomial_2008}, in plane wave basis sets~\cite{babbush_quantum_2019,su_yuan_fault-tolerant_2021,delgado_simulating_2022,zini_quantum_2023,berry_quantum_2024,da_jornada_comprehensive_2025}, and in generic orbital basis sets~\cite{abrams_simulation_1997,toloui_quantum_2013,babbush_exponentially_2018,georges_quantum_2025}. While asymptotic speed-ups over second quantisation for molecular systems have been shown for generic orbital basis sets~\cite{georges_quantum_2025}, it remains unclear whether a practical advantage exists for a basis of realistic size.

In addition, a crucial challenge in developing quantum algorithms for quantum chemistry is the problem of reducing the required qubit resources. While non-Clifford gate counts for benchmark problems have fallen significantly due to algorithmic improvements in recent years~\cite{berry_qubitization_2019,von_burg_quantum_2021,lee_even_2021,caesura_faster_quantum_chemistry_2025,low_fast_2025}, the necessary qubit counts remain daunting when considering the engineering requirements to scale up quantum processors. As such, there have been efforts to significantly reduce the qubit requirements, notably the recent work of Ref.~\cite{low_denser_2026}.

\begin{figure}[!b]
\centering
\includegraphics[width=0.75\textwidth]{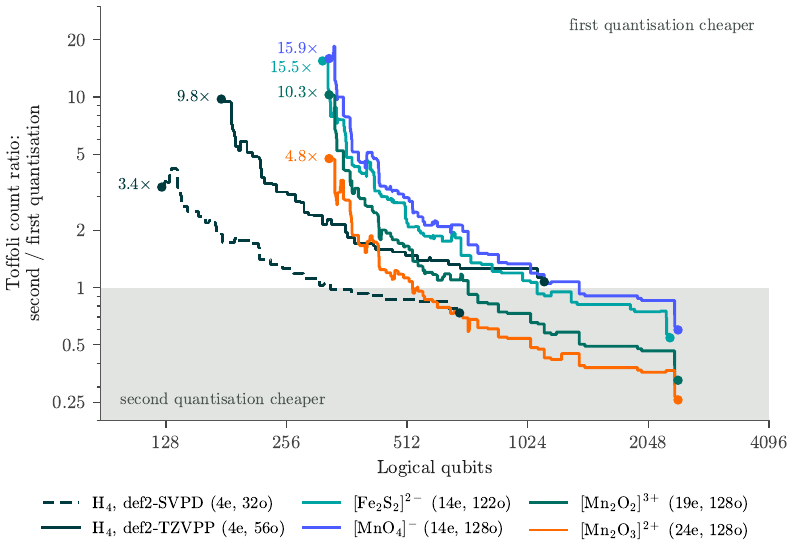}
\caption{Summary of the main result: ratio of the Toffoli counts of quantum phase estimation in second and first quantisation for the six systems of~\cref{sec:resource_estimates} at the same number of logical qubits. Each curve starts at the fewest qubits on which the second-quantised algorithm~\cite{low_denser_2026} can be implemented and ends where it reaches its minimal Toffoli count, beyond which the ratio does not change. In the white region, first quantisation requires fewer Toffoli gates, while second quantisation requires fewer Toffoli gates in the grey region. Note that first quantisation can be implemented using even fewer logical qubits (not shown in this figure), 1.4 to 2.0 times fewer than second quantisation, e.g.\ on 170 instead of 327 logical qubits for [MnO$_4$]$^-$, and it also reaches its minimal Toffoli count on fewer qubits. Both algorithms are compiled at the logical level with the same techniques, assuming the architecture of Ref.~\cite{low_denser_2026}; the underlying costs are shown in~\cref{fig:qpe_fronts_cold}.}
\label{fig:summary}
\end{figure}

In this paper, we develop a first-quantised sum-of-squares spectral amplification (SOSSA) algorithm that works in generic orbital basis sets, and benchmark it against the second-quantised SOSSA algorithm \cite{low_fast_2025,low_denser_2026}, establishing the regime of advantage for both methods. In the important low-qubit and low-filling regime we show that first quantisation can reduce the Toffoli count by an order of magnitude compared to the second-quantised method. Specifically, for six chemical systems with 4 to 24 electrons and up to 128 orbitals, we find that when only a few hundred logical qubits are available, i.e.\ at the fewest logical qubits on which the second-quantised algorithm can be implemented, first quantisation requires 3 to 16 times fewer Toffoli gates at the same number of logical qubits. Part of this advantage comes from the smaller wavefunction register: at the same number of logical qubits, first quantisation has more qubits left for computation, which it uses as ancillae to reduce the Toffoli count. The smaller register also lets first quantisation run on 1.4 to 2 times fewer logical qubits than second quantisation. When each method runs on its minimal number of qubits, first quantisation still requires 1.3 to 3.5 times fewer Toffoli gates. It also remains cheaper up to between several hundred and about 1\,400 logical qubits (\cref{fig:summary}). With more logical qubits, second quantisation keeps trading qubits for Toffoli gates, up to about 2\,400 logical qubits. Its minimal Toffoli count is 1.4 to 3.9 times lower than that of first quantisation, except for H$_4$ in the def2-TZVPP basis set, where first quantisation remains cheaper at every qubit count. For the block-encoding, the advantage of first quantisation at low qubit counts persists up to about 30--40 electrons for 128 orbitals and 60--70 electrons for 256 orbitals.

Our results are achieved by (i) designing factorisation methods of many-body Hamiltonians tailored to first quantisation, (ii) constructing efficient block-encoding strategies, (iii) using a spectral-gap amplification protocol~\cite{zlokapa_hamiltonian_2024,low_fast_2025,king_quantum_2026} and (iv) synthesising the unitaries of the block-encoding as multiplexed unitaries with fewer layers than a full unitary, whose ordering of the layers reduces the Toffoli count. 

While some of these techniques have been explored in second quantisation, extending them to first-quantised Hamiltonians is highly non-trivial. Specifically, in first quantisation, the Hamiltonian is defined on a fixed-particle-number Hilbert space and has a completely different structure to that of a second-quantised Hamiltonian, leading to different factorisations and block-encodings. Applying block-invariant symmetry shifts (BLISS)~\cite{loaiza_block_invariant_symmetry_shift_2023} can result in an indefinite two-body tensor, which a sum of squares with non-negative weights cannot represent. To overcome this, we introduce a \textit{non-convex} representation and accompanying block-encoding strategies. Furthermore, our block-encodings allow for the direct use of the spectral-gap amplification protocol with this non-convex representation, in which some of the squares enter with negative weights. 

Two recent works are related to ours. In Ref.~\cite{dutkiewicz_spectral_2026}, the spectral-gap amplification protocol was applied in first quantisation using an analytical sum-of-squares representation of the Hamiltonian which is specific to the plane-wave basis. Our route is different: we obtain the sum-of-squares representation numerically, by factorising the Hamiltonian in a general orbital basis. Additionally, the method presented in \cite{dutkiewicz_spectral_2026} is all-electron, whereas in an orbital basis we can restrict the calculation to an active space, as is standard in quantum chemistry calculations.
 Very recently, Ref.~\cite{chen_quasiparticle_2026} constructed first-quantised block-encodings in an orbital basis for the Bethe--Salpeter equation, which rely on tensor hypercontraction of the two-body integrals. In contrast, we introduce a new factorisation, the layered-unitary non-convex ansatz (LUNA)~\cref{fig:factorisations}. It builds on double-factorised tensor hypercontraction (DF-THC)~\cite{low_fast_2025}: its rectangular matrices are further factorised by a singular value decomposition whose unitaries are represented by a layered ansatz, and the weights of the squares can have either sign. This factorisation is crucial for an efficient first-quantised algorithm and requires substantially different block-encodings. Unlike Ref.~\cite{chen_quasiparticle_2026}, we combine our block-encodings with the spectral-gap amplification protocol, which reduces the number of calls to the block-encoding in quantum phase estimation (QPE).

The paper is organised as follows. In~\cref{sec:sos}, we introduce the non-convex sum-of-squares representation of the first-quantised Hamiltonian and describe how we obtain it numerically. In~\cref{sec:block-encoding-general}, we construct its block-encoding and combine it with the spectral-gap amplification protocol for QPE. In~\cref{sec:unitary_synthesis}, we describe the synthesis of the multiplexed unitaries, which are the main building block of the block-encoding: their cost model, the orderings of the rotations, and the trainability of unitaries truncated to fewer layers. In~\cref{sec:resource_estimates}, we present resource estimates for six chemical systems and compare them with second quantisation~\cite{low_denser_2026}. In~\cref{sec:conclusions}, discussion and conclusion are presented.

\section{Sum-of-Squares Representation}\label{sec:sos}

Let $N$ be the number of electrons, and $D$ be the number of basis functions. We consider the following Hamiltonian:
\begin{equation}\label{eq:full_ftq_chem_hamiltonian}
    \hat{H}_0 = \alpha  \hat I +  
    \sum_{i=0}^{N-1} \left(\hat T\,\right)_{i}+
    \frac{1}{2}\sum_{i\ne j}^{N-1}\sum_{p,q,r,s=0}^{D-1}h_{pqrs}\left(\hat E_{pq}\right)_{i}  \left(\hat E_{rs}\right)_{j},
\end{equation}
where $\hat I$ is the identity matrix in dimension $D^N$,  $\alpha$ is a constant, $\hat T$ is a $D\times D$ Hermitian matrix, $h_{pqrs}$ is a rank-4 tensor in dimension $D$. A matrix in parentheses with italic subscripts, $i$, means that it is placed in the $i$'th  position in the tensor product with identity matrices in dimension $D$, $\hat I_D$:
\begin{equation}
    (\hat T\,)_i = \hat I_{D}^{\otimes i} \otimes \hat T \otimes \hat I_{D}^{\otimes(N - i - 1)},
\end{equation}
We also introduce the notation for a matrix unit:
\begin{equation}
    \{\hat E_{pq}\}_{rs} = \delta_{rp} \delta_{qs}.
\end{equation}
\Cref{eq:full_ftq_chem_hamiltonian} is the electronic Hamiltonian in a basis of $D$ orthonormal spatial orbitals $\phi_p$. To obtain this, one expands the wavefunction of the $N$ electrons in products of orbitals,
\begin{equation}
    \Psi(\mathbf r_0,\dots,\mathbf r_{N-1})=\sum_{p_0,\dots,p_{N-1}=0}^{D-1} C_{p_0\dots p_{N-1}}\,\phi_{p_0}(\mathbf r_0)\cdots\phi_{p_{N-1}}(\mathbf r_{N-1}),
\end{equation}
and projects the Schr\"odinger equation onto these products. This gives the eigenvalue equation $\hat H_0 C = E\,C$ for the vector of coefficients $C$, of dimension $D^N$~\cite[Sec.~1]{fetter_quantum_1971}. Therefore, for wavefunctions with the fermionic symmetry~\cite[Eq.~(1.6)]{fetter_quantum_1971}, the eigenvalues of~\cref{eq:full_ftq_chem_hamiltonian} are the full configuration interaction energies in this basis, the same as those of the second-quantised Hamiltonian restricted to the $N$-electron symmetry sector.
The matrix $\hat T$ and the tensor $h_{pqrs}=(pq|rs)$ are the conventional one- and two-electron integrals, the same as in second quantisation, and are readily available from quantum chemistry software packages. The constant $\alpha$ is the nuclear repulsion energy; in an active space, the frozen orbitals add their energy to $\alpha$ and their mean-field contribution to $\hat T$. The same representation is used in Ref.~\cite{georges_quantum_2025}. 

The Hamiltonian~\eqref{eq:full_ftq_chem_hamiltonian} has a certain freedom in choosing its matrix elements. We can add to $\hat T$ any potential
\begin{equation}\label{eq:bliss_potential}
    \hat  B = \sum_{pq} b_{pq} \hat E_{pq}
\end{equation}
and subtract a term from $h_{pqrs}$ that compensates for that addition without loss of generality. This is equivalent to the block-invariant symmetry shift (BLISS) used in second quantisation~\cite{loaiza_block_invariant_symmetry_shift_2023,patel_global_minimization_2025}, which leaves the spectrum of the $N$-electron symmetry sector unchanged. This terminology is redundant in first quantisation as we always work with an $N$-particle wavefunction; however, for clarity, we will still refer to this technique as BLISS. Similarly, we shift the Hamiltonian by a constant and compensate for this shift by modifying the two-body term\footnote{One can modify the one-body term, but this would not have an effect on the gap or subnormalisation factor in the block-encoding.}. The Hamiltonian is then
\begin{align}
    \hat H_0 & = (\alpha + \beta N) \hat I + \sum_{i=0}^{N-1} (\hat T + \hat B)_i + \\
    & \frac{1}{2}\sum_{i\neq j}^{N-1}\sum_{pqrs=0}^{D-1} \left(h_{pqrs} -\frac{(b_{pq} + \beta\delta_{pq})\delta_{rs} + \delta_{pq}(b_{rs} + \beta\delta_{rs})}{N-1} \right)(\hat E_{pq})_i (\hat E_{rs})_{j}
\end{align}
Let us also introduce a Hamiltonian with a zero constant shift:
\begin{equation}
    \hat H_1 = \hat{H}_0 - (\alpha + \beta N) \hat I   
\end{equation}
We approximate this Hamiltonian using \textit{a non-convex} sum-of-squares representation:
\begin{equation}
    \label{eq:ham_ansatz}
    \hat H_1 \approx
    \hat H' = -\frac{N^2}{2} \sum_{v=0}^{V-1} \sum_{l=0}^{L-1}\omega_{vl}\alpha_{vl}^2  \hat I + \sum_{j=0}^{N-1} (\hat  W_0)_j  + \frac{1}{2}\sum_{v=0}^{V-1}\sum_{l=0}^{L-1}\omega_{vl} \left(\sum_{j=0}^{N-1} (\hat W^{(v,l)} + \alpha_{vl} \hat I)_{j}\right)^2,
\end{equation}
where $\hat W_0$ is a $D\times D$ matrix,
\begin{align}
    \hat W^{(v,l)} = \sum_{p,q=0}^{D-1} W_{vlpq} \hat E_{pq},\quad W_{vlpq}, \alpha_{vl}, \omega_{vl}\in \mathbb{R}\quad \forall \quad v,l,p,q.
\end{align}
and $V$ and $L$ are some integers. Notice that in~\cref{eq:ham_ansatz} we allow for negative coefficients in front of two-body contributions, unlike previous approaches, because BLISS can make the two-body tensor indefinite.

\begin{figure*}[htb!]
\centering
\definecolor{fdink}{HTML}{003B40}\definecolor{fdorange}{HTML}{FF6A00}\definecolor{fdteal}{HTML}{00A3A3}%
\tikzset{
  fdfull/.style={draw=fdink, fill=white, line width=0.5pt, minimum width=8.4mm, minimum height=5.8mm, inner sep=1pt, font=\footnotesize, text=fdink},
  fdlay/.style={fdfull, fill=fdteal!25},
  fddiag/.style={draw=fdink, fill=white, circle, line width=0.5pt, minimum size=8.6mm, inner sep=0pt, font=\footnotesize, text=fdink},
  fdiD/.style={draw=fdink, line width=0.8pt},
  fdiM/.style={draw=fdorange, line width=0.8pt},
  fdplate/.style={draw=fdink!55, dashed, rounded corners=2mm, line width=0.5pt},
  fdlab/.style={font=\footnotesize, inner sep=1.6pt, text=fdink},
  fdptitle/.style={font=\small, text=fdink},
  fdformula/.style={font=\footnotesize, text=fdink},
}%
\def\fdY{0.8}\def\fdHv{1.62}\def\fdHl{1.36}
\def\fdends#1{
  \node[fdlab, anchor=east] at (0,\fdY) {$p$};  \node[fdlab, anchor=west] at (#1,\fdY) {$q$};
  \node[fdlab, anchor=east] at (0,-\fdY) {$r$}; \node[fdlab, anchor=west] at (#1,-\fdY) {$s$};}%
\begin{tikzpicture}[x=1cm,y=1cm]
\begin{scope}
  \node[fdptitle] at (1.3,2.0) {(a) Exact};
  \fdends{2.6}
  \node[fdfull, minimum height=24mm, minimum width=9mm] (g) at (1.3,0) {$g$};
  \foreach \y in {\fdY,-\fdY} { \draw[fdiD] (0,\y)--(g.west|-0,\y); \draw[fdiD] (g.east|-0,\y)--(2.6,\y); }
  \node[fdformula] at (1.3,-2.1) {$g_{pqrs}$};
\end{scope}
\begin{scope}[xshift=3.6cm]
  \node[fdptitle] at (1.3,2.0) {(b) Single factorisation};
  \fdends{2.6}
  \draw[fdplate] (0.3,-\fdHv) rectangle (2.3,\fdHv);
  \node[fdlab] at (1.3,0) {$\sum_{vl}$};
  \foreach \y in {\fdY,-\fdY} {
    \node[fdfull] (W) at (1.3,\y) {$\hat W^{(v,l)}$};
    \draw[fdiD] (0,\y)--(W.west); \draw[fdiD] (W.east)--(2.6,\y); }
  \node[fdformula] at (1.3,-2.1) {$g_{pqrs}=\sum_{vl}W^{(v,l)}_{pq}W^{(v,l)}_{rs}$};
  \node[fdformula] at (1.3,-2.6) {\cref{eq:single factorisation}};
\end{scope}
\begin{scope}[xshift=7.2cm]
  \node[fdptitle] at (2.2,2.0) {(c) Double factorisation};
  \fdends{4.4}
  \draw[fdplate] (0.3,-\fdHv) rectangle (4.1,\fdHv);
  \node[fdlab] at (2.2,0) {$\sum_{vl}$};
  \foreach \y in {\fdY,-\fdY} {
    \node[fdfull] (A) at (0.95,\y) {$\hat U^{(v,l)}$};
    \node[fddiag] (B) at (2.2,\y) {$\hat F^{(v,l)}$};
    \node[fdfull] (C) at (3.45,\y) {$\hat U^{(v,l)\dagger}$};
    \draw[fdiD] (0,\y)--(A.west) (A.east)--(B.west) (B.east)--(C.west) (C.east)--(4.4,\y); }
  \node[fdformula] at (2.2,-2.1) {$\hat W^{(v,l)}=\hat U^{(v,l)}\hat F^{(v,l)}\hat U^{(v,l)\dagger}$};
\end{scope}
\begin{scope}[xshift=12.6cm]
  \node[fdptitle] at (2.05,2.0) {(d) Tensor hypercontraction};
  \fdends{4.1}
  \draw[fdplate] (1.47,-\fdHl) rectangle (2.63,\fdHl);
  \node[fdlab] at (2.05,0) {$\sum_{vl}$};
  \foreach \y in {\fdY,-\fdY} {
    \node[fdfull] (A) at (0.9,\y) {$\hat R$};
    \node[fddiag] (B) at (2.05,\y) {$\hat F^{(v,l)}$};
    \node[fdfull] (C) at (3.2,\y) {$\hat R^{\dagger}$};
    \draw[fdiD] (0,\y)--(A.west) (C.east)--(4.1,\y);
    \draw[fdiM] (A.east)--(B.west) (B.east)--(C.west); }
  \node[fdformula] at (2.05,-2.1) {$\hat W^{(v,l)}=\hat R\,\hat F^{(v,l)}\hat R^{\dagger}$};
\end{scope}
\begin{scope}[xshift=0.6cm,yshift=-5.5cm]
  \node[fdptitle] at (2.05,2.0) {(e) DF-THC};
  \fdends{4.1}
  \draw[fdplate] (0.3,-\fdHv) rectangle (3.8,\fdHv);
  \node[fdlab, anchor=north west] at (0.36,\fdHv) {$\sum_v$};
  \draw[fdplate] (1.47,-\fdHl) rectangle (2.63,\fdHl);
  \node[fdlab] at (2.05,0) {$\sum_l$};
  \foreach \y in {\fdY,-\fdY} {
    \node[fdfull] (A) at (0.9,\y) {$\hat R^{(v)}$};
    \node[fddiag] (B) at (2.05,\y) {$\hat F^{(v,l)}$};
    \node[fdfull] (C) at (3.2,\y) {$\hat R^{(v)\dagger}$};
    \draw[fdiD] (0,\y)--(A.west) (C.east)--(4.1,\y);
    \draw[fdiM] (A.east)--(B.west) (B.east)--(C.west); }
  \node[fdformula] at (2.05,-2.1) {$\hat W^{(v,l)}=\hat R^{(v)}\hat F^{(v,l)}\hat R^{(v)\dagger}$};
  \node[fdformula] at (2.05,-2.6) {\cref{eq:joint_diagonalization}};
\end{scope}
\begin{scope}[xshift=6.3cm,yshift=-5.5cm]
  \node[fdptitle] at (4.55,2.0) {(f) This work: LUNA};
  \fdends{9.1}
  \draw[fdplate] (0.3,-\fdHv) rectangle (8.8,\fdHv);
  \node[fdlab, anchor=north west] at (0.36,\fdHv) {$\sum_v$};
  \draw[fdplate] (3.87,-\fdHl) rectangle (5.23,\fdHl);
  \node[fdlab] at (4.55,0) {$\sum_l\omega_{vl}$};
  \foreach \y in {\fdY,-\fdY} {
    \node[fdlay]  (A) at (0.95,\y) {$\hat U^{(v)}$};
    \node[fddiag] (B) at (2.15,\y) {$\hat D^{(v)}$};
    \node[fdlay]  (C) at (3.35,\y) {$\hat V^{(v)\dagger}$};
    \node[fddiag] (F) at (4.55,\y) {$\hat F^{(v,l)}$};
    \node[fdlay]  (E) at (5.75,\y) {$\hat V^{(v)}$};
    \node[fddiag] (G) at (6.95,\y) {$\hat D^{(v)}$};
    \node[fdlay]  (H) at (8.15,\y) {$\hat U^{(v)\dagger}$};
    \draw[fdiD] (0,\y)--(A.west) (A.east)--(B.west) (G.east)--(H.west) (H.east)--(9.1,\y);
    \draw[fdiM] (B.east)--(C.west) (C.east)--(F.west) (F.east)--(E.west) (E.east)--(G.west); }
  \draw[decorate, decoration={brace, mirror, amplitude=1.5mm}, draw=fdink!70] (0.5,-1.8)--(3.8,-1.8)
     node[fdlab, midway, below=2mm] {$\hat R^{(v)}=\hat U^{(v)}\hat D^{(v)}\hat V^{(v)\dagger}$};
  \node[fdformula] at (2.15,-2.68) {\cref{eq:R_svd}};
  \draw[decorate, decoration={brace, mirror, amplitude=1.5mm}, draw=fdink!70] (5.3,-1.8)--(8.6,-1.8)
     node[fdlab, midway, below=2mm] {$\hat R^{(v)\dagger}$};
\end{scope}
\begin{scope}[yshift=-9.1cm,
    lab/.style={fdlab, text height=1.6ex, text depth=0.4ex}, 
    slot/.style={inner sep=0pt, outer sep=0pt, minimum width=7mm, minimum height=4.4mm},
  ]
  \def\legsg{1mm} 
  \def\legig{5mm}   

  \node[fdfull, minimum width=6mm, minimum height=4.2mm] (s1) at (0,0) {};
  \node[lab, right=\legsg of s1] (l1) {matrix};

  \node[fddiag, minimum size=4.6mm, right=\legig of l1] (s2) {};
  \node[lab, right=\legsg of s2] (l2) {diagonal matrix};

  \node[fdlay, minimum width=6mm, minimum height=4.2mm, right=\legig of l2] (s3) {};
  \node[lab, right=\legsg of s3] (l3) {unitary of $N_S$ layers};

  \node[slot, right=\legig of l3] (s4) {};
  \draw[fdiD] (s4.west) -- (s4.east);
  \node[lab, right=\legsg of s4] (l4) {index with $D$ values};

  \node[slot, right=\legig of l4] (s5) {};
  \draw[fdiM] (s5.west) -- (s5.east);
  \node[lab, right=\legsg of s5] (l5) {$M$ values};

  \node[slot, minimum width=8mm, right=\legig of l5] (s6) {};
  \draw[fdplate, rounded corners=0.7mm] (s6.south west) rectangle (s6.north east);
  \node[lab, right=\legsg of s6] (l6) {sum};
\end{scope}
\end{tikzpicture}
\caption{Factorisations of the two-body tensor $g_{pqrs}$. A box is an array of
numbers with one index for each line attached to it, so a box with two lines is a matrix. A line that joins two
boxes is summed over. The four free ends are the indices of $g_{pqrs}$: $p$ and $q$ on the upper line belong to
one electron, $r$ and $s$ on the lower line to the other. Read from left to right, each line is a product of
matrices. Circles are diagonal matrices. A dashed frame with a sum sign acts as a bracket with this sum in front
of it: it encloses exactly the matrices that depend on the indices of the sum. For example, panel (e) reads
$g=\sum_v(\hat R^{(v)}\otimes\hat R^{(v)})\bigl[\sum_l\hat F^{(v,l)}\otimes\hat F^{(v,l)}\bigr](\hat R^{(v)}\otimes\hat R^{(v)})^\dagger$,
where the first factor of each product $\otimes$ acts on the upper line and the second on the lower line. (a) The exact tensor. (b) Single factorisation~\cite{berry_qubitization_2019}: a sum of terms, numbered by the pair $(v,l)$,
each a product of one matrix $\hat W^{(v,l)}$ for each electron. (c) Double factorisation~\cite{motta_low_2021,von_burg_quantum_2021}: each
$\hat W^{(v,l)}$ is diagonalised by its own unitary $\hat U^{(v,l)}$, so all matrices are inside the frame. (d)
Tensor hypercontraction~\cite{lee_even_2021}: one rectangular matrix $\hat R$ serves all terms, so only the diagonal matrices
$\hat F^{(v,l)}$ are inside the frame. (e) DF-THC~\cite{low_fast_2025} combines the two: the terms form $V$ groups, and the
terms of a group share one matrix $\hat R^{(v)}$. The sum over $v$ therefore encloses everything, and the sum
over $l$ only $\hat F^{(v,l)}$. (f) This work, the layered-unitary non-convex ansatz (LUNA), builds on DF-THC. Each $\hat R^{(v)}$ of (e) is further factorised by a singular value decomposition,
with the singular values in $\hat D^{(v)}$. Its unitaries (shaded) are sequences of $N_S$ layers of multiplexed
rotations. The weights $\omega_{vl}$ of the inner sum can have either sign and appear only here.}
\label{fig:factorisations}
\end{figure*}
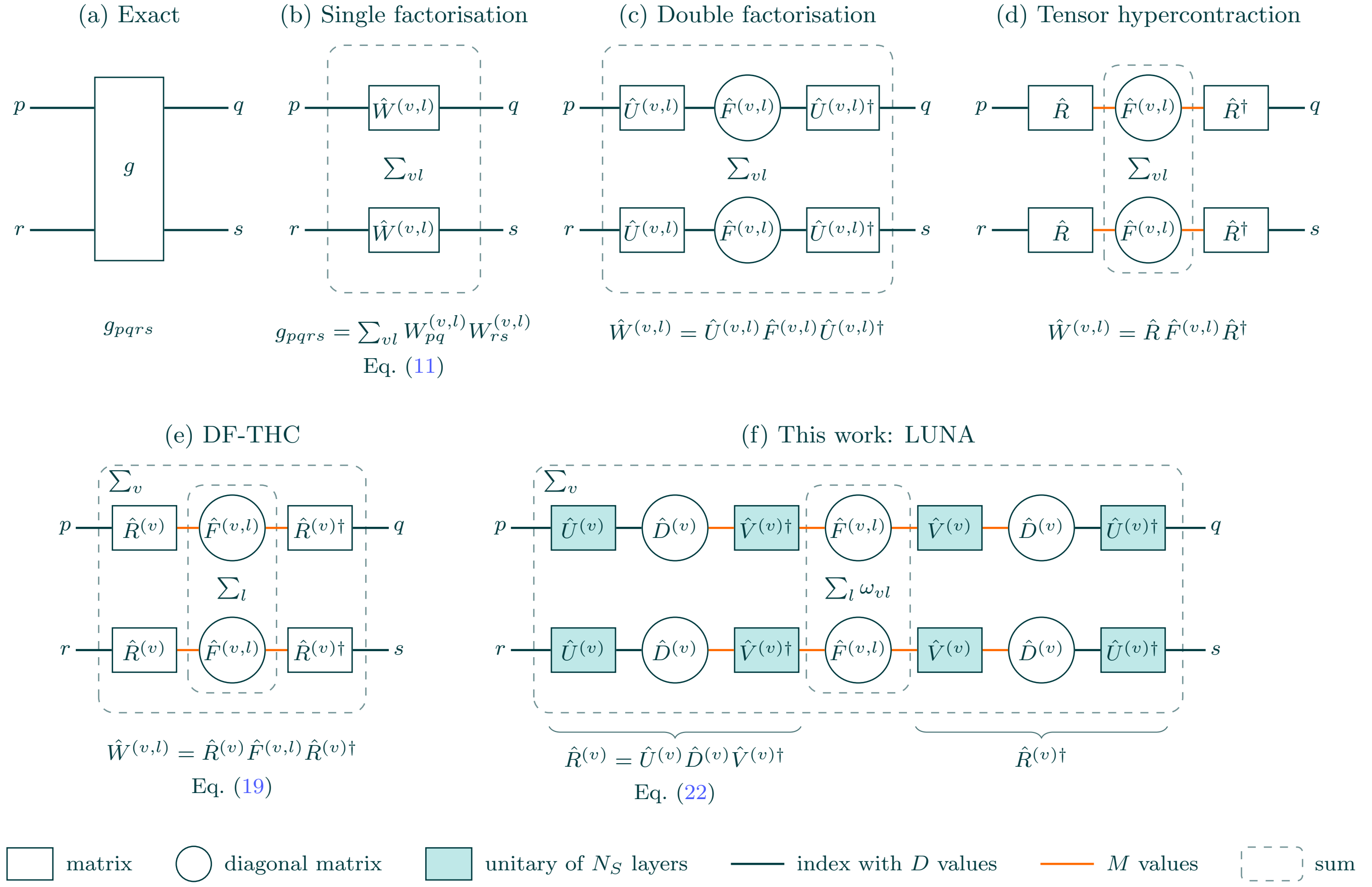

Let us begin by introducing the single factorisation (illustrated in  \Cref{fig:factorisations}a\&b)
\begin{equation}
\label{eq:single factorisation}
    g_{pqrs} = \sum_{vl}^{V,L} \omega_{vl} W_{vlpq} W_{vlrs},
\end{equation}
and expand the last bracket in \eqref{eq:ham_ansatz} to see how it relates to the original Hamiltonian
\begin{align}
    \hat H' & = \sum_{j=0}^{N-1} \left(\hat W_0 + N\sum_{vl}^{V,L}\omega_{vl}\alpha_{vl} \hat W^{(v,l)}  \right)_j + \frac{1}{2}\sum_{ij}^{N}\sum_{pqrs}^{D} g_{pqrs} (\hat E_{pq})_i (\hat E_{rs})_j \\
     & = \sum_{j=0}^{N-1} \left(\hat W_0 + N\sum_{vl}^{V,L}\omega_{vl}\alpha_{vl} \hat W^{(v,l)} + \frac{1}{2} \sum_{pqr}^{D} g_{prrq} \hat E_{pq} \right)_j + \frac{1}{2}\sum_{i\neq j}^{N} \sum_{pqrs}^{D} g_{pqrs} (\hat E_{pq})_i (\hat E_{rs})_j
\end{align}
To approximate $\hat H_1$, we should choose parameters of $\hat H'$ such that:
\begin{align} \label{eq:gpqrs}
    g_{pqrs} & = h_{pqrs}  - \frac{(b_{pq} + \beta\delta_{pq})\delta_{rs} + \delta_{pq}(b_{rs} + \beta\delta_{rs})}{N-1}, \\
    \label{eq:one-body-approximate-W_0}
    \hat W_0 & =  \hat T + \hat B - N \sum_{vl=0}^{V-1,L-1} \omega_{vl}\alpha_{vl} \hat W^{(v,l)} -\frac{1}{2} \sum_{pqr=0}^{D-1} g_{prrq} \hat E_{pq}.
\end{align}
The relevant operator which allows for the gap amplification protocol for qubitisation-based quantum phase estimation~\cite{low_fast_2025} is
\begin{align}
    \frac{\hat H_0 + \Omega}{\Lambda} - 1,
    \label{eq: ShiftedHamiltonian}
\end{align}
where $\Omega$ and $\Lambda$ are  constants. In~\cref{sec:block-encoding-general}, we will introduce the block-encoding of this operator with $\Lambda$ and $\Omega$ defined as:
\begin{align}
    \label{eq:Omega}
    \Omega = & \frac{N^2}{2} \left(\sum_{v,l}^{V-1,L-1}\omega_{vl}\alpha_{vl}^2 + \frac{1}{2}\sum_{v,l=0}^{V-1,L-1}(|\omega_{vl}| - \omega_{vl})\epsilon_{vl}^2\right) - (\delta_0 + \beta) N - \alpha,\\
    \label{eq:Lambda}
    \Lambda = &  \frac{N (\delta_{D-1} - \delta_0)}{2} +  \sum_{vl=0}^{V-1, L-1} \frac{|\omega_{vl}| \epsilon_{vl}^2 N^2}{4},
\end{align}
where  $\epsilon_{vl}$ is the subnormalisation factor in the block-encoding of $\hat W^{(v,l)} + \alpha_{vl}\hat I$ (it can be the maximum singular value); $\delta_0$ and $\delta_{D-1}$ are the lowest and the largest eigenvalues of $\hat W_0$.
Equations~\eqref{eq:Omega} and~\eqref{eq:Lambda} justify adding the BLISS constant, $\beta$, to a two-body term. If it were added to a one-body term instead, it would effectively shift $\delta_0$; however, this shift would be cancelled by the $\beta N$ term in the definition of $\Omega$~\eqref{eq:Omega}, see \eqref{eq:effective_norm}. The one-body contribution to the subnormalisation factor $\Lambda$ depends on the difference between the lowest and highest eigenvalues, which is invariant under a constant shift. Because we modify a two-body term, we can then optimise $\Lambda$ and $\Omega$ with respect to $\beta$.
One needs to find block-encodings of matrices $ \hat W^{(v,l)}$ with optimal subnormalisation factors. If one uses QROM-based approaches, this requires loading an $O(VLD^2)$ amount of information. To reduce this, one can separate matrices into $V$ groups, and perform joint factorisation within each group.
\begin{equation}\label{eq:joint_diagonalization}
    \hat W^{(v,l)} = \hat R^{(v)} \hat F^{(v,l)} \hat R^{(v)\dagger},
\end{equation}
where $\hat F^{(v,l)}$ is diagonal for each $v,l$ 
\begin{equation}
    \hat F^{(v,l)} = \sum_{m=0}^{M-1}f_{vlm}\hat E_{mm},
\end{equation}
and $\hat R^{(v)}$ are rectangular matrices for each $v$:
\begin{equation}
    \hat R^{(v)} = \sum_{p=0}^{D-1}\sum_{m=0}^{M-1}R_{vpm} \hat E_{pm}.
\end{equation}
This would lead to $O(VDM+VLM)$ amount of information, the first contribution coming from $\hat R^{(v)}$ and the second from $\hat F^{(v,l)}$. Generally, a good approximate joint factorisation is possible with $M=O(D)$, which reduces the data loading. This factorisation of the rank four tensor is exactly DF-THC factorisation~\cite{low_fast_2025} and covers double factorisation
(DF)~\cite{motta_low_2021,von_burg_quantum_2021} and its compressed variants~\cite{cohn_quantum_2021,rocca_reducing_the_runtime_2024,oumarou_accelerating_quantum_2024}, and tensor-hypercontraction (THC)~\cite{lee_even_2021} as particular cases. However, the sum-of-squares representation in first quantisation is constructed using tensor product of a Hermitian matrix with identity matrices, and the subnormalisation factor is an upper bound on spectral norm over $N$-particle Hilbert space, while in second quantisation the subnormalisation factor in block-encoding bounds the spectral norm over all possible number-of-electron sectors.
In order to reduce the implementation cost of block-encoding and its subnormalisation factor further, we impose additional structure on the $\hat R^{(v)}$ matrices. Namely, we first consider their explicit SVD decomposition:
\begin{equation}\label{eq:R_svd}
    \hat R^{(v)} = \hat U^{(v)} \hat D^{(v)} \hat V^{(v)\dagger},
\end{equation}
where $\hat U^{(v)}$, and $\hat V^{(v)\dagger}$ are unitary matrices and $\hat D^{(v)}$ is a diagonal rectangular matrix. Since $\hat R^{(v)}$ is real, these unitaries can be chosen real, i.e.\ orthogonal, with determinant $+1$: changing the sign of a column of $\hat U^{(v)}$ or $\hat V^{(v)}$ is compensated by changing the sign of the corresponding diagonal element of $\hat D^{(v)}$, which we therefore allow to be negative.
We use the SVD in~\cref{eq:R_svd} to reduce the cost of the first-quantised block-encoding: In first quantisation, $\hat R^{(v)}$ acts on the register of one electron and has to be block-encoded as a matrix (\cref{sec: be_hamiltonian}). Without the SVD, one could embed $\hat R^{(v)}$ in a larger unitary and implement this unitary directly. For $M\ge D$, the larger unitary has dimension $2M$ and requires $2M(2M-1)/2$ rotation angles. The SVD requires two unitaries of dimensions $D$ and $M$ and $D$ singular values, i.e.\ at most $M(M-1)/2+D(D-1)/2+D \leq M^2$ rotation angles, about a factor of two fewer.  In second quantisation, the block-encodings are different: each column of $\hat R^{(v)}$ defines a rotated orbital, which is implemented by its own sequence of Givens rotations~\cite{von_burg_quantum_2021,lee_even_2021,low_fast_2025}, and $\hat R^{(v)}$ is never block-encoded as a matrix.

To reduce the cost of the unitaries $\hat{U}^{(v)}$ and $\hat{V}^{(v)}$ in \cref{eq:R_svd} further, we use the layered ansatz~\eqref{eq:unitary ansatz} with fewer layers than a general unitary requires (\cref{sec:unitary_synthesis}). 
Any such matrix $\hat U$ can be decomposed into layers of multiplexed single-qubit rotations using, for example, cosine-sine decomposition. Generally, for a $D\times D$ unitary, we need to implement $N_{S}$ layers, where $N_{S}=2^{\lceil\log_2D\rceil}-1$, and to specify $O(D)$ rotation angles for each layer:
\begin{equation}
\label{eq:unitary ansatz}
    \hat U = \prod_{l=0}^{N_S-1} \hat Q_{s(l)}(\vec{\theta}_l),
\end{equation}
where $s(l)$ is a function mapping an index $l$ into a qubit number, and 
\begin{equation}\label{eq:multiplexor}
    \hat Q_{s}(\vec\theta)
    \;=\;
    \sum_{c=0}^{2^{d-1}-1}
    \bigl(\hat R_y(\theta_c)\bigr)_{s}\;\bigl(\ketbra{c}{c}\bigr)_{\bar s}, \quad d = \lceil\log_2D\rceil,
\end{equation}
where $(\,\cdot\,)_{s}$ means the operator acts on qubit $s$ and $(\,\cdot\,)_{\bar s}$ on
the other $d-1$ qubits, and $\hat R_y(\theta)=e^{-i\theta\hat Y}$ rotates the two basis states of qubit $s$ by the angle $\theta\in[0,2\pi)$. Because we approximate the Hamiltonian only to within a certain accuracy, we can treat $N_S$, rotations angles $\vec{\theta}_{l}$ and target-qubit ordering $s(l)$ as free parameters. Thus, the ansatz for matrices $\hat R^{(v)}$ is
\begin{equation}
    \hat R = \left(\prod_{l=0}^{N_S-1} \hat Q_{s(l)}(\vec{\theta}_{l,1})\right)\;\hat D \;\left(\prod_{l=0}^{N_S-1} \hat Q_{s(l)}(\vec{\theta}_{l,2})\right)
\end{equation}
Because the implementation of unitaries $\left(\prod_{l=0}^{N_S-1} \hat Q_{s(l)}(\vec{\theta}_{l,1})\right)$ is the dominant cost in the algorithm, we will discuss it separately in~\cref{sec:unitary_synthesis}. \Cref{fig:factorisations} compares the resulting factorisation with single factorisation, double factorisation, tensor hypercontraction and DF-THC. To our knowledge, this factorisation has not been considered before. We call it the layered-unitary non-convex ansatz (LUNA). It builds on DF-THC: each $\hat R^{(v)}$ is further factorised by an SVD with layered unitaries~\eqref{eq:unitary ansatz}, and the weights $\omega_{vl}$ can have either sign.

\subsection{Employing unitary invariance}
The Hamiltonian is defined up to some unitary conjugation. We shall explore this unitary freedom to reduce the quantum resources for block-encoding. When seeking a factorised Hamiltonian, we can diagonalise the one-body matrix~\eqref{eq:one-body-approximate-W_0} with a unitary $\hat U_{\rm o}$, and use $\hat U_{\rm global} = \bigotimes_{j} \left(\hat U_{\rm o}\right)_j$ to work with the new Hamiltonian $\hat{\tilde{H}} = \hat U_{\rm global}^{\dagger} \hat H \hat U_{\rm global}$ which has diagonal one-body term. This reduces the number of unitaries one needs to implement and simplifies the quantum circuit. If we use $N_S < 2^{\lceil\log_2D\rceil}-1$, then we need to be sure that the unitary matrices we implement have this truncated depth in the new rotated frame. In order to ensure this,  we work in the original frame, but treat $\hat U_{\rm o}$ as a variational orthogonal matrix and write it explicitly in the SVD of $\hat R^{(v)}$~\eqref{eq:R_svd} and in the BLISS potential matrix $\hat B$~\eqref{eq:bliss_potential}:
\begin{align}
    \label{eq:R_rotated}
    \hat R^{(v)} & = \hat U_{\rm o}\,\hat{\tilde U}^{(v)} \hat D^{(v)} \hat V^{(v)\dagger},\\
    \label{eq:B_rotated}
    \hat B & = \hat U_{\rm o}\,\hat{\tilde B}\,\hat U_{\rm o}^{\dagger},
\end{align}
where $\hat{\tilde U}^{(v)} = \hat U_{\rm o}^{\dagger}\hat U^{(v)}$ is the unitary of~\eqref{eq:R_svd} in the rotated frame, and the same $\hat U_{\rm o}$ is used for all $v$. We fix the off-diagonal elements of $\hat{\tilde B}$ such that the one-body term is diagonal in the rotated frame, $\hat U_{\rm o}^{\dagger}\hat W_0\hat U_{\rm o} = {\rm diag}(w_0,\dots,w_{D-1})$:
\begin{equation}\label{eq:B_offdiag}
    \tilde b_{pq} = -\left[\hat U_{\rm o}^{\dagger}\left(\hat T - N \sum_{vl=0}^{V-1,L-1} \omega_{vl}\alpha_{vl} \hat W^{(v,l)} - \frac{1}{2}\hat G\right)\hat U_{\rm o}\right]_{pq}, \quad p\neq q,
\end{equation}
where $\hat G = \sum_{pqr} g_{prrq}\hat E_{pq}$. The constraint is linear in $\hat{\tilde B}$, so it is satisfied exactly at each step of the optimisation. The diagonal elements $\tilde b_{pp}$, $\beta$, $\hat U_{\rm o}$ and the parameters of the factorisation remain free. The number of variational parameters is unchanged: $\hat U_{\rm o}$ adds $D(D-1)/2$ parameters and~\cref{eq:B_offdiag} removes the same number. Then $\delta_0 = \min_p w_p$ and $\delta_{D-1} = \max_p w_p$, and the block-encoding of the one-body term requires no unitary.

We impose the depth $N_S$ on both unitaries of the SVD, $\hat{\tilde U}^{(v)}$ and $\hat V^{(v)}$, which are the ones implemented in the rotated frame, $\hat U_{\rm o}^{\dagger}\hat R^{(v)} = \hat{\tilde U}^{(v)} \hat D^{(v)} \hat V^{(v)\dagger}$. The unitary $\hat U_{\rm o}$ is never used for block-encoding; it only enters the classical preparation of the initial state if it was given in the original frame. For $N_S = 2^{\lceil\log_2D\rceil}-1$, the ansatz is equivalent to the original one: for any solution $\hat R^{(v)} = \hat U^{(v)}\hat D^{(v)}\hat V^{(v)\dagger}$ found in the original frame, we can take $\hat U_{\rm o}$ that diagonalises $\hat W_0$, and then $\hat{\tilde U}^{(v)} = \hat U_{\rm o}^{\dagger}\hat U^{(v)}$ is again a full-depth unitary, only with different angles, and $\hat{\tilde B} = \hat U_{\rm o}^{\dagger}\hat B\hat U_{\rm o}$ satisfies~\cref{eq:B_offdiag}.

\subsection{Finding Optimal Sum-of-Squares Representation}
In order to find the optimal Hamiltonian representation, several objective functions and factorisations were explored in the literature~\cite{lee_even_2021,cohn_quantum_2021,rocca_reducing_the_runtime_2024,oumarou_accelerating_quantum_2024,caesura_faster_quantum_chemistry_2025,low_fast_2025}. We modify the objective function from Ref.~\cite{low_fast_2025} by an additional, smoothed HF term and also modify the $E_{\rm gap} = E_0 + \Omega$ term as follows:
\begin{equation}
\label{eq: objective_function}
    \mathcal{L} = \frac{\|g - h' \|}{\epsilon_{\rm fro}} + \frac{\Lambda}{\epsilon_{\rm \Lambda}} + \frac{{\rm ReLU}(E_{\rm gap})}{\epsilon_{\rm gap}} + \sqrt{\frac{\left(E^{0}_{\rm HF} - E_{\rm HF}\right)^2}{\epsilon^2_{\rm HF}} + \eta_{\rm HF}}.
\end{equation}

The empirical regularisation parameters $\epsilon_{\rm fro}$, $\epsilon_{\Lambda}$, $\epsilon_{\rm gap}$, and $\epsilon_{\rm HF}$ combine different optimisation objectives.
The first term, with
\begin{equation}
    h'_{pqrs} = h_{pqrs} - \frac{(b_{pq} + \beta\delta_{pq})\delta_{rs} + \delta_{pq}(b_{rs} + \beta\delta_{rs})}{N-1},
\end{equation}
favours an accurate factorisation, by aiming to fulfill \eqref{eq:gpqrs}.
The second and third terms favour a low effective subnormalisation $\lambda_\text{eff}\approx\sqrt{2E_\text{gap}\Lambda}$, i.e.~large spectral amplification, see \cref{sec: QPE}. In the optimisation, we use the UCCSD(T) energy for $E_0$ in $E_{\rm gap}$.

In Ref.~\cite{low_fast_2025}, the gap enters the objective function as
\begin{equation}\label{eq:gap_term_low}
    {\rm ReLU}\left(\frac{E_{\rm gap}-E_{\rm reg}}{E_{\rm reg}}\right),
\end{equation}
which vanishes once $E_{\rm gap}$ is below the target $E_{\rm reg}$. There, $E_{\rm reg}$ is equal to, or slightly larger than, the gap computed beforehand, or it is scanned over a range of values when that gap is not available. Our third term has no target, so that no knowledge of the gap is needed before the optimisation.

The last term is a smoothed absolute value of the difference between the exact Hartree--Fock (HF) energy, $E^{0}_{\rm HF}$, and the HF energy, $E_{\rm HF}$, estimated from approximate integrals but with the exact HF orbitals. Here, the exact HF orbitals are the occupied orbitals of the spin-unrestricted HF solution of the unfactorised Hamiltonian, which we keep fixed during the optimisation, and $E^{0}_{\rm HF}$ is the corresponding HF energy. $E_{\rm HF}$ can be explicitly calculated using the conventional spin-unrestricted equation~\cite{mayer2003simple}; we give its expression in terms of the factorisation parameters in~\cref{app:hf_energy}. The smoothing parameter $\eta_{\rm HF}>0$ serves to replace $|x|$ by $\sqrt{x^2+\eta_{\rm HF}}$, whose gradient vanishes at $x=0$. Without it, the gradient of the HF term does not decrease as $E_{\rm HF}$ approaches $E^0_{\rm HF}$, and it dominates the gradient of the objective even when the HF energy is already accurate. We found that the last term is crucial for finding physical solutions. Previous work focused on estimating errors purely from the correlation energy approximated at a certain level of theory. However, QPE estimates the total energy, and as discussed in Ref.~\cite{low_fast_2025}, solutions obtained from the original minimisation procedure ``may be representing a Hamiltonian with fundamentally different physics''~\cite{low_fast_2025} as the error in the total energy can be an order of magnitude larger than required for chemical accuracy (see Table~VI of Ref.~\cite{low_fast_2025} and the discussion there). This additional term in the objective function circumvents this problem. We test this on the Fe$_2$S$_2$ Hamiltonian of Ref.~\cite{li_spin-projected_2017}, which is used in Ref.~\cite{low_fast_2025}, in~\cref{app:sq_validation_low}.

For the rotated ansatz,~\cref{eq:R_rotated,eq:B_rotated}, we evaluate all terms of~\cref{eq: objective_function} in the original frame: we build $\hat R^{(v)}$ and $\hat B$ from the rotated-frame parameters, with the off-diagonal elements of $\hat{\tilde B}$ given by~\cref{eq:B_offdiag}, so that $h'$, $\Lambda$, $E_{\rm gap}$ and $E_{\rm HF}$ are computed as before and the HF orbitals do not need to be rotated. We parametrise $\hat U_{\rm o} = e^{\hat K}\hat U_{\rm ref}$, where $\hat K$ is a real antisymmetric matrix with $\hat K=0$ at the start of the optimisation, and $\hat U_{\rm ref}$ is a fixed orthogonal matrix, the eigenvectors of $\hat W_0$ of the full-depth solution we start from. The variational parameters are $\beta$, $\tilde b_{pp}$, $\alpha_{vl}$, $\omega_{vl}$, $f_{vlm}$, the rotation angles and singular values of the SVD of $\hat R^{(v)}$, and $\hat K$.

We minimise $\mathcal{L}$ with the Adam optimiser~\cite{kingma_adam_2015}. The learning-rate schedule, the initial guess and the stages of the optimisation, which we refer to as our production protocol, are given in~\cref{app:factorisation_search}.

\section{Block-Encoding}\label{sec:block-encoding-general}

\subsection{Quantum Phase Estimation with Spectral Amplification}\label{sec: QPE}
In order to estimate an eigenvalue of the Hamiltonian, we will use the quantum phase estimation algorithm~\cite{Kitaev1995} with qubitisation~\cite{berry_improved_techniques_preparing_2018,poulin_quantum_2018,low_hamiltonian_simulation_qubitization_2019}. Recently it was shown that if the Hamiltonian is positive semi-definite and one has oracle access to the ``square root'' of the Hamiltonian, one can use the spectral amplification protocol~\cite{zlokapa_hamiltonian_2024,low_fast_2025,king_quantum_2026}: QPE on the walk operator resulting from qubitising the shifted Hamiltonian's block encoding \eqref{eq: ShiftedHamiltonian} estimates $x=\arccos\left(\frac{E+\Omega}{\Lambda}-1\right)$ for an energy $E$. Based on a targeted state with energy $E_0$, the ``effective normalisation''
\begin{equation}
\label{eq:effective_norm}
    \lambda_\text{eff} := \left|\frac{dE}{dx}\right| = \sqrt{E_\text{gap}(2\Lambda-E_\text{gap})},\qquad E_\text{gap} = E_0+\Omega>0
\end{equation}
determines how accurately $x$ must be measured to determine $E$. Using the derivative at point $E_0$ to estimate the error propagation is justified when the desired energy error $\sigma_\text{QPE}\ll E_\text{gap}$.
Then, the required number of calls $N_W$ to the quantum walk operator for a desired $\sigma_\text{QPE}$ standard deviation uncertainty of the energy is~\cite{babbush_encoding_2018}
\begin{equation}\label{eq: no qpe calls}
    N_W = \frac{\pi\lambda_\text{eff}}{2\sigma_\text{QPE}}.
\end{equation}

As noted in Refs.~\cite{low_fast_2025,king_quantum_2026}, the amplification arises from the non-linearity of the eigenphases $x(E)$ of the walk operator near $-1$, so spectral amplification becomes stronger for block-encodings of $(\hat H+\Omega)/\Lambda-\hat I$ with small ratio
\begin{equation}
\label{eq: small ratio}
 E_{\rm gap}/\Lambda \ll 1.
\end{equation}
In previous work, this block-encoding was constructed from a sum of squares with non-negative weights, by applying a Chebyshev polynomial of second order. Our key observation is that negative weights can be included: the square of a block-encoded operator enters with a negative weight by changing the sign of its block-encoding~\cite{von_burg_quantum_2021,rocca_reducing_the_runtime_2024,ivanov_quantum_computation_of_electronic_2025}, and the result is still a block-encoding of the form~\eqref{eq: ShiftedHamiltonian}, at the cost of an additional shift in $\Omega$. This allows us to use the non-convex sum-of-squares representation of~\cref{eq:ham_ansatz}, which is needed when BLISS makes the two-body tensor indefinite. Moreover, factorisations with non-negative weights are a special case of this representation, and the additional shift due to negative weights enters $\Omega$ and hence $\lambda_{\rm eff}$, which we minimise. Therefore, by the variational principle, the optimal non-convex representation is never worse than the optimal one with non-negative weights, although a numerical optimisation is only guaranteed to find a local minimum. We outline such a block-encoding in the next subsection.

\subsection{Block-Encoding of the Hamiltonian}\label{sec: be_hamiltonian}
We block-encode the Hamiltonian in the rotated frame,~\cref{eq:R_rotated,eq:B_rotated}, in which the one-body term is diagonal. To simplify the notation, in this section we drop the tildes: $\hat R^{(v)}=\hat U^{(v)}\hat D^{(v)}\hat V^{(v)\dagger}$ stands for the rotated-frame factor $\hat U_{\rm o}^{\dagger}\hat R^{(v)}=\hat{\tilde U}^{(v)}\hat D^{(v)}\hat V^{(v)\dagger}$, and $\hat W_0={\rm diag}(w_0,\dots,w_{D-1})$ for the diagonal one-body matrix $\hat U_{\rm o}^{\dagger}\hat W_0\hat U_{\rm o}$. Its diagonal elements $w_p$ are the eigenvalues of $\hat W_0$ in the order of the rotated orbitals, which is in general not ascending, while $\delta_0$ and $\delta_{D-1}$ of~\cref{eq:Omega,eq:Lambda} are the lowest and the largest of them.

For the actual implementation of the block-encoding, we absorb absolute values of $\omega_{vl}$, and the shift $\alpha_{vl}$ into the definition of $\hat W^{(v,l)}$, and leave only the phases $(-1)^{\theta_{vl}}={\rm sign}(\omega_{vl})$ in front of the sum of squares. That is, in this section $\hat W^{(v,l)}$ stands for $\sqrt{|\omega_{vl}|}\,(\hat W^{(v,l)}+\alpha_{vl}\hat I)$ of~\cref{eq:ham_ansatz}, and $\hat F^{(v,l)}$ and $\alpha_{vl}$ stand for $\sqrt{|\omega_{vl}|}\,\hat F^{(v,l)}$ and $\sqrt{|\omega_{vl}|}\,\alpha_{vl}$. Thus, our task is to block-encode the factorised indefinite Hamiltonian:
\begin{align}
    \label{eq: hamiltonian_for_be}
    & \hat H'' = \sum_{v=0}^{V-1}\sum_{l=0}^{L-1} \frac{(-1)^{\theta_{vl}}}{2}\left(\sum_{j=0}^{N-1}\left(\hat W^{(v,l)} \right)_j \right)^2 + \sum_{j=0}^{N-1}\left(\hat W_0\right)_j, \\
    \label{eq: factorizations_two_body}
    &\hat W^{(v,l)} = \hat R^{(v)} \hat F^{(v,l)} \hat R^{(v)\dagger} +\alpha_{vl}\hat I, \quad (v, l) \in [V-1]_0 \times [L-1]_0.
\end{align}
Because only the square of $\hat W^{(v,l)}$ enters $\hat H''$, we can change its sign, $\hat F^{(v,l)}\rightarrow-\hat F^{(v,l)}$ and $\alpha_{vl}\rightarrow-\alpha_{vl}$, and therefore assume $\alpha_{vl}\geq0$. Further, we define
\begin{equation}
    \hat  W^{(V,0)} := 2\frac{\hat W_0 - \delta_0}{\delta_{D-1} - \delta_0}  - \hat I = \hat F^{(V,0)},
    \label{eq: factorizations_one_body}
\end{equation}
where $\delta_0=\min_p w_p$ and $\delta_{D-1}=\max_p w_p$ are the lowest and the largest eigenvalues of $\hat W_0$, respectively. In the rotated frame, $\hat W_0$ is diagonal, so $\hat W^{(V,0)}$ is the diagonal matrix $\hat F^{(V,0)}$ with elements $f_{V0p}=2(w_p-\delta_0)/(\delta_{D-1}-\delta_0)-1\in[-1,1]$, and it has the form of~\cref{eq: factorizations_two_body} with $\hat R^{(V)}=\hat I$ and $\alpha_{V0}=0$. Therefore, the one-body term requires no unitary. Define $r_{v}=\max_i |d_{vi}|$ to be the largest absolute value of the elements $d_{vi}$ of $\hat D^{(v)}$ in the SVD~\eqref{eq:R_svd} of $\hat R^{(v)}$, and $q_{vl} = r_v^2 \max_{m}\left(|f_{vlm}|\right)$. The subnormalisation factor of the block-encoding of $\hat W^{(v,l)}$ is then
\begin{equation}
    \epsilon_{vl} = q_{vl} + \alpha_{vl},
\end{equation}
and $\epsilon_{V0}=1$ for the one-body term. Let $\rm PREP$ be
\begin{align}
    \label{eq: PREP}
    &{\rm PREP}\ket{0}_b\ket{0}_{\rm outer} \rightarrow \frac{1}{\sqrt{2\Lambda}}\left(\ket{0}_b \sum_{v=0}^{V-1}\sum_{l=0}^{L-1}\frac{N\epsilon_{vl}}{\sqrt{2}}\ket{v,l} + \ket{1}_b \sqrt{N(\delta_{D-1}-\delta_0)}\ket{V,0}\right),\\
    \label{eq: Lambda'}
    & \Lambda =\sum_{v=0}^{V-1}\sum_{l=0}^{L-1} \frac{N^2\epsilon^2_{vl}}{4} + \frac{N (\delta_{D-1} - \delta_{0})}{2}.
\end{align}
With $\sqrt{|\omega_{vl}|}$ absorbed into $\epsilon_{vl}$, $\Lambda$ coincides with~\cref{eq:Lambda}.

\begin{figure}[htb!]
    \scalebox{0.88}{%
\begin{tikzpicture}
\begin{yquant}
qubit {control} control;
qubit {$\ket{0}_{\theta}$} theta;
qubit {$\ket{0}_{\rm b}$} flag_body;
qubit {$\ket{0}_{\rm outer}$} outer_iter;
qubit {$\ket{0}_{\rm ne}$} num_elec;
qubit {$\ket{0}_{\rm flag}$} ancila_flag;
qubit {$\ket \Psi$} system;

["north:$\lceil \log_2N\rceil$" {font=\protect\footnotesize, inner sep=2pt}]
slash num_elec;

["north:$N \lceil \log_2D\rceil$" {font=\protect\footnotesize, inner sep=2pt}]
slash system;

["north:$4$" {font=\protect\footnotesize, inner sep=2pt}]
slash ancila_flag;

["north:$\lceil \log_2(V+1)\rceil+\lceil \log_2L\rceil$" {font=\protect\footnotesize, inner sep=2pt}]
slash outer_iter;

box {PREP} (outer_iter, flag_body);

[multictrl] box {BE$\left[\frac{\sum_j(\hat W^{(v,l)})_j}{N\epsilon_{vl}}\right]$} (num_elec, ancila_flag, system) | control ~ outer_iter;
[multictrl] box {data: $\theta_{vl}$} theta ~ outer_iter;
box {REF} (num_elec, ancila_flag) | control ~ flag_body;
z theta | control;
[multictrl] box {data$^{\dagger}$: $\theta_{vl}$} theta ~ outer_iter;
[multictrl, name=be2] box {BE$\left[\frac{\sum_j(\hat W^{(v,l)})_j}{N\epsilon_{vl}}\right]$} (num_elec, ancila_flag, system) | control, flag_body ~ outer_iter;

box {UNPREP} (outer_iter, flag_body);
\end{yquant}
\opencontrol{be2-p1}
\end{tikzpicture}
}
\caption{Block-encoding of $(\hat H'' + \Omega'')/\Lambda - \hat I$, where $\Omega'' = -\delta_0 N + \sum_{vl:\theta_{vl}=1} \frac{N^2\epsilon_{vl}^2}{2}$. Hamiltonian $\hat H''$ defined in~\cref{eq: hamiltonian_for_be}, and $\Lambda$ is given in~\cref{eq: Lambda'}. The outer register holds $\ket{v,l}$ with $v=0,\dots,V$; a slashed circle denotes a multiplexed control, i.e.\ a different operation for each value of the register. The register $\ket{0}_{\rm flag}$ holds the four qubits that flag the block-encoding of $\hat W^{(v,l)}/\epsilon_{vl}$ in~\cref{fig: quantum_circuit_be_multiplex_be}: the little-LCU qubit and ${\rm flag}_0$, ${\rm flag}_1$, ${\rm flag}_2$.}
\label{fig: quantum_circuit_be_full}
\end{figure}
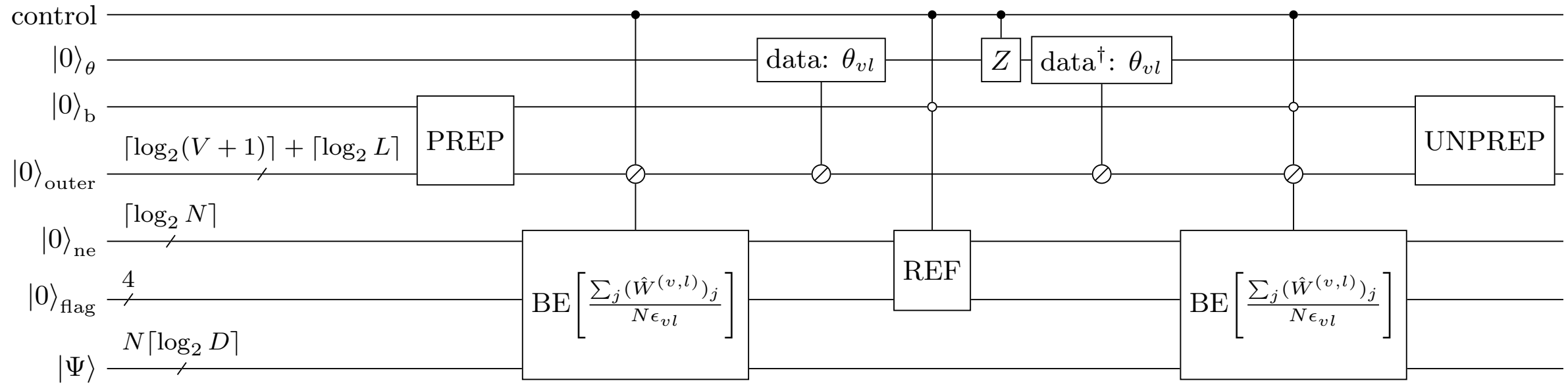

Then, we can block-encode the Hamiltonian~\cref{eq: hamiltonian_for_be} using the quantum circuit shown in~\cref{fig: quantum_circuit_be_full}. Because we block-encode $\sum_{j=0}^{N-1}(\hat W^{(v,l)})_j$ with a circuit that is its own inverse, and apply a reflection only on the two-body term, according to~\cite{low_hamiltonian_simulation_qubitization_2019} we implement the Chebyshev polynomial of second order, and because the one-body contribution $\hat W^{(V,0)}$ is correctly normalised and shifted ($\epsilon_{V,0}=1$), we achieve the desired block-encoding. Notice that because we introduced the implementation of phases $\theta_{vl}$, the block-encoding embeds $(\hat H'' + \Omega'')/\Lambda - 1$, and $\hat H'' + \Omega''$ becomes automatically positive semidefinite because $\Omega''$ will be given by $\Omega'' = -\delta_0 N + \sum_{vl:\theta_{vl}=1} \frac{N^2\epsilon_{vl}^2}{2}$ from the construction of the quantum circuit (\cref{app:be_derivation}). Since $\hat H''=\hat H_0-\alpha-\beta N+\frac{N^2}{2}\sum_{vl}\omega_{vl}\alpha_{vl}^2$, with $\omega_{vl}$ and $\alpha_{vl}$ of~\cref{eq:ham_ansatz}, this is the block-encoding of~\cref{eq: ShiftedHamiltonian} with $\Omega$ of~\cref{eq:Omega}.

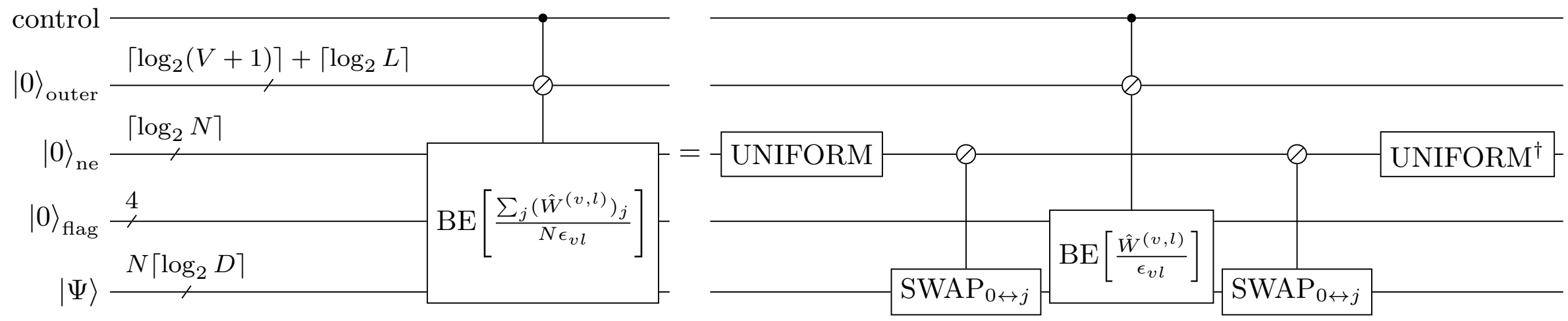
\begin{figure}[htb!]
\scalebox{0.9}{%
\begin{tikzpicture}
\begin{yquant}
qubit {control} control;
qubit {$\ket{0}_{\rm outer}$} outer_iter;
qubit {$\ket{0}_{\rm ne}$} num_elec;
qubit {$\ket{0}_{\rm flag}$} ancila_flag;
qubit {$\ket \Psi$} system;

["north:$\lceil \log_2N\rceil$" {font=\protect\footnotesize, inner sep=2pt}]
slash num_elec;

["north:$N \lceil \log_2D\rceil$" {font=\protect\footnotesize, inner sep=2pt}]
slash system;

["north:$4$" {font=\protect\footnotesize, inner sep=2pt}]
slash ancila_flag;

["north:$\lceil \log_2(V+1)\rceil+\lceil \log_2L\rceil$" {font=\protect\footnotesize, inner sep=2pt}]
slash outer_iter;

[multictrl] box {BE$\left[\frac{\sum_j(\hat W^{(v,l)})_j}{N\epsilon_{vl}}\right]$} (num_elec, ancila_flag, system) | control ~ outer_iter;

text {$=$} (-);

box {UNIFORM} num_elec;
[multictrl] box {SWAP$_{0\leftrightarrow j}$} system ~ num_elec;
[multictrl] box {BE$\left[\frac{\hat W^{(v,l)}}{\epsilon_{vl}}\right]$} (ancila_flag, system) | control ~ outer_iter;
[multictrl] box {SWAP$_{0\leftrightarrow j}$} system ~ num_elec;
box {UNIFORM$^{\dagger}$} num_elec;
\end{yquant}
\end{tikzpicture}
}
\caption{Block-encoding of $\sum_{j}(\hat W^{(v,l)})_j/(N\epsilon_{vl})$. The block-encoding will be flagged by $\ket{0}_{\rm flag}\ket{0}_{\rm ne}$ qubits, where $\ket{0}_{\rm flag}$ holds the little-LCU qubit and ${\rm flag}_{0,1,2}$ of~\cref{fig: quantum_circuit_be_multiplex_be}.}
\label{fig: quantum_circuit_be_sum_over_electrons}
\end{figure}

In order to implement block-encodings of $\sum_{j=0}^{N-1}(\hat W^{(v,l)})_j$, we use the quantum circuit depicted in~\cref{fig: quantum_circuit_be_sum_over_electrons}. There, we swap all electronic states, $j$, into the first register and then implement the block-encoding of $\hat W^{(v,l)}$ on that register. UNIFORM creates a uniform superposition over $N$ electrons.

In order to implement $\text{BE}\left[\hat W^{(v,l)}/\epsilon_{vl}\right]$, we make use of the fact that they are already factorised according to~\cref{eq: factorizations_one_body,eq: factorizations_two_body}. In this case, we can implement block-encodings of $\hat W^{(v,l)}$ through LCU and products of block-encodings. The block-encoding for diagonal $\hat F^{(v,l)}$ can be implemented using the phase gradient approach~\cite{sanders2020compilation}.

Meanwhile, the block-encoding of the $D\times M$ rectangular matrix, $\hat R^{(v)}$, can be implemented using its SVD~\eqref{eq:R_svd}, where $\hat U^{(v)}$ is a $D\times D$ unitary, $\hat V^{(v)\dagger}$ is an $M\times M$ unitary and $\hat D^{(v)}$ is a real rectangular diagonal matrix, whose elements are loaded as the rotation angles $\chi_{vp}=\arccos(d_{vp}/r_v)$. We denote the multiplexed block-encoding of $\hat R^{(v)\dagger}/r_v$ by
\begin{equation}\label{eq: multiplexed_R_be}
    \hat{\mathcal R} = \sum_{v=0}^{V}\ketbra{v}{v}\otimes{\rm BE}\left[\hat R^{(v)\dagger}/r_v\right],
\end{equation}
and show its quantum circuit in~\cref{fig: quantum_circuit_be_rectungular_matrices}. For $M\neq D$, we pad the singular values with zeros, $d_{vi}=0$ for $\min(D,M)\le i<\max(D,M)$, so that the lookup loads the angle $\pi/2$ whenever the register $\ket{p,m}$ holds such a value $i$. The rotation then removes the components of the state outside the first $\min(D,M)$ basis states, which arise in $\hat{\mathcal R}$ for $M<D$ and in $\hat{\mathcal R}^{\dagger}$ for $M>D$. The padded angles are equal, so the lookup need not resolve their addresses, as in the QROM with variable spacing of Ref.~\cite{sanders2020compilation}. The unitaries $\hat U^{(v)\dagger}$ and $\hat V^{(v)}$ are multiplexed over $v$, and we implement them as the multiplexed unitaries of~\cref{sec:unitary_synthesis}. For the one-body term, $v=V$, $\hat R^{(V)}=\hat I$, so all its rotation angles are zero and the lookup skips them: the cost of $\hat{\mathcal R}$ is that of the $V$ two-body unitaries, and the one-body term only adds a value to the register $\ket{v}$. After these block-encodings are specified we can define a multiplexed block-encoding of $\hat W^{(v,l)}$ as is shown in~\cref{fig: quantum_circuit_be_multiplex_be}. The little-LCU qubit selects between $\alpha_{vl}\hat I$ and $\hat R^{(v)}\hat F^{(v,l)}\hat R^{(v)\dagger}$: it is rotated by $\hat R_y(\phi_{vl})$, with $\hat R_y(\phi)=e^{-i\phi\hat Y}$ as in~\cref{eq:multiplexor} and
\begin{equation}\label{eq:little_lcu_angle}
    \phi_{vl} = \arccos\sqrt{\frac{\alpha_{vl}}{\epsilon_{vl}}},
\end{equation}
and $\phi_{V0}=\pi/2$ for the one-body term. The block-encoding of $\hat F^{(v,l)}$ is a single lookup multiplexed over $(v,l,m)$ followed by a rotation of $\ket{0}_{\rm flag_2}$ into the phase-gradient state. A rotation $\hat R_y(\chi)$ is not its own inverse, so we apply a $Z$ gate on $\ket{\cdot}_{\rm flag_2}$ controlled by the little-LCU qubit: $\hat Z\hat R_y(\chi)$ is a reflection with the same matrix element $\cos\chi$. All other rotations of~\cref{fig: quantum_circuit_be_multiplex_be} are applied together with their inverses, so that the circuit is its own inverse. A Hermitian/self-inverse block encoding is required to apply the Chebyshev polynomial in~\cref{fig: quantum_circuit_be_full}, and also for use in QPE. The controlled $Z$ gate is a Clifford gate and does not change the cost. We note that such a block-encoding does not require application of controlled $\hat{\mathcal R}$: if the little-LCU qubit stays in $\ket{0}$, neither the Toffoli gate on $\ket{0}_{\rm flag_1}$ nor the block-encoding of $\hat F^{(v,l)}$ acts, and $\hat{\mathcal R}$ and $\hat{\mathcal R}^{\dagger}$ cancel. Therefore, to control the whole block-encoding, it is sufficient to control the rotations of the little-LCU qubit.

\begin{figure}[htb!]
\centering
\scalebox{0.9}{%
\begin{tikzpicture}
\begin{yquant}
qubit {$\ket{v}$} v;
qubit {$\ket{p}$} system;
qubit {$\ket{m}$} additional;
qubit {$\ket{0}_{\rm angles}$} angles;
qubit {$\ket{0}_{\rm flag_0}$} q;
qubit {$\ket{\mathcal F_b}$} phase_grad;

["north:$\lceil \log_2 (V+1)\rceil$" {font=\protect\footnotesize, inner sep=2pt}]
slash v;
["north:$\lceil \log_2 D\rceil$" {font=\protect\footnotesize, inner sep=2pt}]
slash system;
["north:$\lceil \log_2 M \rceil - \lceil \log_2 D \rceil$" {font=\protect\footnotesize, inner sep=2pt}]
slash additional;
["north:$b$" {font=\protect\footnotesize, inner sep=2pt}]
slash angles;
["north:$b$" {font=\protect\footnotesize, inner sep=2pt}]
slash phase_grad;

[multictrl] box {BE$\left[\frac{\hat R^{(v)\dagger}}{r_v}\right]$} (system, additional, angles, q, phase_grad) ~ v;
text {$=$} (-);

[multictrl] box {$\hat U^{(v)\dagger}$} system ~ v;
[multictrl] box {data: $\chi_{vp}$} angles ~ v, system, additional;
[name=rot] box {\Ifnum\idx<1 $\chi_{vp}$\Else\Ifnum\idx<2 $R_y$\Else \addpg\Fi\Fi} angles, q, phase_grad;
[multictrl] box {data$^{\dagger}$: $\chi_{vp}$} angles ~ v, system, additional;
[multictrl] box {$\hat V^{(v)}$} (system, additional) ~ v;

\end{yquant}
\draw (rot-0) -- (rot-1) -- (rot-2);
\end{tikzpicture}
}
\caption{
    Multiplexed block-encoding $\hat{\mathcal R}$,~\cref{eq: multiplexed_R_be}, of $\hat R^{(v)\dagger}/r_{v}$ through SVD, $\hat R^{(v)\dagger}/r_v = \hat V^{(v)} (\hat D^{(v)\dagger}/r_v) \hat U^{(v)\dagger}$. The lookup loads the $b$-bit angle $\chi_{vp}=\arccos(d_{vp}/r_v)$, and \addpg{} rotates $\ket{0}_{\rm flag_0}$ by $\hat R_y(\chi_{vp})$ by adding the angle into the $b$-qubit phase-gradient state $\ket{\mathcal F_b}$ (three connected boxes, as in~\cref{fig:window_select_xor}). \addpg{} costs $b-1$ Toffolis with $b-1$ clean ancillae, or $2b-2$ Toffolis without ancillae,~\cref{eq:element_costs}. Preparing a phase gradient state is $O(b \log(b/\epsilon_{ph.grad.}))$, but because it does not depend on the outer loop $vl$ and is a catalytic state, it can be prepared once and reused for whole algorithm, so its cost is negligible and we therefore ignore it. For $v=V$, $\hat U^{(V)}=\hat V^{(V)}=\hat I$ and $\chi_{Vp}=0$. The figure shows the case $M\ge D$, otherwise the register splitting is opposite.
}
\label{fig: quantum_circuit_be_rectungular_matrices}
\end{figure}
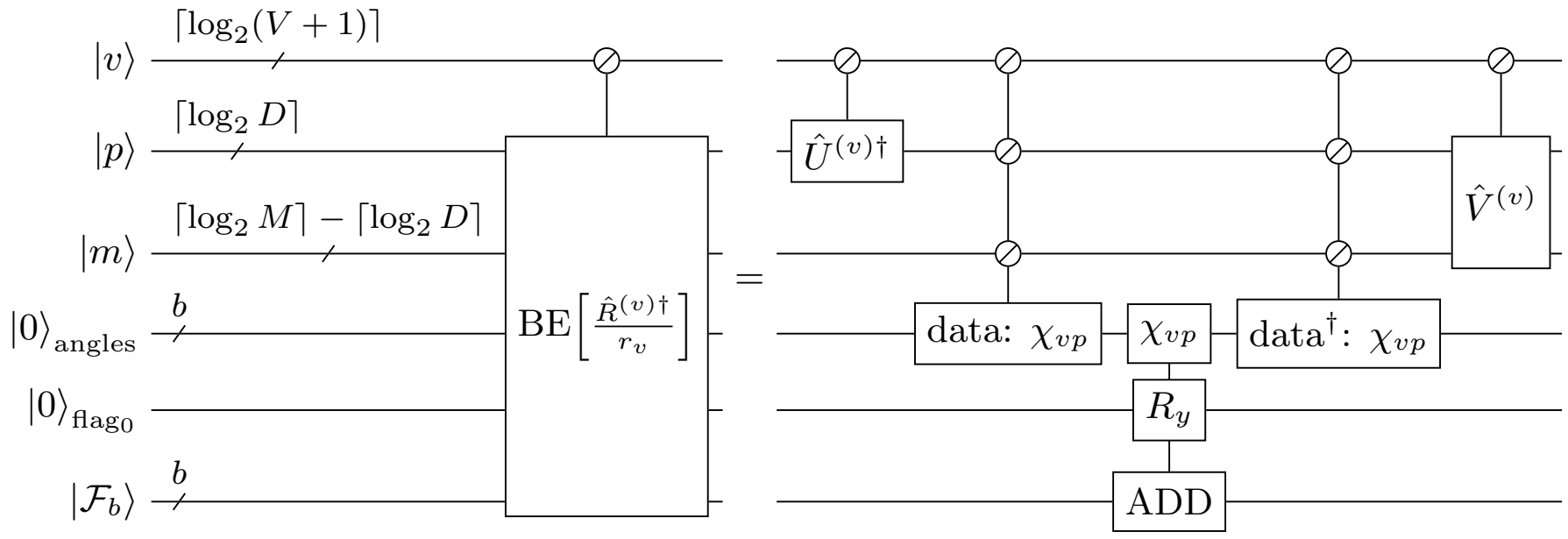

\begin{figure}[htb!]
\centering
\scalebox{1}{%
\begin{tikzpicture}
\begin{yquant}
qubit {control} control;
qubit {$\ket{l}$} l;
qubit {$\ket{v}$} v;
qubit {$\ket{0}_{\text{little LCU}}$} lcu;
qubit {$\ket{p,m}$} pm;
qubit {$\ket{0}_{\rm flag_0}$} flag0;
qubit {$\ket{0}_{\rm flag_1}$} flag1;
qubit {$\ket{0}_{\rm flag_2}$} flag2;
qubit {$\ket{\mathcal F_b}$} phase_grad;
qubit {$\ket{0}_{\rm angles}$} angles;

["north:$\lceil \log_2 L\rceil$" {font=\protect\footnotesize, inner sep=2pt}]
slash l;
["north:$\lceil \log_2 (V+1)\rceil$" {font=\protect\footnotesize, inner sep=2pt}]
slash v;
["north:$\lceil \log_2 M\rceil$" {font=\protect\footnotesize, inner sep=2pt}]
slash pm;
["north:$b$" {font=\protect\footnotesize, inner sep=2pt}]
slash phase_grad;
["north:$b$" {font=\protect\footnotesize, inner sep=2pt}]
slash angles;

[multictrl] box {BE$\left[\frac{\hat W^{(v,l)}}{\epsilon_{vl}}\right]$} (lcu, pm, flag0, flag1, flag2, phase_grad, angles) | control ~ l, v;
text {$=$} (-);

[multictrl] box {BE$\left[\frac{\hat R^{(v)\dagger}}{r_v}\right]$} (pm, flag0) ~ v;
[multictrl] box {$\hat R_y(\phi_{vl})$} lcu | control ~ l, v;
cnot flag1 | lcu, flag0;
[multictrl] box {BE$\left[\frac{\hat F^{(v,l)}}{\max_m|f_{vlm}|}\right]$} (flag2, phase_grad, angles) | lcu ~ l, v, pm;
z flag2 | lcu;
[multictrl] box {$\hat R_y^{\dagger}(\phi_{vl})$} lcu | control ~ l, v;
[multictrl] box {BE$\left[\frac{\hat R^{(v)\dagger}}{r_v}\right]^{\dagger}$} (pm, flag0) ~ v;
\end{yquant}
\end{tikzpicture}
}
\caption{
    Multiplexed block-encoding of $\hat W^{(v,l)}/\epsilon_{vl}$, including the one-body term $(v,l)=(V,0)$, for which $\hat{\mathcal R}$ acts as the identity, $\phi_{V0}=\pi/2$ and $\hat F^{(V,0)}$ is the diagonal one-body matrix~\cref{eq: factorizations_one_body}. The multiplexed ${\rm BE}[\hat R^{(v)\dagger}/r_v]$ is $\hat{\mathcal R}$ of~\cref{fig: quantum_circuit_be_rectungular_matrices}, which also uses the angle and phase-gradient registers, and its adjoint $\hat{\mathcal R}^{\dagger}$ block-encodes $\hat R^{(v)}/r_v$. ${\rm BE}[\hat F^{(v,l)}/\max_m|f_{vlm}|]$ loads the angle $\arccos\left(f_{vlm}/\max_{m'}|f_{vlm'}|\right)$ and rotates $\ket{0}_{\rm flag_2}$ with the phase-gradient state. The controlled-$Z$ gate that follows turns this rotation into a reflection, so that the whole circuit is its own inverse. The Toffoli gate flips $\ket{0}_{\rm flag_1}$ if $\hat{\mathcal R}$ fails, $\ket{1}_{\rm flag_0}$, on the branch where $\hat F^{(v,l)}$ is applied. Thereby flag1 induces a product of block encoded matrices $\hat R^{(v)}\hat F^{(v,l)}\hat R^{(v)\dagger}$ on that branch. The block-encoding is flagged by $\ket{0}_{\text{little LCU}}\ket{0}_{\rm flag_0}\ket{0}_{\rm flag_1}\ket{0}_{\rm flag_2}$.
}
\label{fig: quantum_circuit_be_multiplex_be}
\end{figure}
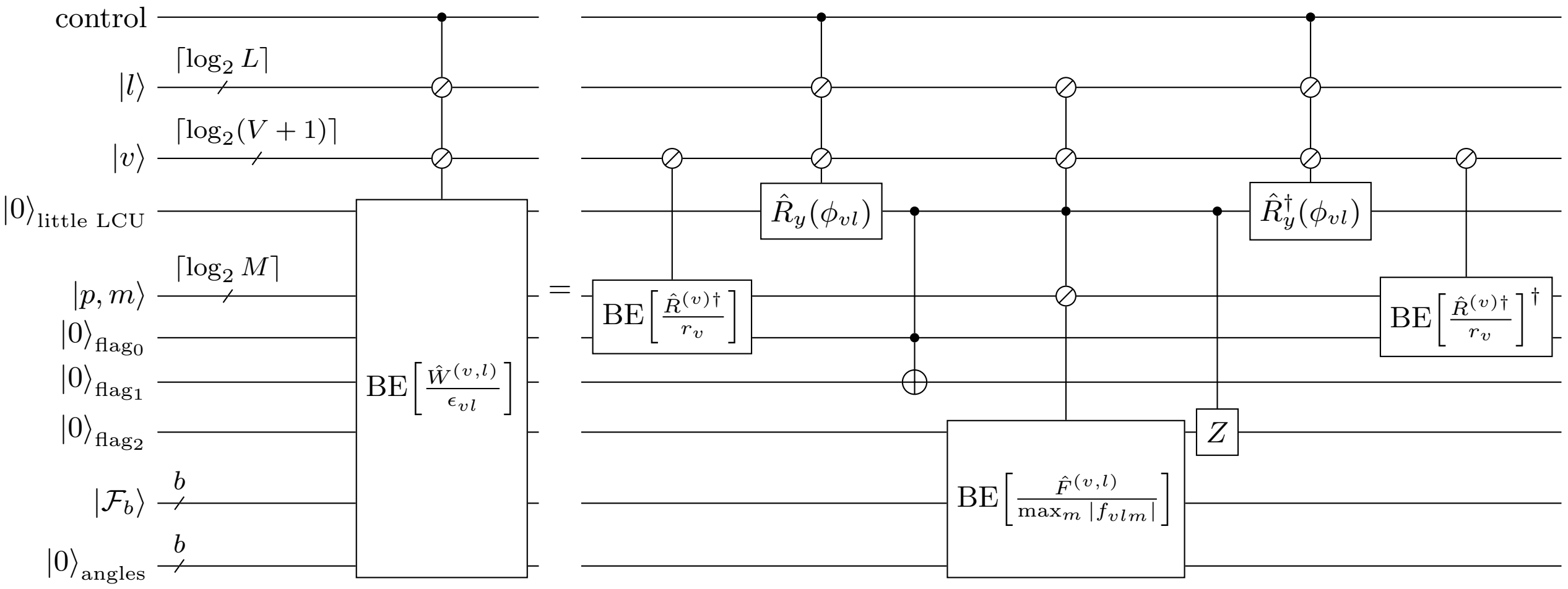

\section{Unitary Synthesis}\label{sec:unitary_synthesis}
The dominant cost of the block-encoding in~\cref{sec:block-encoding-general} is the multiplexed block-encoding $\hat{\mathcal R}$,~\cref{fig: quantum_circuit_be_rectungular_matrices}, which consists of the multiplexed unitaries $\sum_{v}\ketbra{v}{v}\otimes\hat U^{(v)\dagger}$ and $\sum_{v}\ketbra{v}{v}\otimes\hat V^{(v)}$. In this section, we describe how we synthesise the multiplexed unitaries. We represent a unitary as a product of multiplexed rotations. As we are interested in quantum chemistry applications, we assume the unitaries to be real. Let the unitary be of dimension $D\times D$, and introduce $d=\lceil\log_2D\rceil$. We first assume that $D=2^d$, and discuss other values of $D$ in~\cref{app:nonpow2}. For $D=2^d$, the number of angles per layer is $\dim(\vec\theta)=D/2$, and the cosine-sine decomposition (CSD) represents any such unitary exactly with $N_S=D-1$ layers, which provide its $D(D-1)/2$ real parameters. We use this as the maximal number of layers for all orderings $s(l)$. For a general ordering, the same number of parameters does not guarantee that every unitary can be represented. We therefore tested the orderings considered in this work (\cref{sec:us_orderings}) numerically at full depth, $N_S=D-1$, for $D$ up to 128. The Jacobian of the ansatz with respect to the angles has full rank $D(D-1)/2$ at random angles, so that no angle is redundant, and Haar-random real unitaries with determinant $+1$ are reproduced to machine precision. Our goal, however, is to approximate the Hamiltonian, not arbitrary unitaries, and with fewer layers, $N_S<D-1$, and as we will show later, the orderings approximates the Hamiltonians of our examples to the target accuracy (\cref{sec:us_trainability,sec:resource_estimates}). A general theory of the expressivity of such an ansatz is beyond the scope of this paper.

The remaining question is the choice of the function $s(l)$, which affects both the total Toffoli cost and the trainability of the ansatz \eqref{eq:unitary ansatz}. In~\cref{sec:us_cost}, we explain how a sequence of multiplexed rotations is implemented for an \textit{arbitrary} $s(l)$ and define its cost. In~\cref{sec:us_orderings}, we compare different choices of $s(l)$ and their Toffoli counts for a given ancilla budget, and in~\cref{sec:us_trainability}, we discuss the trainability of the resulting ansatz. The compilation of the whole block-encoding for a given ancilla budget is described in~\cref{app:cost_model}.
\subsection{Cost model}\label{sec:us_cost}
\begin{figure*}[htb!]
\centering
\scalebox{0.9}{%
\begin{tikzpicture}
\begin{yquant}
qubit {$\ket{p}_{0}$} p0;
qubit {$\ket{p}_{1}$} p1;
qubit {$\ket{p}_{2}$} p2;
qubit {$\ket{p}_{3}$} p3;
qubit {$\ket{0}^{\otimes b}_{0,0}$} a00;
qubit {$\ket{0}^{\otimes b}_{0,1}$} a01;
qubit {$\ket{0}^{\otimes b}_{0,2}$} a02;
qubit {$\ket{0}^{\otimes b}_{0,3}$} a03;
qubit {$\ket{0}^{\otimes b}_{1,0}$} a10;
qubit {$\ket{0}^{\otimes b}_{1,1}$} a11;
qubit {$\ket{0}^{\otimes b}_{1,2}$} a12;
qubit {$\ket{0}^{\otimes b}_{1,3}$} a13;
qubit {$\ket{\mathcal F_b}$} f;

slash a00;
slash a01;
slash a02;
slash a03;
slash a10;
slash a11;
slash a12;
slash a13;
slash f;

[name=sel0] box {$\Ifcase\idx \tilde\theta^{(0)}_{0,0}\Or \tilde\theta^{(0)}_{1,0}\Or \tilde\theta^{(0)}_{2,0}\Or \tilde\theta^{(0)}_{3,0}\Or \tilde\theta^{(1)}_{0,0}\Or \tilde\theta^{(1)}_{1,0}\Or \tilde\theta^{(1)}_{2,0}\Or \tilde\theta^{(1)}_{3,0}\Fi$} a00, a01, a02, a03, a10, a11, a12, a13 ~ p0;
[name=sel1] box {$\Ifcase\idx \tilde\theta^{(0)}_{0,1}\Or \tilde\theta^{(0)}_{1,1}\Or \tilde\theta^{(0)}_{2,1}\Or \tilde\theta^{(0)}_{3,1}\Or \tilde\theta^{(1)}_{0,1}\Or \tilde\theta^{(1)}_{1,1}\Or \tilde\theta^{(1)}_{2,1}\Or \tilde\theta^{(1)}_{3,1}\Fi$} a00, a01, a02, a03, a10, a11, a12, a13 | p0;
align -;

[name=x01] cnot a00 | p3, a01 ~ p2;
[name=x02] cnot a00 | p2, a02 ~ p3;
[name=x03] cnot a00 | p2, p3, a03;
[name=rot0] box {\Ifnum\idx<1 $R_y$\Else\Ifnum\idx<2 $a$\Else ADD\Fi\Fi} p1, a00, f;
h a00;
measure a00;
[name=z01] box {$Z^{\mu}$} a01 | p3 ~ p2;
[name=z02] box {$Z^{\mu}$} a02 | p2 ~ p3;
[name=z03] box {$Z^{\mu}$} a03 | p2, p3;
discard a00;
align -;

[name=x11] cnot a10 | p3, a11 ~ p1;
[name=x12] cnot a10 | p1, a12 ~ p3;
[name=x13] cnot a10 | p1, p3, a13;
[name=rot1] box {\Ifnum\idx<1 $R_y$\Else\Ifnum\idx<2 $a$\Else ADD\Fi\Fi} p2, a10, f;
h a10;
measure a10;
[name=z11] box {$Z^{\mu}$} a11 | p3 ~ p1;
[name=z12] box {$Z^{\mu}$} a12 | p1 ~ p3;
[name=z13] box {$Z^{\mu}$} a13 | p1, p3;
discard a10;
align -;

[name=qdag] box {\Ifnum\idx<1 In\Else QROAM$^\dagger$\Fi} (p0), (a01, a02, a03, a11, a12, a13);
\end{yquant}
\draw (rot0-0) -- (rot0-1) -- (rot0-2);
\draw (rot1-0) -- (rot1-1) -- (rot1-2);
\draw (qdag-0) -- (qdag-1);
\coordinate (W) at ([xshift=-3pt]current bounding box.west);
\draw[decorate, decoration={brace, amplitude=4pt}] (W |- qdag-0.south) -- (W |- qdag-0.north)
      node[midway, left=5pt, font=\footnotesize] {$j_{\rm s}$};
\coordinate (T3) at ([yshift=3pt]sel0-0.north);   
\draw[decorate, decoration={brace, amplitude=4pt}] (W |- T3) -- (W |- rot0-0.north)
      node[midway, left=5pt, font=\footnotesize, align=right] {targets};
\draw[decorate, decoration={brace, amplitude=4pt}] (W |- sel0-3.south) -- (W |- sel0-0.north)
      node[midway, left=5pt, font=\footnotesize] {$r=0$};
\draw[decorate, decoration={brace, amplitude=4pt}] (W |- sel0-7.south) -- (W |- sel0-4.north)
      node[midway, left=5pt, font=\footnotesize] {$r=1$};
\node[draw, dashed, rounded corners, inner sep=3pt, fit=(sel0-0) (sel1-7),
      label={[font=\footnotesize, align=center, label distance=0.75cm]below:\qroam: column $j_{\rm s}$\\of all layers}] {};
\node[draw, dashed, rounded corners, inner sep=3pt, fit=(rot0-0) (rot0-2) (x01-0) (z03-0),
      label={[font=\footnotesize, align=center]below:layer $r=0$\\\selxor, \addpg, \selz}] {};
\node[draw, dashed, rounded corners, inner sep=3pt, fit=(rot1-0) (rot1-2) (x11-0) (z13-0),
      label={[font=\footnotesize, align=center]below:layer $r=1$\\\selxor, \addpg, \selz}] {};
\end{tikzpicture}
}
\caption{Quantum circuit example for synthesis of three consecutive multiplexed rotations acting on 4 qubits $(p_0, p_1, p_2, p_3)$ and sharing a single control $p_0$; only the first two of them and their registers are shown. In the first step, the angles $\tilde\theta^{(r)}_{j_{\rm t},j_{\rm s}}$ are loaded and selected by their column index, $j_{\rm s}$, which in this case is just the value of $p_0$. For every layer  $r$ (a multiplexed rotation), we apply \selxor{}, \addpg{} (rotations using addition into phase gradient state), a measurement in the $X$ basis and \selz{}.
\selxor{} is a unary iterator on input row registers ($p_2, p_3$ for layer $r=0$) which XORs row $j_{\rm t}$ of the \qroam{} output into the first row of the \qroam{} output. The resulting value is then used in \addpg{} block to perform rotation on the target qubit ($p_1$ for layer $r=0$). Then the first row of \qroam{} output for a given layer is uncomputed by a measurement in the $X$ basis, with outcome $\mu$, followed by \selz{} (the $Z^\mu$ boxes), which applies $Z$ corrections that depend on $\mu$ to the other rows of this layer, controlled by the same row registers as \selxor{}. This is repeated for all layers (further layers are added in the same way) and in the end the initial \qroam{} is uncomputed to free the remaining data qubits.}
\label{fig:window_select_xor}
\end{figure*}
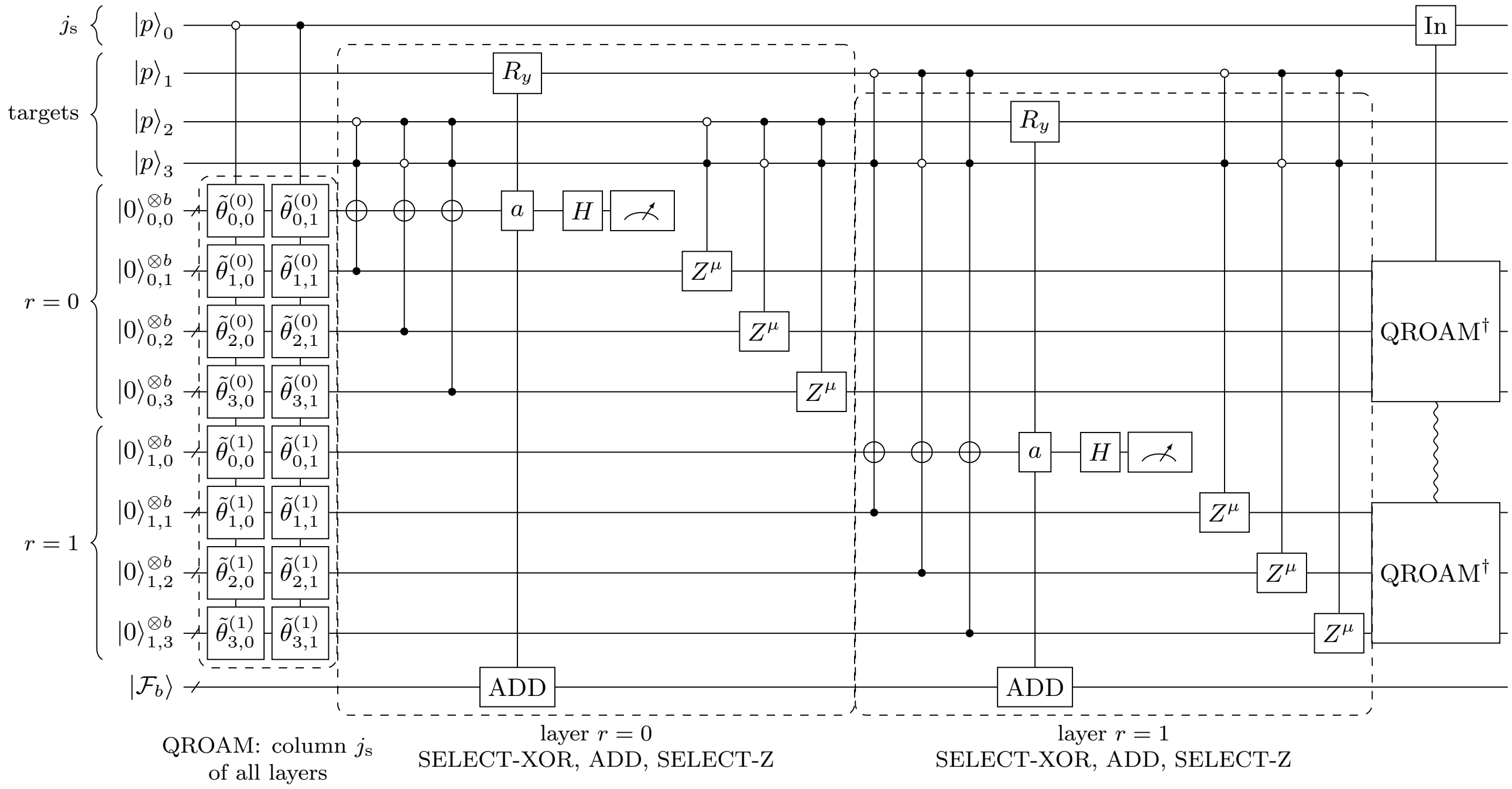

Our ansatz generalises the idea of Tan~\cite{tan2025unitarysynthesisfewert}, who synthesised a unitary
from its cosine-sine decomposition by splitting the sequence of multiplexed rotations into blocks of
consecutive rotations that target only a few qubits. We call such a block with consecutive rotations a window, and a single multiplexed rotation in this window a layer. In a window, the layers can share the control qubits whose state does not change throughout the window. The values of the other control qubits change between layers because they are targeted by rotations. Let $r=0,\dots,\ell-1$ enumerate the layers in the window. Each layer needs $2^{d-1}$ angles, which we denote by $\{\theta^{(r)}_{j}\}$ where $j \in \{0,\dots 2^{d-1}-1\}$. These angles are loaded in an initial lookup table at the beginning of each window. We can split $j$ into indices which are addressed by shared controls between layers, $j_{\rm s}$, and the rest $j_{\rm t}$, so our lookup table is $j_{\rm s} \mapsto \{\theta^{(r)}_{j_{\rm t},j_{\rm s}}\}$. Here $j_{\rm t}$ is a $g-1$-bit integer and $j_{\rm s}$ is a $d-g$ bit integer. To query the $j=(j_{\rm t},j_{\rm s})$ angle, we first address the lookup by its column index, $j_{\rm s}$ and load all angles for all $r, j_{\rm t}$. Then immediately before each rotation we select the angle by its row index $j_{\rm t}$, and we repeat this for each layer $r$. \Cref{fig:window_select_xor} shows the first two layers of a window with three layers on four qubits, $(p_0,p_1,p_2,p_3)$, which share the control $p_0$: the column index $j_{\rm s}$ is the value of $p_0$, layer $r$ targets $p_{r+1}$, and the row index $j_{\rm t}$ is the value of the other two qubits of $(p_1,p_2,p_3)$.  

In Tan's construction, the windows are the aligned blocks of $2^k-1$ layers of the CSD ordering, all with the same $k$, and the single layers between them. 
This construction provides an exact implementation of the unitary, up to the finite precision of the rotation angles. Other orderings at full depth, $N_S=D-1$, also represent the unitaries accurately (see the beginning of~\cref{sec:unitary_synthesis}). The approximation in our ansatz is the number of layers: we truncate the sequence to $N_S<D-1$ layers and treat the ordering $s(l)$ as part of the ansatz. For an arbitrary ordering, the windows are chosen to minimise the Toffoli cost within a given ancilla budget. We also implement the windows with different gadgets: \qroam~\cite{babbush_encoding_2018,berry_qubitization_2019,low_trading_2024} and \selxor, which load and select the angles, respectively, and additions into a phase-gradient state~\cite{gidney_halving_2018,sanders2020compilation,low_denser_2026}.

More specifically, we assume the angles are specified to $b$-bit precision, $\theta=2\pi a/2^b$ with an integer $a\in\{0,\dots,2^b-1\}$, so that each rotation is implemented by adding $a$ into a $b$-qubit phase-gradient state, and we start our construction by building a lookup table which stores differences between rows of the matrix $\theta^{(r)}_{j_{\rm t}, j_{\rm s}}$ w.r.t.\ its first row:
\begin{align}
\tilde{\theta}^{(r)}_{0, j_{\rm s}} & = \theta^{(r)}_{0, j_{\rm s}},  \\
\tilde{\theta}^{(r)}_{j_{\rm t},j_{\rm s}} & =  \theta^{(r)}_{0,j_{\rm s}} \oplus \theta ^{(r)}_{j_{\rm t},j_{\rm s}}, \quad j_{\rm t} > 0 .
\end{align}
For a given input $j=(j_{\rm t}, j_{\rm s})$, we load the column $j_{\rm s}$ of the lookup table, i.e.\
$\tilde{\theta}^{(r)}_{i,j_{\rm s}}$ for all layers $r$ and rows $i=0,\dots,2^{g-1}-1$, into the output registers
$(r,i)$ of $\qroam_r$. Then, for layer $r$, we apply a SELECT operator whose selection register is the row register $\ket{j_{\rm t}}$ and whose target operation XORs the $b$-bit word in register $(r,i)$ into register $(r,0)$:
\begin{equation}\label{eq:selxor_def}
    \selxor_r := \ketbra{0}{0}_{\rm t}\otimes\hat I
    + \sum_{i=1}^{2^{g-1}-1}\ketbra{i}{i}_{\rm t}\otimes{\rm CNOT}^{\otimes b}_{(r,i)\to(r,0)},
    \quad
    {\rm CNOT}^{\otimes b}_{(r,i)\to(r,0)} := \prod_{k=0}^{b-1}{\rm CNOT}_{(r,i,k)\to(r,0,k)},
\end{equation}
where ${\rm CNOT}_{(r,i,k)\to(r,0,k)}$ has qubit $k$ of register $(r,i)$ as control and qubit $k$ of
register $(r,0)$ as target. On computational basis states,
\begin{equation}
    \selxor_r\,\ket{j_{\rm t}}\ket{x_0}_{r,0}\ket{x_1}_{r,1}\cdots
    = \ket{j_{\rm t}}\ket{x_0\oplus x_{j_{\rm t}}}_{r,0}\ket{x_1}_{r,1}\cdots
    \quad (j_{\rm t}>0),
\end{equation}
and $\selxor_r$ acts trivially for $j_{\rm t}=0$. With $x_i=\tilde\theta^{(r)}_{i,j_{\rm s}}$, register $(r,0)$ therefore holds $\theta^{(r)}_{j_{\rm t},j_{\rm s}}$ for every value of $j_{\rm t}$.  Then we rotate the target qubit by adding register $(r,0)$ into the phase-gradient state~\cite{gidney_halving_2018,low_denser_2026} (\addpg), which leaves register $(r,0)$ unchanged. To uncompute $\selxor_r$, we use measurement-based uncomputation~\cite{berry_qubitization_2019}: measuring register $(r,0)$ releases it and, for the outcome $\mu\in\{0,1\}^b$, leaves the phase $(-1)^{\mu\cdot\theta^{(r)}_{j_{\rm t},j_{\rm s}}}$, where $\mu\cdot x$ denotes the parity of the bitwise AND of $\mu$ and $x$. Since $\mu\cdot(x\oplus y)=\mu\cdot x\oplus\mu\cdot y$, the definition of $\tilde\theta^{(r)}_{j_{\rm t},j_{\rm s}}$ gives
\begin{equation}\label{eq:meas_phase}
    (-1)^{\mu\cdot\theta^{(r)}_{j_{\rm t},j_{\rm s}}}
    =(-1)^{\mu\cdot\tilde\theta^{(r)}_{0,j_{\rm s}}}\times
    \begin{cases}
        1, & j_{\rm t}=0,\\
        (-1)^{\mu\cdot\tilde\theta^{(r)}_{j_{\rm t},j_{\rm s}}}, & j_{\rm t}>0.
    \end{cases}
\end{equation}
The first factor depends only on $j_{\rm s}$, which does not change during the window. Therefore, it commutes with the rest of the window, and we remove it in the uncomputation of \qroam{}. The second factor is removed by a SELECT operator with the same selection register, whose target operation applies $\hat Z$ gates to register $(r,i)$ according to the outcome $\mu$:
\begin{equation}\label{eq:selz_def}
    \selz_r := \ketbra{0}{0}_{\rm t}\otimes\hat I
    + \sum_{i=1}^{2^{g-1}-1}\ketbra{i}{i}_{\rm t}\otimes\hat Z^{\mu}_{(r,i)},
    \quad
    \hat Z^{\mu}_{(r,i)} := \prod_{k=0}^{b-1}\big(\hat Z_{(r,i,k)}\big)^{\mu_k},
\end{equation}
where $\hat Z_{(r,i,k)}$ acts on qubit $k$ of register $(r,i)$, so that $\hat Z^{\mu}_{(r,i)}\ket{x_i}_{r,i}=(-1)^{\mu\cdot x_i}\ket{x_i}_{r,i}$ with $x_i=\tilde\theta^{(r)}_{i,j_{\rm s}}$. The measurement and $\selz_r$ must be applied before the next layer rotates the qubits of the row register $\ket{j_{\rm t}}$. Therefore, for each layer $r$, we apply $\selxor_r$, \addpg, the measurement of register $(r,0)$ and $\selz_r$. After the last layer, we uncompute all other output registers of \qroam{} by measurement-based uncomputation~\cite{berry_qubitization_2019}, whose phase correction is a function of $j_{\rm s}$ and therefore also includes the first factor of~\cref{eq:meas_phase} for every layer. We then repeat this procedure for all windows. The circuit for a single window is shown in~\cref{fig:window_select_xor}.

We propose SELECT-XOR with SELECT-Z uncomputation as a new routine. Conventionally, as in Tan \cite{tan2025unitarysynthesisfewert}, for each layer, ancilla qubits would have been swapped according to $j_t$, and the rotation performed, and qubits swapped back according to $j_t$.  Usage of SELECT-XOR removes the need for the unswapping, replacing it by SELECT-Z, which is cheaper because its cost only depends on the register size of $j_t$ and not the number of ancilla qubits.

Thus, the Toffoli cost of the sequence of multiplexed rotations is
\begin{equation}\label{eq:total_toffoli}
    T=\sum_{\omega=0}^{W-1}\Big[\qroam(L_{{\rm s},\omega},B_\omega)
    +\ell_\omega\big(\selxor(L_{{\rm t},\omega},b)+\addpg(b)+\selz(L_{{\rm t},\omega})\big)
    +\qroam^\dagger(L_{{\rm s},\omega})\Big],
\end{equation}
where window $\omega$ has $\ell_\omega$ layers that target $g_\omega$ qubits. Its lookup table therefore has $L_{{\rm s},\omega}=2^{d-g_\omega}$ columns, one for each value of the shared controls $j_{\rm s}$, and $L_{{\rm t},\omega}=2^{g_\omega-1}$ rows, one for each value of the row register $j_{\rm t}$, and \qroam{} loads $B_\omega=b\,\ell_\omega L_{{\rm t},\omega}$ bits. The elements are:
\begin{itemize}
    \item $\qroam(L_{\rm s},B)$: loads one of $L_{\rm s}$ columns of $B$ bits, addressed by the shared controls $\ket{j_{\rm s}}$, with block size $\lambda$, a power of two~\cite{berry_qubitization_2019,low_trading_2024};
    \item $\qroam^\dagger(L_{\rm s})$: its measurement-based uncomputation, with block size $\lambda'$~\cite{berry_qubitization_2019};
    \item $\selxor(L_{\rm t},b)$ and $\selz(L_{\rm t})$: the SELECT operators~\eqref{eq:selxor_def} and~\eqref{eq:selz_def} over the row register $\ket{j_{\rm t}}$ with $L_{\rm t}$ values, implemented by unary iteration~\cite{babbush_encoding_2018};
    \item $\addpg(b)$: the addition of a $b$-bit register into the phase-gradient state~\cite{gidney_halving_2018,low_denser_2026}.
\end{itemize}
Their costs~\cite{gidney_halving_2018,babbush_encoding_2018,berry_qubitization_2019,low_trading_2024,low_denser_2026} are
\begin{equation}\label{eq:element_costs}
\begin{array}{lll}
    & \text{Toffolis} & \text{clean ancillae}\\[2pt]
    \qroam(L_{\rm s},B) & \lceil L_{\rm s}/\lambda\rceil+B(\lambda-1) & B\lambda+\lceil\log_2(L_{\rm s}/\lambda)\rceil\\
    \qroam^\dagger(L_{\rm s}) & \lceil L_{\rm s}/\lambda'\rceil+\lambda' & \lambda'+\lceil\log_2(L_{\rm s}/\lambda')\rceil\\
    \selxor(L_{\rm t},b) & b(L_{\rm t}-1)+L_{\rm t}-2 & \lceil\log_2L_{\rm t}\rceil-1\\
    \selz(L_{\rm t}) & L_{\rm t}-2 & \lceil\log_2L_{\rm t}\rceil-1\\
    \addpg(b) & b-1\ \text{ or }\ 2b-2 & b-1\ \text{ or }\ 0
\end{array}
\end{equation}
The costs of $\selxor$ and $\selz$ follow from unary iteration~\cite{babbush_encoding_2018}, which implements $\sum_{i}\ketbra{i}{i}\otimes\hat U_i$ over the $L_{\rm t}$ values of $j_{\rm t}$ with $L_{\rm t}-2$ Toffolis and $\lceil\log_2L_{\rm t}\rceil-1$ clean ancillae by applying each $\hat U_i$ controlled by a single flag qubit; this is one Toffoli and one ancilla fewer than in Ref.~\cite{babbush_encoding_2018}, because our SELECT operators are not controlled by an additional qubit. For $\selxor$, $\hat U_i={\rm CNOT}^{\otimes b}_{(r,i)\to(r,0)}$, and each controlled CNOT becomes a Toffoli, i.e.\ $b$ Toffolis for every $i\ge1$. For $\selz$, $\hat U_i=\hat Z^{\mu}_{(r,i)}$, and each controlled $\hat Z$ becomes a CZ gate, which is Clifford. Both are trivial for $L_{\rm t}=1$, i.e.\ for $g_\omega=1$. For \addpg, we use either the adder with $b-1$ clean ancillae~\cite{gidney_halving_2018} or the adder without ancillae~\cite{low_denser_2026}. The $B$-qubit output register of \qroam{}, which we count as clean ancillae, is held until $\qroam^\dagger$. The $(\lambda-1)B$ qubits of junk~\cite{berry_qubitization_2019} can be measured in the $X$ basis right after \qroam{}: the resulting phase depends only on the address, commutes with every element of the window, and is corrected together with the phase of the output register by $\qroam^\dagger$ at no extra cost. The junk qubits, like the unary-iteration qubits of each element, are therefore released for reuse. Therefore window $\omega$ needs
\begin{equation}\label{eq:window_ancilla}
    S_\omega=\max\Big(B_\omega\lambda_\omega+\lceil\log_2(L_{{\rm s},\omega}/\lambda_\omega)\rceil,\;
    B_\omega+\max\big(A_{\addpg},\,\lceil\log_2L_{{\rm t},\omega}\rceil-1\big),\;
    \lambda'_\omega+\lceil\log_2(L_{{\rm s},\omega}/\lambda'_\omega)\rceil\Big)
\end{equation}
clean ancillae, where $A_{\addpg}$ is the ancilla count of \addpg. The windows are executed one after another, so the whole circuit needs $S=\max_\omega S_\omega$ clean ancillae, in addition to the $d$ system qubits and the $b$ qubits of the phase-gradient state.

\paragraph{Multiplexed unitaries.}  Generalisation of this approach to $V$ multiplexed unitaries is straightforward. Multiplexing increases the number of shared control qubits by $\lceil\log_2V\rceil$ and thus, the way the cost changes is by replacing address register of QROAM with:
\begin{equation}
    L_{{\rm s},\omega} \rightarrow V L_{{\rm s},\omega}.
\end{equation}

\paragraph{Asymptotic cost.} To obtain the asymptotic Toffoli cost, we consider the case where all windows have the same number of layers and of target qubits, $(\ell,g)$, and we keep $\lambda=1$. Using~\cref{eq:total_toffoli,eq:element_costs}, the leading Toffoli cost of a window per layer, is
\begin{equation}\label{eq:per_layer}
    \frac{T_\omega}{\ell}\simeq\frac{V2^{d-g}}{\ell}+b\,2^{g-1},
\end{equation}
where the first term is the \qroam{} cost, shared by the $\ell$ layers of the window, and the second is $\selxor$; \addpg{}, $\selz$ and $\qroam^\dagger$ are subdominant. Adding a qubit to the targets halves the address space of \qroam{} but doubles the cost of $\selxor$, so the cost is minimal at an intermediate
$g$, and it is lower the more layers a window has for a given $g$. The ordering $s(l)$ enters the cost only through $\ell(g)$, the largest number of consecutive layers that target $g$ qubits. We discuss different orderings and their asymptotic costs in the next section. 

\subsection{Orderings of multiplexed rotations}\label{sec:us_orderings}
\begin{figure*}[htb!]
\centering
\textit{(a) Cyclic}\\[2pt]
\scalebox{1}{%
\begin{tikzpicture}
\begin{yquant}
qubit {$\ket{p}_{0}$} p0;
qubit {$\ket{p}_{1}$} p1;
qubit {$\ket{p}_{2}$} p2;
qubit {$\ket{p}_{3}$} p3;
[multictrl, name=cyc0] box {$R_y$} p0 ~ p1, p2, p3;
[multictrl, name=cyc1] box {$R_y$} p1 ~ p0, p2, p3;
[multictrl, name=cyc2] box {$R_y$} p2 ~ p0, p1, p3;
[multictrl, name=cyc3] box {$R_y$} p3 ~ p0, p1, p2;
[multictrl, name=cyc4] box {$R_y$} p0 ~ p1, p2, p3;
[multictrl, name=cyc5] box {$R_y$} p1 ~ p0, p2, p3;
[multictrl, name=cyc6] box {$R_y$} p2 ~ p0, p1, p3;
[multictrl, name=cyc7] box {$R_y$} p3 ~ p0, p1, p2;
[multictrl, name=cyc8] box {$R_y$} p0 ~ p1, p2, p3;
[multictrl, name=cyc9] box {$R_y$} p1 ~ p0, p2, p3;
[multictrl, name=cyc10] box {$R_y$} p2 ~ p0, p1, p3;
[multictrl, name=cyc11] box {$R_y$} p3 ~ p0, p1, p2;
[multictrl, name=cyc12] box {$R_y$} p0 ~ p1, p2, p3;
[multictrl, name=cyc13] box {$R_y$} p1 ~ p0, p2, p3;
[multictrl, name=cyc14] box {$R_y$} p2 ~ p0, p1, p3;
\end{yquant}
\node[draw, dashed, rounded corners, inner sep=2pt, fit=(cyc0) (cyc1)] {};
\node[draw, dashed, rounded corners, inner sep=2pt, fit=(cyc2) (cyc3)] {};
\node[draw, dashed, rounded corners, inner sep=2pt, fit=(cyc4) (cyc5)] {};
\node[draw, dashed, rounded corners, inner sep=2pt, fit=(cyc6) (cyc7)] {};
\node[draw, dashed, rounded corners, inner sep=2pt, fit=(cyc8) (cyc9)] {};
\node[draw, dashed, rounded corners, inner sep=2pt, fit=(cyc10) (cyc11)] {};
\node[draw, dashed, rounded corners, inner sep=2pt, fit=(cyc12) (cyc13)] {};
\node[draw, dashed, rounded corners, inner sep=2pt, fit=(cyc14)] {};
\end{tikzpicture}
}\\[4pt]
\textit{(b) Triangle wave}\\[2pt]
\scalebox{1}{%
\begin{tikzpicture}
\begin{yquant}
qubit {$\ket{p}_{0}$} p0;
qubit {$\ket{p}_{1}$} p1;
qubit {$\ket{p}_{2}$} p2;
qubit {$\ket{p}_{3}$} p3;
[multictrl, name=tri0] box {$R_y$} p1 ~ p0, p2, p3;
[multictrl, name=tri1] box {$R_y$} p0 ~ p1, p2, p3;
[multictrl, name=tri2] box {$R_y$} p1 ~ p0, p2, p3;
[multictrl, name=tri3] box {$R_y$} p2 ~ p0, p1, p3;
[multictrl, name=tri4] box {$R_y$} p3 ~ p0, p1, p2;
[multictrl, name=tri5] box {$R_y$} p2 ~ p0, p1, p3;
[multictrl, name=tri6] box {$R_y$} p1 ~ p0, p2, p3;
[multictrl, name=tri7] box {$R_y$} p0 ~ p1, p2, p3;
[multictrl, name=tri8] box {$R_y$} p1 ~ p0, p2, p3;
[multictrl, name=tri9] box {$R_y$} p2 ~ p0, p1, p3;
[multictrl, name=tri10] box {$R_y$} p3 ~ p0, p1, p2;
[multictrl, name=tri11] box {$R_y$} p2 ~ p0, p1, p3;
[multictrl, name=tri12] box {$R_y$} p1 ~ p0, p2, p3;
[multictrl, name=tri13] box {$R_y$} p0 ~ p1, p2, p3;
[multictrl, name=tri14] box {$R_y$} p1 ~ p0, p2, p3;
\end{yquant}
\node[draw, dashed, rounded corners, inner sep=2pt, fit=(tri0) (tri1) (tri2)] {};
\node[draw, dashed, rounded corners, inner sep=2pt, fit=(tri3) (tri4) (tri5)] {};
\node[draw, dashed, rounded corners, inner sep=2pt, fit=(tri6) (tri7) (tri8)] {};
\node[draw, dashed, rounded corners, inner sep=2pt, fit=(tri9) (tri10) (tri11)] {};
\node[draw, dashed, rounded corners, inner sep=2pt, fit=(tri12) (tri13) (tri14)] {};
\end{tikzpicture}
}\\[4pt]
\textit{(c) CSD}\\[2pt]
\scalebox{1}{%
\begin{tikzpicture}
\begin{yquant}
qubit {$\ket{p}_{0}$} p0;
qubit {$\ket{p}_{1}$} p1;
qubit {$\ket{p}_{2}$} p2;
qubit {$\ket{p}_{3}$} p3;
[multictrl, name=csd0] box {$R_y$} p3 ~ p0, p1, p2;
[multictrl, name=csd1] box {$R_y$} p2 ~ p0, p1, p3;
[multictrl, name=csd2] box {$R_y$} p3 ~ p0, p1, p2;
[multictrl, name=csd3] box {$R_y$} p1 ~ p0, p2, p3;
[multictrl, name=csd4] box {$R_y$} p3 ~ p0, p1, p2;
[multictrl, name=csd5] box {$R_y$} p2 ~ p0, p1, p3;
[multictrl, name=csd6] box {$R_y$} p3 ~ p0, p1, p2;
[multictrl, name=csd7] box {$R_y$} p0 ~ p1, p2, p3;
[multictrl, name=csd8] box {$R_y$} p3 ~ p0, p1, p2;
[multictrl, name=csd9] box {$R_y$} p2 ~ p0, p1, p3;
[multictrl, name=csd10] box {$R_y$} p3 ~ p0, p1, p2;
[multictrl, name=csd11] box {$R_y$} p1 ~ p0, p2, p3;
[multictrl, name=csd12] box {$R_y$} p3 ~ p0, p1, p2;
[multictrl, name=csd13] box {$R_y$} p2 ~ p0, p1, p3;
[multictrl, name=csd14] box {$R_y$} p3 ~ p0, p1, p2;
\end{yquant}
\node[draw, dashed, rounded corners, inner sep=2pt, fit=(csd0) (csd1) (csd2)] {};
\node[draw, dashed, rounded corners, inner sep=2pt, fit=(csd3)] {};
\node[draw, dashed, rounded corners, inner sep=2pt, fit=(csd4) (csd5) (csd6)] {};
\node[draw, dashed, rounded corners, inner sep=2pt, fit=(csd7)] {};
\node[draw, dashed, rounded corners, inner sep=2pt, fit=(csd8) (csd9) (csd10)] {};
\node[draw, dashed, rounded corners, inner sep=2pt, fit=(csd11)] {};
\node[draw, dashed, rounded corners, inner sep=2pt, fit=(csd12) (csd13) (csd14)] {};
\end{tikzpicture}
}\\[4pt]
\textit{(d) Triangle wave + CSD}\\[2pt]
\scalebox{1}{%
\begin{tikzpicture}
\begin{yquant}
qubit {$\ket{p}_{0}$} p0;
qubit {$\ket{p}_{1}$} p1;
qubit {$\ket{p}_{2}$} p2;
qubit {$\ket{p}_{3}$} p3;
[multictrl, name=nst0] box {$R_y$} p1 ~ p0, p2, p3;
[multictrl, name=nst1] box {$R_y$} p0 ~ p1, p2, p3;
[multictrl, name=nst2] box {$R_y$} p1 ~ p0, p2, p3;
[multictrl, name=nst3] box {$R_y$} p3 ~ p0, p1, p2;
[multictrl, name=nst4] box {$R_y$} p2 ~ p0, p1, p3;
[multictrl, name=nst5] box {$R_y$} p3 ~ p0, p1, p2;
[multictrl, name=nst6] box {$R_y$} p1 ~ p0, p2, p3;
[multictrl, name=nst7] box {$R_y$} p0 ~ p1, p2, p3;
[multictrl, name=nst8] box {$R_y$} p1 ~ p0, p2, p3;
[multictrl, name=nst9] box {$R_y$} p3 ~ p0, p1, p2;
[multictrl, name=nst10] box {$R_y$} p2 ~ p0, p1, p3;
[multictrl, name=nst11] box {$R_y$} p3 ~ p0, p1, p2;
[multictrl, name=nst12] box {$R_y$} p1 ~ p0, p2, p3;
[multictrl, name=nst13] box {$R_y$} p0 ~ p1, p2, p3;
[multictrl, name=nst14] box {$R_y$} p1 ~ p0, p2, p3;
\end{yquant}
\node[draw, dashed, rounded corners, inner sep=2pt, fit=(nst0) (nst1) (nst2)] {};
\node[draw, dashed, rounded corners, inner sep=2pt, fit=(nst3) (nst4) (nst5)] {};
\node[draw, dashed, rounded corners, inner sep=2pt, fit=(nst6) (nst7) (nst8)] {};
\node[draw, dashed, rounded corners, inner sep=2pt, fit=(nst9) (nst10) (nst11)] {};
\node[draw, dashed, rounded corners, inner sep=2pt, fit=(nst12) (nst13) (nst14)] {};
\end{tikzpicture}
}\\[4pt]
\textit{(e) Cyclic + CSD}\\[2pt]
\scalebox{1}{%
\begin{tikzpicture}
\begin{yquant}
qubit {$\ket{p}_{0}$} p0;
qubit {$\ket{p}_{1}$} p1;
qubit {$\ket{p}_{2}$} p2;
qubit {$\ket{p}_{3}$} p3;
[multictrl, name=ccs0] box {$R_y$} p0 ~ p1, p2, p3;
[multictrl, name=ccs1] box {$R_y$} p1 ~ p0, p2, p3;
[multictrl, name=ccs2] box {$R_y$} p3 ~ p0, p1, p2;
[multictrl, name=ccs3] box {$R_y$} p2 ~ p0, p1, p3;
[multictrl, name=ccs4] box {$R_y$} p3 ~ p0, p1, p2;
[multictrl, name=ccs5] box {$R_y$} p0 ~ p1, p2, p3;
[multictrl, name=ccs6] box {$R_y$} p1 ~ p0, p2, p3;
[multictrl, name=ccs7] box {$R_y$} p3 ~ p0, p1, p2;
[multictrl, name=ccs8] box {$R_y$} p2 ~ p0, p1, p3;
[multictrl, name=ccs9] box {$R_y$} p3 ~ p0, p1, p2;
[multictrl, name=ccs10] box {$R_y$} p0 ~ p1, p2, p3;
[multictrl, name=ccs11] box {$R_y$} p1 ~ p0, p2, p3;
[multictrl, name=ccs12] box {$R_y$} p3 ~ p0, p1, p2;
[multictrl, name=ccs13] box {$R_y$} p2 ~ p0, p1, p3;
[multictrl, name=ccs14] box {$R_y$} p3 ~ p0, p1, p2;
\end{yquant}
\node[draw, dashed, rounded corners, inner sep=2pt, fit=(ccs0) (ccs1)] {};
\node[draw, dashed, rounded corners, inner sep=2pt, fit=(ccs2) (ccs3) (ccs4)] {};
\node[draw, dashed, rounded corners, inner sep=2pt, fit=(ccs5) (ccs6)] {};
\node[draw, dashed, rounded corners, inner sep=2pt, fit=(ccs7) (ccs8) (ccs9)] {};
\node[draw, dashed, rounded corners, inner sep=2pt, fit=(ccs10) (ccs11)] {};
\node[draw, dashed, rounded corners, inner sep=2pt, fit=(ccs12) (ccs13) (ccs14)] {};
\end{tikzpicture}
}\\[4pt]
\caption{The five orderings for $d=4$ with $N_S=15$ layers. Dashed boxes show a tiling of each sequence into windows with at most $g=2$ target qubits; the width of a box is $\ell_\omega$ and its height is $g_\omega$. (a) Cyclic: windows with $\ell_\omega=g_\omega=2$. (b) Triangle wave: every window is centred on a reflection, $\ell_\omega=2g_\omega-1=3$. (c) CSD: aligned windows with $\ell_\omega=2^{g_\omega}-1=3$, separated by single layers on qubits $0$ and $1$. (d) Triangle wave + CSD with groups $\{0,1\}$ and $\{2,3\}$: every block is a window with $\ell_\omega=3$. (e) Cyclic + CSD with groups $\{0\}$, $\{1\}$ and $\{2,3\}$: the blocks on $\{2,3\}$ are windows with $\ell_\omega=3$, and each pair of single-qubit blocks forms a window with $\ell_\omega=g_\omega=2$; with the two groups of (d), it would give the same sequence as (d). For $g=2$, the triangle wave reaches the same $\ell_\omega=3$ as CSD because $2g-1=2^g-1$; for $g\ge3$ it falls behind ($5$ against $7$ for $g=3$).}
\label{fig:orderings_n4}
\end{figure*}
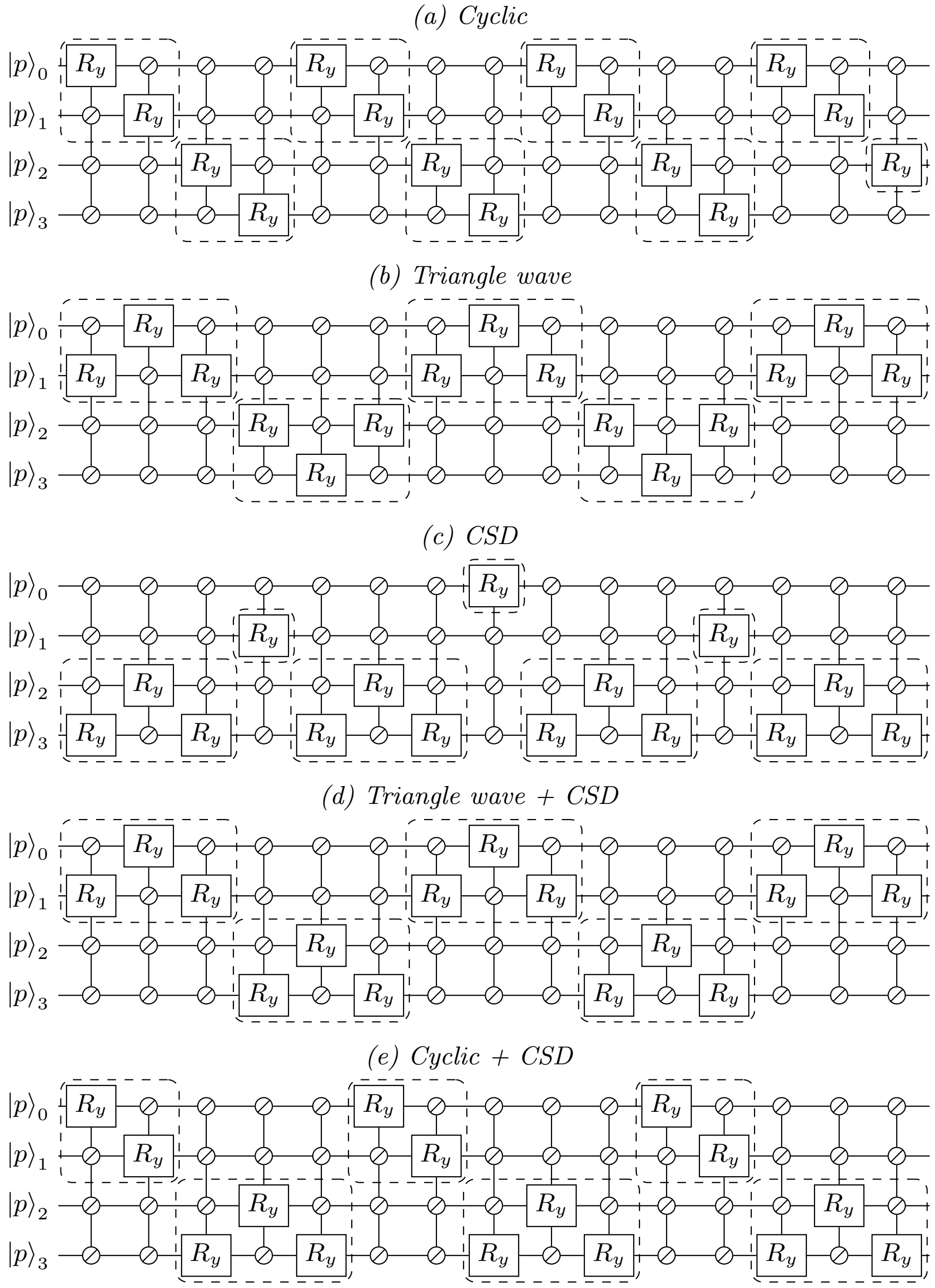

We consider five orderings $s(l)$, shown in~\cref{fig:orderings_n4} for $d=4$. They differ in two respects. The first is the length $\ell(g)$ of the longest windows with $g$ target qubits, which sets the asymptotic
Toffoli cost, obtained by minimising~\eqref{eq:per_layer} over $g$. The number of layers $N_S$ only multiplies the cost per layer, so it does not change the optimal $g$, as long as a window fits into the sequence, $N_S\ge\ell(g)$. Therefore, in this estimate, the window lengths are the same for full-depth and truncated unitaries, and we set $N_S\propto D$, as at full depth, $N_S=D-1$, only to express the total cost in terms of $D$. Since a window acts on its targets as an orthogonal matrix of dimension $2^g$, the number of its layers is bounded by $\ell(g)\le2^g-1$ (see the beginning of~\cref{sec:unitary_synthesis}). The second is how the rotations are
distributed over the qubits, which affects the trainability of the ansatz, i.e. how hard it is to find the optimal angles. We compile each unitary for a given ancilla budget. For very large ancilla budgets, the Toffoli counts of the orderings differ. However, as we will see, at the ancilla budget that minimises the quantum volume, all orderings give very similar quantum volumes, so the choice of the ansatz in practice for this regime is driven by its trainability. The asymptotic cost compares the orderings at a fixed number of layers $N_S$. However, the number of layers needed to reach a given accuracy also depends on the ordering, so an ordering with a lower cost per layer is not necessarily the cheapest in practice. In our quantum-chemistry examples, at the same number of layers, CSD reaches the target accuracy less often than the cyclic ordering, whereas the cyclic ordering and triangle wave + CSD reach it equally often, the former with a somewhat lower $\lambda_{\rm eff}$ in the larger systems.

\begin{enumerate}
\item \textit{Cyclic.} $s(l)=l\bmod d$. Any $\ell<d$ consecutive layers target $\ell$ different qubits, so $\ell(g)=g$. Every qubit is targeted once in every $d$ consecutive layers, so the rotations are distributed uniformly, $N_S/d$ per qubit. The optimum is at $2^{2g}\simeq2VD/(bg)$, so
\begin{equation}\label{eq:cyclic_scaling}
    T_{\rm cyc}\simeq N_S\sqrt{\frac{2VbD}{g}},\qquad g\simeq\tfrac12\log_2\frac{2VD}{b},
\end{equation}
i.e.\ $T_{\rm cyc}=O\big(D^{3/2}\sqrt{Vb/\log(VD/b)}\big)$.
\item \textit{Triangle wave.} The target advances by one qubit and reflects at qubits $0$ and $d-1$. When we choose the number of layers, $N_S < D$, we use a symmetric triangular wave so the number of layers is always odd. The rotations are distributed almost uniformly. A window centred on a reflection of the wave has $\ell(g)=2g-1$, all others $\ell(g)=g$. Since reflections occur only once every $d-1$ layers, the scaling is that of~\eqref{eq:cyclic_scaling}, and the prefactor improves by at most $\sqrt2$. We include the triangle wave mainly as an ingredient of the nested ordering triangle wave + CSD below: on its own, it has the cost scaling of the cyclic ordering, and the nested orderings are cheaper.
\item \textit{Cosine-sine decomposition (CSD).} Distribution of rotations and its angles can be obtained from exact cosine-sine decomposition. It follows the tree structure where the top qubit generates two children on lower qubits. This leads to highly unbalanced distribution: qubit $q$ is targeted $2^q$ times, so qubit $0$ has a single rotation.
When the number of layers is lower than $D$, one cannot find angles using cosine-sine decomposition and the training becomes hard due to highly unbalanced distribution of rotations. The longest window size is $\ell(g)=2^g-1$, which saturates the bound, and consecutive windows are separated by a single layer.  The Toffoli minimum is at $2^g\simeq(4VD/b)^{1/3}$, so
\begin{equation}\label{eq:csd_scaling}
    T_{\rm CSD}\simeq3\cdot2^{-4/3}N_S\,b^{2/3}(VD)^{1/3}=O\big(b^{2/3}V^{1/3}D^{4/3}\big).
\end{equation}
Again we present this ordering as an ingredient to the next two nested orderings.
\item \textit{Triangle wave + CSD.} We combine consecutive qubits into non-overlapping groups. Within each group, we apply the CSD ordering, i.e.\ a \textit{block} of $2^{g}-1$ consecutive layers on the $g$ qubits of the group, and the groups follow each other in triangle-wave order. Every block in this nested ordering is a window with $\ell(g)=2^g-1$, and, unlike for CSD, no separating layers or alignment are needed. Therefore the scaling is~\eqref{eq:csd_scaling}, with groups of size $g\simeq\frac13\log_2(4VD/b)$, provided that $g\le d/2$. We label a nested ordering by its partition $g_1$-$g_2$-$\dots$-$g_m$ -- the sizes of the groups listed from qubit $0$ up to qubit $d-1$, with $\sum_i g_i=d$. For example, for $d=4$ the partition 2-2 gives the groups $\{0,1\}$ and $\{2,3\}$ (\cref{fig:orderings_n4}(d)), and 1-1-2 gives $\{0\}$, $\{1\}$ and $\{2,3\}$ (\cref{fig:orderings_n4}(e)). The partition is fixed before training, and we choose it to minimise the cost at the given ancilla budget. 
\item \textit{Cyclic + CSD.} The same blocks visit the groups in cyclic order, and we keep the first $N_S$ layers. It has the same $\ell(g)=2^g-1$ and the same scaling~\eqref{eq:csd_scaling} as triangle wave + CSD, however rotations are distributed more uniformly here.
\end{enumerate}
The nested orderings are our main motivation for extending Tan's construction to arbitrary orderings. CSD is the nested ordering with a single group, and using several groups improves on it in two ways. (i) Every block is a window with $\ell(g)=2^g-1$, and no separating single layers are needed, which reduces the constant factors of the Toffoli cost. (ii) The rotations are distributed much more uniformly over the qubits, which we expect to make the ansatz easier to train (\cref{sec:us_trainability}).

\begin{figure*}[htb!]
\centering
\includegraphics[width=\textwidth]{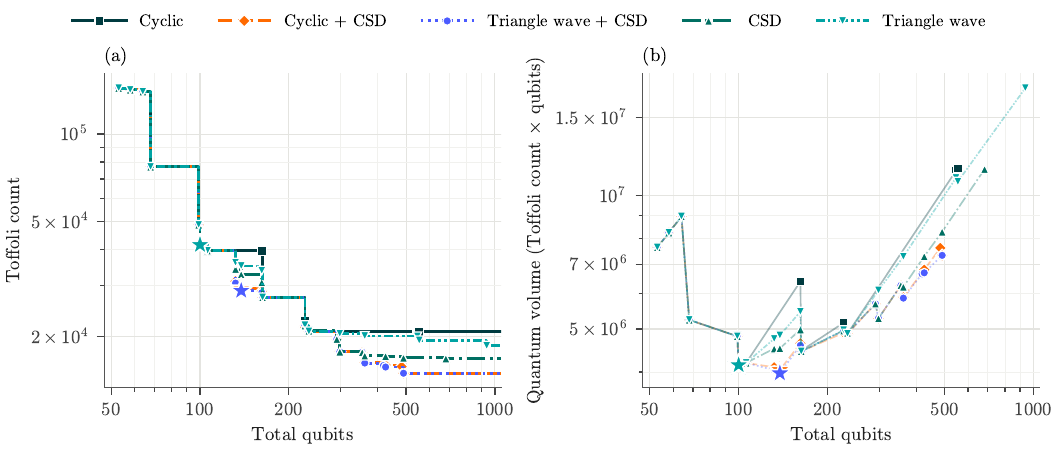}
\caption{(a) Toffoli count and (b) quantum volume, i.e.\ Toffoli count times total qubits, of one multiplexed unitary with $D=128$, $V=16$, $b=16$ and $N_S=127$ layers, against the total number of qubits. The total includes $d+\lceil\log_2V\rceil=11$ system qubits and $b=16$ qubits of the phase-gradient state; the remaining qubits are clean ancillae. Each point is the Toffoli-optimal compilation within the corresponding ancilla budget,~\cref{eq:total_toffoli,eq:window_ancilla}. For the nested orderings, the cheapest partition into groups is chosen at every budget. Stars mark the minimum quantum volume of each ordering; those of the cyclic, triangle-wave and CSD orderings coincide.}
\label{fig:ordering_fronts}
\end{figure*}

\Cref{fig:ordering_fronts} shows the Toffoli count and the quantum volume of the five orderings as functions of the total number of qubits, for $D=128$, $V=16$ and $b=16$. The minimal quantum volumes of all orderings are within $5\%$ of each other. The minimal Toffoli counts differ more: that of the nested orderings is $28\%$ lower than that of the cyclic ordering and $11\%$ lower than that of CSD. The orderings differ because the layers of a window share one \qroam{} read, and the cost of this read grows with $V$. For $V=1$, all orderings have the same minimal quantum volume, and the nested orderings reduce the minimal Toffoli count by only $11\%$ for $b=16$ and $7\%$ for $b=32$ compared with the cyclic ordering.

\subsection{Trainability}\label{sec:us_trainability}
Our goal is to approximate the Hamiltonian with a factorised representation, not to approximate arbitrary given unitaries. Therefore, we evaluate the orderings based on (1) the accuracy of the energy, (2) the effective subnormalisation factor $\lambda_{\rm eff}$ of the block-encoding, (3) the cost of the block-encoding, and (4) the classical computational effort needed to approximate the Hamiltonian within a given accuracy. For the latter, we run the optimisation starting from several initial guesses and with different numbers of iterations, and count how many runs reach the target accuracy. The general trainability of these orderings, e.g. for approximating a given unitary, is beyond the scope of this paper.

As a representative example, we consider the 14 valence electrons of permanganate, [MnO$_4$]$^-$, in $D=64$ orbitals. First, we optimise the factorisation with $V=16$, $L=64$ and $M=64$ for 25k steps, without representing $\hat R^{(v)}$ by its SVD~\eqref{eq:R_svd}. In this step, the elements of $\hat R^{(v)}$ are the variational parameters, so it involves neither rotation angles nor an ordering. This provides the best solutions with unitaries at full depth, $N_S=D-1$. We then compute the SVD of $\hat U_{\rm ref}^{\dagger}\hat R^{(v)}$ and fit the angles of $N_S=41$ layers of a given ordering to each of its two unitaries, $\hat{\tilde U}^{(v)}$ and $\hat V^{(v)}$, by least squares. For $N_S<D-1$, the fit is only approximate, so we take the fitted angles as the initial guess and re-optimise the factorisation in the rotated frame,~\cref{eq:R_rotated}. All orderings start from the same full-depth solutions and have the same number of angles, $N_SD/2$ per unitary, so the classical cost of a run is practically the same for all of them. In both steps, we use the regularisation parameters $\epsilon_{\rm fro}=10^{-3}$, $\epsilon_{\Lambda}=8$, $\epsilon_{\rm gap}=0.1$ and $\epsilon_{\rm HF}=5\times10^{-3}$ (all in Ha), and $\eta_{\rm HF}=10^{-4}$ in~\cref{eq: objective_function}.

\begin{figure*}[htb!]
\centering
\includegraphics[width=\textwidth]{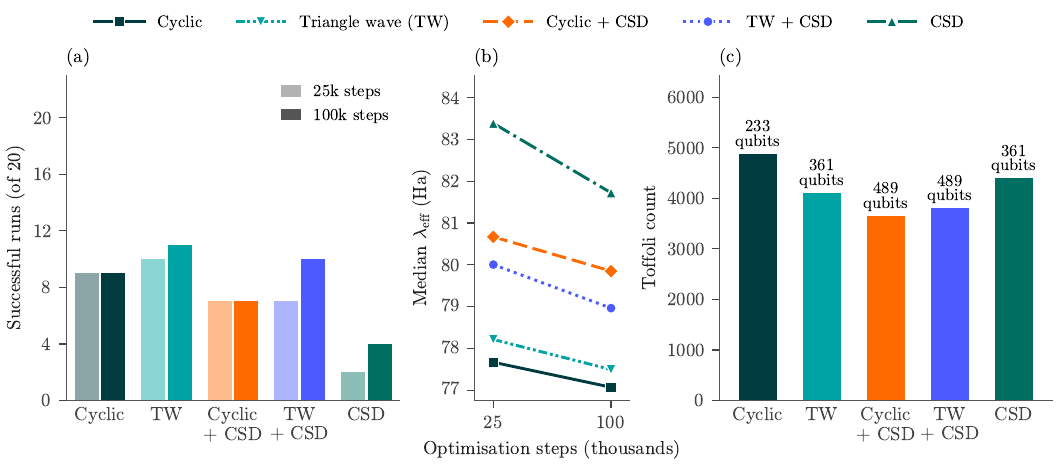}
\caption{Performance of the orderings when the factorised Hamiltonian of [MnO$_4$]$^-$ ($D=64$) is approximated with unitaries truncated to $N_S=41$ layers; 20 runs per ordering, each starting from a different full-depth solution. (a) Number of runs with $\max(|\Delta E_{\rm tot}|,|\Delta E_{\rm corr}|)<0.3$~mHa after 25k (light) and 100k (dark) optimisation steps. (b) Median $\lambda_{\rm eff}$ of all runs. (c) Minimal Toffoli count of a single multiplexed unitary with $D=64$, $V=16$, $b=16$ and $N_S=41$; the numbers above the bars are the total numbers of qubits of the multiplexed unitary. Cyclic + CSD and triangle wave + CSD use the Toffoli-optimal partitions 3-3 and 1-2-2-1, respectively.}
\label{fig:trainability}
\end{figure*}

The results are shown in~\cref{fig:trainability}. For each ordering, we perform 20 runs, each starting from a different full-depth solution, and count a run as successful if $\max(|\Delta E_{\rm tot}|,|\Delta E_{\rm corr}|)<0.3$~mHa, i.e.\ if both the total-energy and the correlation-energy errors are below $0.3$~mHa. We estimate the energy errors with UCCSD(T) on an unrestricted Hartree--Fock reference, computed with PySCF~\cite{sun_python_2026}. After 25k optimisation steps (\cref{fig:trainability}(a)), the cyclic, triangle-wave, cyclic + CSD and triangle wave + CSD orderings succeed in 7--10 runs out of 20, whereas CSD succeeds in 2. With 100k steps, triangle wave + CSD improves to 10 and CSD to 4 successful runs, while the other orderings change by at most one run. Therefore, CSD is the hardest ordering to train. 

Among the solutions that reach the target accuracy, we choose the ordering by its cost, which is determined by $\lambda_{\rm eff}$ and the Toffoli and qubit counts. As seen in~\cref{fig:trainability}(b,c), the cyclic + CSD ordering has the lowest Toffoli count, $25\%$ below the cyclic ordering, while its $\lambda_{\rm eff}$ is within $4\%$ of that of the cyclic ordering. Since the total cost is proportional to the product of the Toffoli count and $\lambda_{\rm eff}$, cyclic + CSD is $22\%$ cheaper than cyclic and $8\%$ cheaper than the triangle wave. A higher success rate than CSD for the same number of optimisation steps, and a lower Toffoli count compared with the uniform orderings (cyclic and triangle wave), were the main motivation for developing the nested orderings.

Finding optimal factorisations requires substantial classical resources, especially for larger basis sets, $D>100$. For example, for [MnO$_4$]$^-$ with $D=128$, a single run of 50k steps in the rotated frame takes about 29 minutes on one NVIDIA H100 GPU (about 73 minutes on an L40), so 10 runs for each of six combinations of ordering and $N_S$ take about 30 GPU-hours for a single set of regularisation parameters, and in practice several regularisation parameters have to be explored. Therefore, an ordering that reaches the desired accuracy in fewer optimisation steps is important.

\section{Resource Estimates}\label{sec:resource_estimates}
The overall cost of the quantum phase estimation algorithm is given by the product of the cost of the block-encoding, $C_{\rm B.E.}$, and the number of calls, $N_{W}$, to the quantum walk operator. In order to compare the resource requirements of the algorithm to the second-quantised counterpart, we consider several chemical examples: an H$_4$ molecule in def2-SVPD and def2-TZVPP basis sets, and four transition-metal complexes, [Fe$_2$S$_2$]$^{2-}$, [MnO$_4$]$^-$, [Mn$_2$O$_2$]$^{3+}$ and [Mn$_2$O$_3$]$^{2+}$, with 14, 14, 19 and 24 active electrons, respectively, in 122, 128, 128 and 128 orbitals.
The three manganese complexes are high-valent manganese--oxo oxidants: permanganate and the bis-$\mu$-oxo dimer oxidise C--H bonds of hydrocarbons~\cite{gardner_mayer_1995,wang_mayer_1997,larsen_2002}, and oxo-bridged manganese dimers are synthetic models of the oxygen-evolving complex of photosystem~II~\cite{wieghardt_1988,roemelt_2018}. They are also established test cases for electron correlation: the absorption spectrum of permanganate requires multiconfigurational methods~\cite{olsen_modeling_2017}, and the exchange coupling between the antiferromagnetically coupled manganese centres of the dimers is sensitive to both static and dynamical correlation~\cite{orio_2009,roemelt_2018}. The molecular models and the choice of active spaces are described in~\cref{app:chemical_systems}.

To separate the contributions to the total QPE cost, we analyse (i) the cost of the block-encoding, which does not depend on $\lambda_{\rm eff}$ and depends only on the number of electrons, $N$, the number of orbitals, $D$, and the factorisation parameters $(V,L,M)$; (ii) the effective subnormalisation factor $\lambda_{\rm eff}$, which is system specific and determines $N_W$,~\cref{eq: no qpe calls}; (iii) the total cost of QPE which is the product of (i) and (ii) for a given ancilla budget.

Although we do not perform detailed resource estimates for quantum error correction, we assume the architecture of Ref.~\cite{low_denser_2026}, which is based on a denser planar surface code. In this architecture, the least frequently accessed qubits are stored in a dense but slow memory region, the cold storage, and therefore they cannot be borrowed as dirty ancillae. The other logical qubits are kept in a faster but less dense hot storage, from which they are accessed by the compute region, where lattice surgery computation is performed and magic states are produced and consumed. In our cost model, the cold storage holds the following registers. In second quantisation, as in Ref.~\cite{low_denser_2026}, these are the two QPE qubits, the $2D$ qubits of the spin orbitals and the two spin indices of the block-encoding. In first quantisation, these are the two QPE qubits, the electron registers in positions $1,\dots,N-1$ and the electron index $\ket{j}_{ne}$ (\cref{fig: quantum_circuit_be_sum_over_electrons}). The electron registers are accessed only by the SWAP network. The electron index, like the spin index that selects the spin sector in Ref.~\cite{low_denser_2026}, is accessed only by UNIFORM, the SWAP network and the reflection of the walk operator. All other qubits are kept in the hot storage. These include the electron register in position 0, since the block-encoding of $\hat W^{(v,l)}$ acts on it. This placement of the registers is the only feature of the architecture that enters our estimates, apart from the relative cost of the qubits in the cold storage in~\cref{sec:qpe_cold_weight}. We count logical qubits and Toffoli gates, and we do not model the lattice surgery or the latency of accessing the cold storage. In both approaches, we compile the table lookups with clean ancillae and with dirty ancillae following Ref.~\cite{low_denser_2026}. The cost model is summarised in~\cref{app:cost_model}.

In second quantisation, we use the sum-of-squares representation of Ref.~\cite{low_fast_2025}; our implementation is described and validated in~\cref{app:sq_implementation}. Both block encodings are compiled with the same techniques: table lookups with measurement-based uncomputation~\cite{berry_qubitization_2019,low_trading_2024}, rotations by addition into a phase-gradient state~\cite{gidney_halving_2018,low_denser_2026}, and the same rules for ancilla qubits. Our cost model of the second-quantised block-encoding reproduces the costs reported in Ref.~\cite{low_denser_2026} (\cref{fig:sq_validation}). We do not consider the most qubit-frugal compilations of Ref.~\cite{low_denser_2026}, which synthesise every rotation from Clifford+T gates and use ancilla-free arithmetic throughout (of the latter, we include the ancilla-free adders). They would extend both fronts to fewer qubits, at several times the Toffoli count, but they would not lower the fronts shown here. The ratios quoted at the fewest qubits of the second-quantised algorithm therefore compare the two methods compiled alike; below that point, the second-quantised algorithm can run only with those further compilations, at a higher Toffoli count, while the first-quantised front continues to lower qubit counts. All qubit counts in this work are logical; in~\cref{sec:qpe_cold_weight}, we also use an effective qubit count, in which those in the cold storage are weighted to illustrate the effect of its denser encoding. The number of physical qubits and the run time depend on the error-correcting code and on how the algorithm is compiled into it, e.g.\ on the layout and the magic-state factories, which can differ between the two algorithms; we leave this compilation to future work.

\subsection{Cost of the block-encoding}
\begin{figure*}[htb!]
\centering
\includegraphics[width=\textwidth]{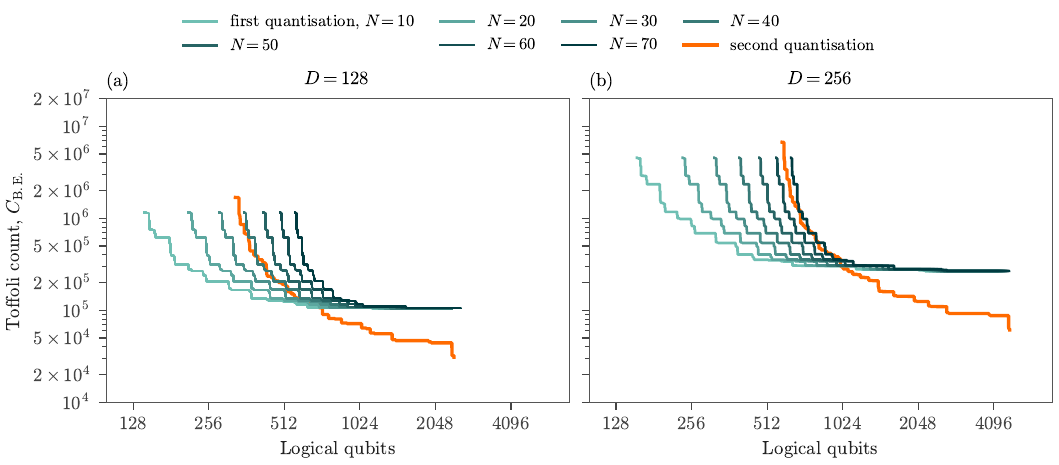}
\caption{Toffoli count of the block-encoding, $C_{\rm B.E.}$, against the number of logical qubits for (a) $D=128$ and (b) $D=256$ orbitals and $(V,L,M)=(16,D/2,D)$. $C_{\rm B.E.}$ is the cost of the block-encoding, and the number of qubits includes the two QPE qubits. In first quantisation, the multiplexed unitaries $\hat U^{(v)}$ and $\hat V^{(v)}$ use the cyclic + CSD ordering with $N_S=3D/4$ layers; the second-quantised block-encoding uses the same factorisation ranks and does not depend on $N$; $b=16$ in both representations. Each point is the Toffoli-optimal compilation at a given qubit budget, with clean and dirty ancillae in the cold-storage architecture.}
\label{fig:be_vs_electrons_D}
\end{figure*}

The wavefunction register requires $2D$ qubits in second quantisation and $N\lceil\log_2D\rceil$ qubits in first quantisation. For active spaces where $D$ is comparable to $N$, second quantisation requires fewer qubits; here, we establish for which numbers of electrons and orbitals first quantisation is advantageous. In~\cref{fig:be_vs_electrons_D}, we compare the cost of the block-encoding for representative factorisation parameters $(V,L,M)=(16,D/2,D)$, which for $D=128$ are those of the [MnO$_4$]$^-$ factorisations in~\cref{tab:factorisations}, and the varying number of electrons. The second-quantised block-encoding does not depend on $N$. In first quantisation, $N$ enters only through the electron registers, the SWAP network and the uniform superposition over electrons (\cref{fig: quantum_circuit_be_sum_over_electrons}). Therefore, for $D=128$, the minimal Toffoli count changes by less than $2\%$ between $N=10$ and $N=70$, whereas the front moves to larger qubit counts by $\lceil\log_2D\rceil=7$ qubits per electron. At the fewest qubits on which the second-quantised block-encoding can be implemented, 327, the first-quantised block-encoding requires 10.1 to 4.4 times fewer Toffoli gates for $N=10$ to $30$, and it remains cheaper up to 600--700 logical qubits. For $N\ge40$, the first-quantised front starts at more qubits than the second-quantised one, and second quantisation is cheaper at every qubit count: the advantage at low qubit counts disappears near the point where the two wavefunction registers become equal in size ($N\approx37$ for $D=128$ gives $2D\approx N\lceil\log_2D\rceil$).

The advantage at low qubit counts grows with the basis size, since doubling $D$ adds only $N$ qubits to the first-quantised register and $2D$ qubits to the second-quantised one, so for $N<D$ this provides advantage to first quantisation approach. For $D=256$, the first-quantised block-encoding requires 19.7 to 2.9 times fewer Toffoli gates for $N=10$ to $60$ at the fewest qubits of the second-quantised one, 587, and it remains cheaper up to about 1\,000 logical qubits, whereas for $N=70$, beyond the point where the two wavefunction registers become equal ($N=64$), second quantisation is cheaper at every qubit count. Therefore, one can expect an advantage at low qubit counts for up to $N\approx30$--$40$ electrons at $D=128$, and for $N\approx60$--$70$ at $D=256$. However, finding a factorisation requires GPU calculations that are already demanding at $D=128$ (\cref{sec:us_trainability}), and at $D=256$ the two-electron tensor alone grows from 2 to 34~GB, so every evaluation of the objective function~\eqref{eq: objective_function} is at least 16 times more expensive. Therefore, we consider $D=128$ with the numbers of electrons of the systems listed above to determine the full cost of the QPE circuit.

\begin{figure*}[htb!]
\centering
\includegraphics[width=\textwidth]{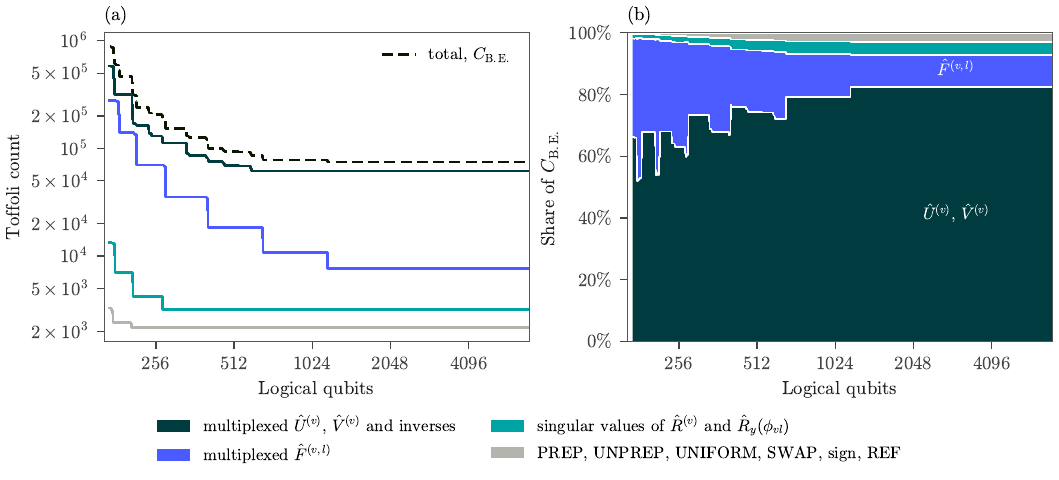}
\caption{Breakdown of the Toffoli count of the first-quantised block-encoding, $C_{\rm B.E.}$, for [MnO$_4$]$^-$ against the number of logical qubits: (a) Toffoli count of each part and (b) its share of $C_{\rm B.E.}$. The circuit is that of the first-quantised factorisation at the Toffoli-optimal end of~\cref{fig:qpe_fronts_cold}(d), with $(V,L,M)=(16,64,128)$ and the cyclic + CSD ordering with $N_S=65$ layers (\cref{tab:factorisations}). The parts are those of~\cref{fig: quantum_circuit_be_full,fig: quantum_circuit_be_sum_over_electrons,fig: quantum_circuit_be_rectungular_matrices,fig: quantum_circuit_be_multiplex_be}; the last group also contains the reflection and the remaining Toffoli gates of the walk operator.}
\label{fig:be_breakdown}
\end{figure*}

The first-quantised fronts flatten at much lower qubit counts than the second-quantised one, so that at the Toffoli-optimal end second quantisation requires 3.4 ($D=128$) and 4.4 ($D=256$) times fewer Toffoli gates. The reason is the unitary synthesis, which does not allow the same space--time trade-offs as the table lookups that dominate the cost of the second-quantised block-encoding. For [MnO$_4$]$^-$, at the Toffoli-optimal point (2\,195 logical qubits), the four multiplexed unitaries $\hat U^{(v)}$, $\hat V^{(v)}$ and their inverses account for $83\%$ of the Toffoli count (\cref{fig:be_breakdown}); the multiplexed unitaries reach their minimal Toffoli count already at about 600 logical qubits. Consequently, the last part of the first-quantised front reduces only the last few percent of the cost at the expense of a large number of qubits: from about 660 to 2\,195 logical qubits, the Toffoli count decreases by only $4\%$, whereas the second-quantised one in~\cref{fig:be_vs_electrons_D}(a), with the same factorisation ranks, still decreases by a factor of 3.7 from about 660 to 2\,430 logical qubits.

The multiplexed unitaries stop improving at about 600 logical qubits because of the trade-off between \qroam{} and \selxor{} in~\eqref{eq:per_layer}. At the fewest qubits, every window of $\hat U^{(v)}$ has a single layer, and \qroam{} and its uncomputation account for $99\%$ of its Toffoli count. Additional ancilla qubits allow longer windows (\cref{eq:window_ancilla}), in which several layers share one \qroam{} read. At the minimal Toffoli count of $\hat U^{(v)}$, most windows have $\ell=7$ layers that target $g=3$ qubits, and \qroam{} and its uncomputation account for $45\%$, \selxor{} for $40\%$ and \addpg{} for $13\%$. Windows with $g=4$ would reduce the \qroam{} cost but increase that of \selxor{} by more, so additional ancilla qubits cannot reduce the cost further.

\subsection{Scaling of the effective subnormalisation factor}
\begin{figure*}[htb!]
\centering
\includegraphics[width=\textwidth]{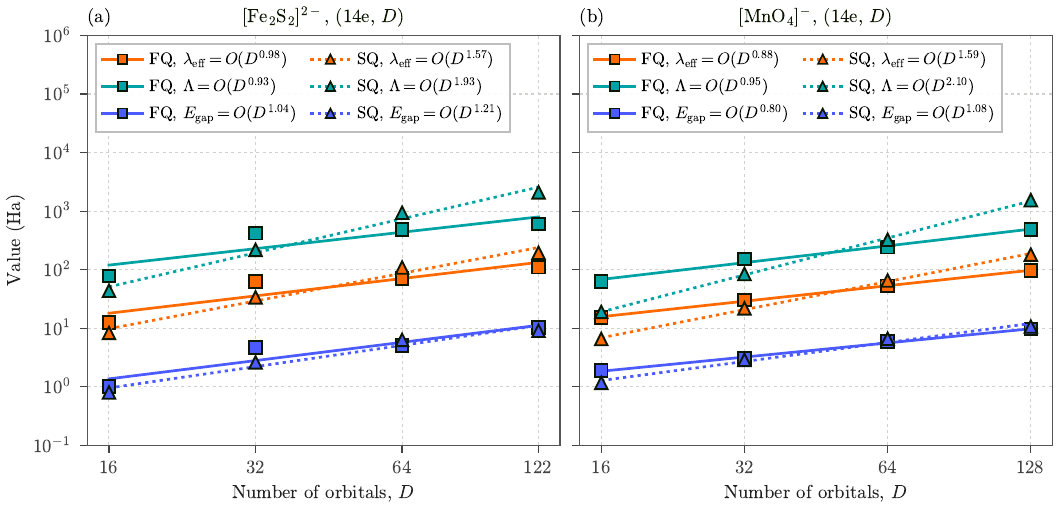}
\caption{$\lambda_{\rm eff}$ (orange), $\Lambda$ (teal) and $E_{\rm gap}$ (violet) in first (squares, solid lines) and second (triangles, dotted lines) quantisation as functions of the number of orbitals for (a) [Fe$_2$S$_2$]$^{2-}$ and (b) [MnO$_4$]$^-$ with 14 active electrons. Lines are power-law fits. Markers: the factorisation with the smallest $\lambda_{\rm eff}$ among those with $\max(|\Delta E_{\rm tot}|,|\Delta E_{\rm corr}|)<0.3$~mHa. Energy errors are estimated with UCCSD(T) on an unrestricted Hartree--Fock reference.}
\label{fig:lambda_vs_D}
\end{figure*}

In~\cref{fig:lambda_vs_D}, we plot $\lambda_{\rm eff}$ (together with $\Lambda$ and $E_{\rm gap}$) as a function of the number of orbitals, $D$, for a fixed number of electrons for [Fe$_2$S$_2$]$^{2-}$ and [MnO$_4$]$^-$. In first quantisation, we use the full-depth factorisations, i.e.\ the unitaries are not truncated to $N_S$ layers. For each $D$, we take the factorisation with the smallest $\lambda_{\rm eff}$ among those with $\max(|\Delta E_{\rm tot}|,|\Delta E_{\rm corr}|)<0.3$~mHa.

In first quantisation, $\lambda_{\rm eff}$ grows approximately linearly with the number of orbitals, as $O(D^{0.98})$ for [Fe$_2$S$_2$]$^{2-}$ and $O(D^{0.88})$ for [MnO$_4$]$^-$, whereas in second quantisation it grows as $O(D^{1.57})$ and $O(D^{1.59})$, respectively. The difference comes from $\Lambda$, which grows approximately as $D$ in first quantisation and as $D^2$ in second quantisation, while $E_{\rm gap}$ grows approximately linearly in both. Consequently, for small basis sets, $D\le32$, $\lambda_{\rm eff}$ is smaller in second quantisation, the fitted lines cross at $D\approx45$--$51$, and from $D=64$ the $\lambda_{\rm eff}$ is smaller in first quantisation, by factors of 1.8 and 1.9 at $D=122$ and $D=128$ for [Fe$_2$S$_2$]$^{2-}$ and [MnO$_4$]$^-$, respectively. Since $N_W\propto\lambda_{\rm eff}$,~\cref{eq: no qpe calls}, first quantisation requires about half as many calls to the quantum walk operator for $D\approx128$.

\Cref{tab:factorisations} lists $\lambda_{\rm eff}$ of the factorisations used for the resource estimates below. These factorisations use truncated unitaries and are chosen for the fewest QPE Toffoli gates rather than the smallest $\lambda_{\rm eff}$. For [Fe$_2$S$_2$]$^{2-}$ and [MnO$_4$]$^-$, $\lambda_{\rm eff}$ is again smaller in first quantisation. For H$_4$, it is smaller in first quantisation already for $D=32$, since $\Lambda$ contains the factor $N^2$ in first quantisation,~\cref{eq:Lambda}, and H$_4$ has only four electrons. For the same reason, the ratio $\lambda^{\rm SQ}_{\rm eff}/\lambda^{\rm FQ}_{\rm eff}$ decreases with the number of electrons for the three manganese systems, which all have $D=128$. For [Mn$_2$O$_2$]$^{3+}$, with 19 electrons, $\lambda_{\rm eff}$ is only slightly smaller in first quantisation, and for [Mn$_2$O$_3$]$^{2+}$, with 24 electrons, it is larger.

Having discussed the cost of the block-encoding and the scaling of the effective subnormalisation factor, we can now estimate the total cost of QPE.

\subsection{Total QPE cost}

\subsubsection{Logical resource estimates}

We perform the resource estimates for an architecture in which the system register, apart from the electron register in position 0 in first quantisation, the QPE qubits and the electron or spin indices are kept in the cold storage, a dense but slow memory region, so that they cannot be borrowed as dirty ancillae, as in Ref.~\cite{low_denser_2026}. For completeness, the QPE cost in a monolithic architecture, where every idle qubit, including those of the system register, can be borrowed as dirty ancillae, is given in~\cref{app:monolithic}.

\begin{figure*}[htb!]
\centering
\includegraphics[width=\textwidth]{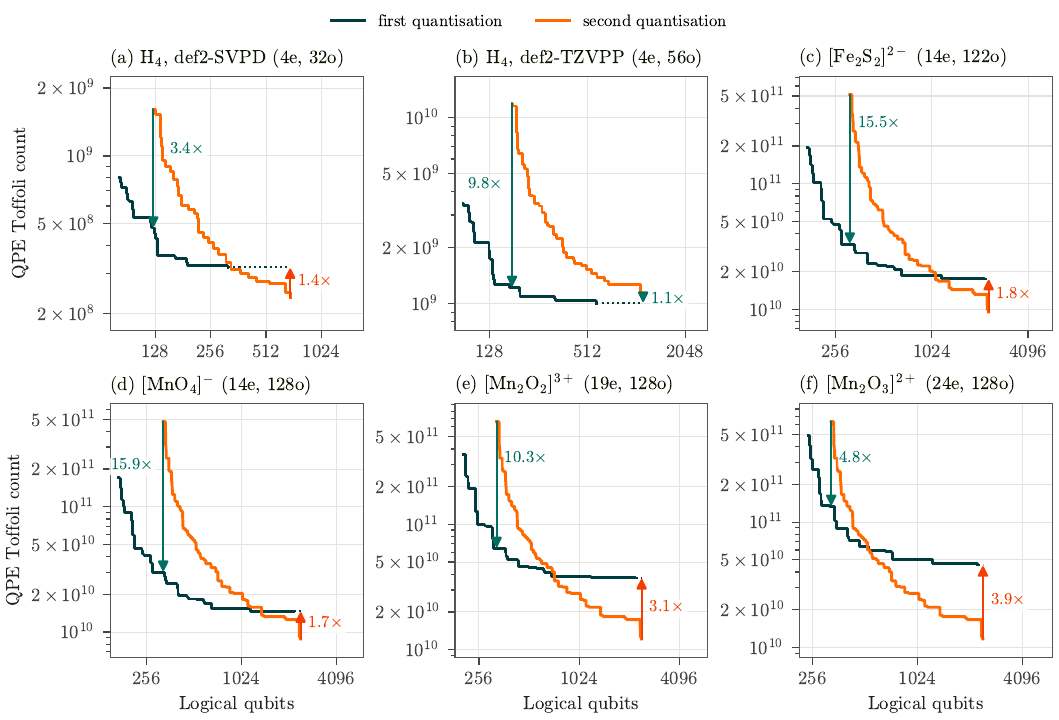}
\caption{Toffoli count of QPE, $N_W C_{\rm B.E.}$ with $\sigma_{\rm QPE}=1.0$~mHa, against the number of logical qubits in first (dark green) and second (orange) quantisation, in the cold-storage architecture. The number of qubits includes the two QPE qubits as in Ref.~\cite{low_denser_2026}. Each point is the Toffoli-optimal compilation at a given qubit budget, with clean and dirty ancillae. Green arrows: ratio of the Toffoli counts at the fewest qubits of the second-quantised QPE. Red arrows: ratio of the minimal Toffoli counts (green in (b), where first quantisation is cheaper); the dotted line extends the minimal Toffoli count of first quantisation to larger qubit counts. At every qubit count, each method uses its factorisation with the fewest QPE Toffoli gates among those with $\max(|\Delta E_{\rm tot}|,|\Delta E_{\rm corr}|)<0.3$~mHa (\cref{tab:factorisations}); $b=16$ in both representations. Second quantisation: the block-encoding of Ref.~\cite{low_denser_2026}.}
\label{fig:qpe_fronts_cold}
\end{figure*}

In~\cref{fig:qpe_fronts_cold}, we compare the total cost of QPE, $N_W \times C_{\rm B.E.}$, in first and second quantisation for the six systems, in the cold-storage architecture as in Ref.~\cite{low_denser_2026}. As in Refs.~\cite{low_fast_2025,low_denser_2026}, we use $\sigma_{\rm QPE}=1.0$~mHa and allow an error of $0.3$~mHa for the approximation of the Hamiltonian. However, our accuracy criterion is different and stricter. In Ref.~\cite{low_fast_2025}, whose factorisations are used in Ref.~\cite{low_denser_2026}, only the error of the high-spin CCSD(T) correlation energy must be below $0.3$~mHa. We require both the total and the correlation energies to be within $0.3$~mHa, $\max(|\Delta E_{\rm tot}|,|\Delta E_{\rm corr}|)<0.3$~mHa, with the errors estimated by UCCSD(T). For each system and method, at every number of logical qubits, we take the factorisation with the fewest QPE Toffoli gates among those that meet this criterion, as one would do when compiling for a given number of logical qubits. Such factorisations exist for all six systems in both representations. More factorisations are available in first quantisation, since there we explored several orderings and numbers of layers of the unitaries, while the second-quantised factorisations are, in our experience, harder to bring to the target accuracy (\cref{app:factorisation_search}). The first-quantised factorisations are optimised in the rotated frame with unitaries truncated to $N_S<2^{\lceil\log_2D\rceil}-1$ layers. The factorisations used are listed in~\cref{tab:factorisations}: two to four along each first-quantised front and one along each second-quantised front. Some of the weights $\omega_{vl}$ in~\cref{eq:ham_ansatz} are indeed negative, though not many of them: 4 of the $VL=224$ weights in the cyclic factorisation of H$_4$ in def2-TZVPP and 8 to 25 of the $VL=1\,365$ weights in the three factorisations of [Fe$_2$S$_2$]$^{2-}$, while in the other first-quantised factorisations the weights are positive, apart from at most one negative weight of magnitude below $10^{-5}$.

At the fewest qubits on which the second-quantised QPE can be implemented, first quantisation requires 3.4 and 9.8 times fewer Toffoli gates for H$_4$ in the def2-SVPD and def2-TZVPP basis sets, 15.5 and 15.9 times fewer for [Fe$_2$S$_2$]$^{2-}$ and [MnO$_4$]$^-$, and 10.3 and 4.8 times fewer for [Mn$_2$O$_2$]$^{3+}$ and [Mn$_2$O$_3$]$^{2+}$ (green arrows). It remains cheaper up to 330 logical qubits for H$_4$ in def2-SVPD, 1\,081 for [Fe$_2$S$_2$]$^{2-}$, 1\,377 for [MnO$_4$]$^-$, 727 for [Mn$_2$O$_2$]$^{3+}$ and 528 for [Mn$_2$O$_3$]$^{2+}$, and at every qubit count for H$_4$ in def2-TZVPP. At the Toffoli-optimal end (red arrows), second quantisation requires 1.4 to 3.9 times fewer Toffoli gates, except for H$_4$ in def2-TZVPP, where first quantisation requires 1.1 times fewer. The advantage at low qubit counts is smaller for the manganese dimers than for [MnO$_4$]$^-$. The three manganese systems have the same number of orbitals and the same second-quantised ranks $(V,L,M)$, so at every number of logical qubits their second-quantised block-encodings cost the same, and their QPE costs differ only through $\lambda_{\rm eff}$. The ratio $\lambda^{\rm SQ}_{\rm eff}/\lambda^{\rm FQ}_{\rm eff}$ is smaller for the dimers than for [MnO$_4$]$^-$. Since $\lambda_{\rm eff}\approx\sqrt{2\Lambda E_{\rm gap}}$ for $\Lambda\gg E_{\rm gap}$,~\cref{eq:effective_norm}, and $E_{\rm gap}$ is similar in the two representations (\cref{tab:factorisations}), this ratio is approximately $\sqrt{\Lambda^{\rm SQ}/\Lambda^{\rm FQ}}$. In first quantisation, $\Lambda$ grows with the number of electrons through the factor $N^2$ in~\cref{eq:Lambda}, whereas in second quantisation it changes little between the three systems. In first quantisation, the cost of the block-encoding also differs between the systems. First, each electron adds a register of $\lceil\log_2D\rceil$ qubits, so the front moves to larger qubit counts as the number of electrons grows (\cref{fig:be_vs_electrons_D}). Second, we explored several orderings and numbers of layers of the unitaries, and the choice with the fewest QPE Toffoli gates differ between the systems (\cref{tab:factorisations}).

\subsubsection{Effect of the hot and cold storage}\label{sec:qpe_cold_weight} The ratios above compare the two algorithms at the same number of logical qubits. In the architecture of Ref.~\cite{low_denser_2026}, however, a logical qubit in the cold storage requires fewer physical qubits than one outside it, and the two algorithms keep very different numbers of qubits in this cold storage. At the lowest qubit count of each algorithm, the qubits outside the cold storage are almost equal in number, 63 to 74 in first and 57 to 67 in second quantisation, since both consist mainly of the phase-gradient state and the ancillae of the table lookups. The cold storage, in contrast, holds $(N-1)\lceil\log_2D\rceil+\lceil\log_2N\rceil+2$ qubits in first quantisation, 19 to 168 for our systems, and $2D+4$ qubits, 68 to 260, in second quantisation. At the same number of logical qubits, first quantisation therefore has more qubits in the hot storage, which it uses as ancillae of the table lookups. For example, [MnO$_4$]$^-$ requires at least 170 logical qubits in first and 327 in second quantisation. At these minimum qubit counts, first quantisation requires 2.8 times fewer Toffoli gates than second quantisation. At 327 logical qubits, first quantisation has 157 more qubits than its minimal requirement, all in the hot storage (230 against 67 in second quantisation). Using these qubits as ancillae reduces the Toffoli count by a factor of 5.7, so that the ratio between first and second quantisation at 327 logical qubits is $2.8\times5.7=15.9$.

The number of Toffoli gates also affects the compilation to the error-correcting code. Since first quantisation requires several times fewer Toffoli gates than second quantisation, its computation is shorter, such that the first-quantised algorithm can potentially use smaller code distances. Alternatively, it could potentially use a slower but denser hot storage and still finish in the same time. In both cases, each logical qubit requires fewer physical qubits, so a detailed compilation of both algorithms might favour first quantisation by an even larger factor. Here, we ignore this effect, since we only aim to show that the advantage is robust. We give a simplified estimate of how much advantage remains when the denser encoding of the cold storage is taken into account. The estimate uses a simple model that accounts for the different densities of the hot and cold storage (\cref{app:cold_weight}). In this model, the two algorithms are compared at the same effective number of logical qubits, $n_{\rm eff}=n_{\rm hot}+\rho\,n_{\rm cold}$, in which a qubit in the cold storage counts as a fraction $\rho$ of a qubit in the hot storage. The value of $\rho$ depends on the layout and on the code distances. In the layouts of Ref.~\cite{low_denser_2026} that, as in our compilation, perform rotations by addition into a phase-gradient state, $\rho=0.26$--$0.40$ with up to 400 logical qubits (\cref{app:cold_weight}). Since second quantisation keeps more of its qubits in the cold storage, a smaller $\rho$ favours second quantisation. For $\rho=0.3$, at the fewest effective qubits on which second quantisation can be implemented, first quantisation requires 2.4 to 8.1 times fewer Toffoli gates, and it remains cheaper up to between about 270 and 1\,200 effective qubits (\cref{fig:cold_weight}(a) in~\cref{app:cold_weight}). Even if the cold storage were free, $\rho=0$, first quantisation requires 1.3 to 3.4 times fewer Toffoli gates at the same number of qubits in the hot storage (left end of~\cref{fig:cold_weight}(b)). Moreover, when each algorithm runs on its smallest logical qubit count, first quantisation requires 1.3 to 3.5 times fewer Toffoli gates and 1.4 to 2.0 times fewer logical qubits than second quantisation. Exact physical qubit counts require a compilation of the first-quantised algorithm to the error-correcting code, with its own layout and magic state factories, similarly to the work on the second-quantised algorithm in Ref.~\cite{low_denser_2026}; we leave this for future work. However, these estimates show that the advantage of first quantisation at small qubit counts is robust: it is reduced but persists for every value of $\rho$.

\begin{table*}[htb!]
\centering
\small
\renewcommand{\arraystretch}{1.2}
\setlength{\tabcolsep}{4pt}
\begin{tabular}{l c c c l r r r r r}
\hline\hline
System & $(N, D)$ & & $(V, L, M)$ & Unitaries, $N_S$ & $\Lambda$ & $E_{\rm gap}$ & $\lambda_{\rm eff}$ & $\Delta E_{\rm tot}$ & $\Delta E_{\rm corr}$ \\
 & & & & & (Ha) & (Ha) & (Ha) & (mHa) & (mHa) \\
\hline
\multirow{3}{*}{H$_4$, def2-SVPD} & \multirow{3}{*}{(4, 32)} & FQ & (4, 32, 32) & cyclic, $25$ & $29.79$ & $4.30$ & $15.41$ & $0.278$ & $0.193$ \\
 &  & FQ & (4, 32, 32) & TW + CSD 1-1-2-1, $25$ & $29.84$ & $4.46$ & $15.70$ & $0.017$ & $-0.049$ \\
 &  & SQ & (4, 32, 32) & -- & $108.70$ & $2.35$ & $22.49$ & $0.232$ & $0.020$ \\
\hline
\multirow{4}{*}{H$_4$, def2-TZVPP} & \multirow{4}{*}{(4, 56)} & FQ & (4, 56, 56) & cyclic, $29$ & $56.79$ & $9.28$ & $31.11$ & $0.085$ & $-0.263$ \\
 &  & FQ & (4, 56, 56) & TW + CSD 1-1-2-1-1, $43$ & $65.26$ & $5.63$ & $26.52$ & $-0.013$ & $-0.031$ \\
 &  & FQ & (7, 56, 64) & TW + CSD 1-1-2-1-1, $29$ & $91.41$ & $3.92$ & $26.50$ & $-0.019$ & $-0.116$ \\
 &  & SQ & (7, 37, 37) & -- & $662.01$ & $2.85$ & $61.36$ & $0.177$ & $0.017$ \\
\hline
\multirow{4}{*}{[Fe$_2$S$_2$]$^{2-}$} & \multirow{4}{*}{(14, 122)} & FQ & (15, 91, 128) & cyclic, $61$ & $733.00$ & $12.84$ & $136.62$ & $0.207$ & $-0.231$ \\
 &  & FQ & (15, 91, 128) & TW + CSD 1-2-2-1-1, $77$ & $710.46$ & $11.70$ & $128.40$ & $-0.237$ & $-0.021$ \\
 &  & FQ & (15, 91, 128) & TW + CSD 1-2-3-1, $77$ & $684.80$ & $12.01$ & $127.71$ & $0.056$ & $0.027$ \\
 &  & SQ & (15, 91, 122) & -- & $2095.19$ & $9.09$ & $194.97$ & $-0.265$ & $-0.142$ \\
\hline
\multirow{3}{*}{[MnO$_4$]$^{-}$} & \multirow{3}{*}{(14, 128)} & FQ & (16, 64, 128) & cyclic, $81$ & $582.73$ & $9.71$ & $105.96$ & $0.198$ & $0.171$ \\
 &  & FQ & (16, 64, 128) & cyclic + CSD 3-4, $65$ & $802.04$ & $9.88$ & $125.48$ & $-0.231$ & $-0.234$ \\
 &  & SQ & (16, 64, 128) & -- & $1537.07$ & $10.71$ & $181.17$ & $-0.137$ & $0.185$ \\
\hline
\multirow{3}{*}{[Mn$_2$O$_2$]$^{3+}$} & \multirow{3}{*}{(19, 128)} & FQ & (16, 64, 128) & cyclic, $81$ & $1340.23$ & $19.28$ & $226.52$ & $-0.039$ & $0.064$ \\
 &  & FQ & (16, 64, 128) & cyclic + CSD 3-4, $97$ & $1309.89$ & $19.55$ & $225.46$ & $-0.265$ & $0.178$ \\
 &  & SQ & (16, 64, 128) & -- & $1696.67$ & $18.56$ & $250.27$ & $-0.215$ & $0.240$ \\
\hline
\multirow{5}{*}{[Mn$_2$O$_3$]$^{2+}$} & \multirow{5}{*}{(24, 128)} & FQ & (16, 64, 128) & cyclic, $97$ & $1933.52$ & $18.87$ & $269.51$ & $0.142$ & $0.013$ \\
 &  & FQ & (16, 64, 128) & TW + CSD 1-3-2-1, $97$ & $1974.64$ & $25.20$ & $314.45$ & $-0.137$ & $-0.136$ \\
 &  & FQ & (16, 128, 128) & cyclic + CSD 2-2-3, $81$ & $2508.63$ & $16.86$ & $290.39$ & $0.088$ & $0.088$ \\
 &  & FQ & (16, 128, 128) & cyclic + CSD 3-4, $81$ & $2790.07$ & $17.05$ & $307.98$ & $-0.198$ & $0.012$ \\
 &  & SQ & (16, 64, 128) & -- & $1550.89$ & $18.71$ & $240.19$ & $-0.239$ & $0.111$ \\
\hline\hline
\end{tabular}
\caption{Factorisations used in~\cref{fig:qpe_fronts_cold,fig:qpe_fronts_monolithic}: number of electrons and orbitals, $(N, D)$, factorisation ranks, $(V, L, M)$, ordering (with the partition of the nested orderings,~\cref{sec:us_orderings}) and number of layers of the first-quantised unitaries, subnormalisation factor, $\Lambda$, $E_{\rm gap}$, effective subnormalisation factor, $\lambda_{\rm eff}$, and the errors of the total and correlation energies estimated with UCCSD(T) on an unrestricted Hartree--Fock reference, for first (FQ) and second (SQ) quantisation. At every number of logical qubits, each method uses the factorisation with the fewest QPE Toffoli gates among those with $\max(|\Delta E_{\rm tot}|,|\Delta E_{\rm corr}|)<0.3$~mHa; listed are all factorisations used somewhere on the fronts, in the order in which they appear as the number of qubits grows. TW: triangle wave.}
\label{tab:factorisations}
\end{table*}

\section{Discussion and conclusions}\label{sec:conclusions}
We have shown that the spectral-gap amplification protocol~\cite{zlokapa_hamiltonian_2024,low_fast_2025,king_quantum_2026} does not require a sum of squares with non-negative weights. Going beyond convex sum-of-squares, we have presented a non-convex sum-of-squares representation of the first-quantised electronic Hamiltonian in a general orbital basis that still allows spectral-gap amplification, together with explicit block-encodings.
Previous work on this protocol uses sums of squares with non-negative weights~\cite{zlokapa_hamiltonian_2024,low_fast_2025,king_quantum_2026,dutkiewicz_spectral_2026,rubin_near-frustration-free_2026}. Loading the sign of a term, including that of a block-encoded square, is a known technique~\cite{von_burg_quantum_2021,rocca_reducing_the_runtime_2024,ivanov_quantum_computation_of_electronic_2025}. Our observation is that the block-encoding with these signs still has the form~\eqref{eq: ShiftedHamiltonian}, with a positive semidefinite $\hat H_0+\Omega$. The price is an additional shift: each square with a negative weight adds $N^2|\omega_{vl}|\epsilon_{vl}^2/2$ to $\Omega$. This is the last term in the bracket of~\eqref{eq:Omega}, which vanishes for non-negative weights. Spectral amplification therefore remains effective as long as the squares with negative weights carry little weight, so that the condition~\eqref{eq: small ratio} still holds. To our knowledge, neither this statement nor the optimisation of $\lambda_{\rm eff}$ over signed weights has been discussed before. We have shown it constructively, with the circuit depicted in~\cref{fig: quantum_circuit_be_full} and the derivation in~\cref{app:be_derivation}. This derivation does not use the form of the operators that are squared. Spectral amplification can therefore be applied directly to a Hamiltonian that is already given as a sum of squares with weights of both signs, without additional classical preprocessing. Such sums appear, for example, in the double factorisation with symmetry shifts~\cite{rocca_reducing_the_runtime_2024} and in the projector augmented-wave contribution to the electron--electron interaction~\cite{ivanov_quantum_computation_of_electronic_2025}. This does not guarantee a small $\lambda_{\rm eff}$, which we have not evaluated for these representations. \Cref{eq:effective_norm} only ensures $\lambda_{\rm eff}\le\Lambda$, and a substantial reduction requires the condition~\eqref{eq: small ratio}. In general, this condition has to be reached by optimising the representation, as we do in this work.

We have benchmarked our approach against the sum-of-squares spectral amplification approach in second quantisation~\cite{low_fast_2025}, with both approaches compiled at the logical level with the same techniques~\cite{low_denser_2026}, for six chemical systems with 4 to 24 electrons and up to 128 orbitals. When only a few hundred logical qubits are available, first quantisation requires 3 to 16 times fewer Toffoli gates at the same number of logical qubits, and it remains cheaper up to between several hundred and about 1\,400 logical qubits, whereas with more logical qubits second quantisation reaches a lower minimal Toffoli count, except for H$_4$ in def2-TZVPP. For the block-encoding, the advantage at low qubit counts persists up to about 30--40 electrons for 128 orbitals and 60--70 electrons for 256 orbitals (\cref{fig:be_vs_electrons_D}). For QPE, the advantage also grows with the number of orbitals, since $\lambda_{\rm eff}$ grows approximately linearly with $D$ in first quantisation and approximately as $D^{1.6}$ in second quantisation. Therefore, first quantisation is best suited to active spaces with several times more orbitals than electrons, as needed to account for dynamical correlation, which is crucial to describe real chemistry.
The six systems are chemically motivated models of this regime rather than targets chosen for their classical difficulty. The models consist of the valence electrons and the virtual MP2 natural orbitals with the largest occupation numbers (\cref{app:chemical_systems}), so that part of the dynamical correlation is described within the active space. For the four transition-metal complexes, the full configuration interaction space contains between about $10^{22}$ and $10^{33}$ determinants, far beyond exact diagonalisation, although approximate classical methods can be applied to them. The two H$_4$ examples are small test cases, though in realistic basis sets, with 8 and 14 times more orbitals than electrons, rather than in a minimal basis set such as STO-3G. Problems with more electrons stay in the regime where first quantisation is advantageous only if the number of orbitals grows as well, since the advantage of the block-encoding disappears near the point where the two wavefunction registers become equal (\cref{fig:be_vs_electrons_D}), and $\lambda_{\rm eff}\approx\sqrt{2\Lambda E_{\rm gap}}$ grows with the number of electrons. For $D\ge256$, however, finding the factorisations becomes the bottleneck (\cref{sec:resource_estimates}), so progress in the numerical factorisation is needed to reach such problems.

Unlike second quantisation, first quantisation does not allow large space--time trade-offs: additional logical qubits reduce the Toffoli count of the first-quantised block-encoding only up to about a thousand logical qubits, whereas the table lookups that dominate the second-quantised block-encoding continue to trade qubits for Toffoli gates up to thousands of logical qubits. The trade-off in first quantisation is limited by the synthesis of the multiplexed unitaries, which is the main bottleneck of the first-quantised block-encoding: at the Toffoli-optimal point of [MnO$_4$]$^-$, they account for $83\%$ of the Toffoli count (\cref{fig:be_breakdown}). Recently, new algorithms for the synthesis of an arbitrary $D\times D$ unitary have reduced the asymptotic T-count from $O(D^{4/3})$ to $O(D^{5/4})$~\cite{yuan_quantum_2026} and $\tilde O(D^{8/7})$~\cite{liu_t-efficient_2026}. Although their asymptotic cost is better, their constant factors are yet to be determined in practice. Moreover, to be used here, they would have to be extended to unitaries multiplexed over $v$, and they would need to take into account the truncation of the unitaries to fewer layers.  Nevertheless, such algorithms may further reduce the cost of the first-quantised block-encoding.

How large the advantage in physical resources is depends on the layout of the architecture. Second quantisation keeps more of its qubits in the dense cold storage. Yet, a simple model of the denser memory (\cref{sec:qpe_cold_weight}) indicates that a substantial advantage is kept, with 2.4 to 8.1 times fewer Toffoli gates at small qubit counts, and that it does not disappear even if the cold storage is free. We have not considered the most qubit-frugal compilations of Ref.~\cite{low_denser_2026}, with Clifford+T rotations and ancilla-free arithmetic. This approach does not lower the fronts shown here but extends them to fewer qubits at a higher Toffoli count; for FeMoco, they extend the second-quantised front from about 170 to 126 logical qubits at about seven times the Toffoli count~\cite{low_denser_2026}. We expect this compilation approach to increase the advantage of first quantisation at small qubit counts. It saves ancillae, but cannot reduce the $2D$ qubits of the second-quantised system register, and for five of our six systems the first-quantised fronts shown here already start below $2D$ logical qubits (88 to 241, against $2D=112$ to 256). Therefore, second quantisation could run on fewer qubits than considered here only at a higher Toffoli count, whereas first quantisation would not need these compilations at these qubit counts. Comparisons including Clifford+T rotations and ancilla-free arithmetic in both algorithms is left to future work.

We have compiled both algorithms at the logical level, assuming the architecture of Ref.~\cite{low_denser_2026}, which determines which qubits can be borrowed as dirty ancillae. The two algorithms access the cold storage differently. In first quantisation, the multiplexed unitaries, which dominate the Toffoli count, act only on the electron register in position 0, which is kept in the hot storage, and the cold storage is accessed only by the SWAP networks, with $4(N-1)\lceil\log_2D\rceil$ controlled swaps per walk step, and by UNIFORM and the reflection on the electron index. In second quantisation, $4D$ controlled swaps and $4(D-1)$ Givens rotations act on qubits in the cold storage in every walk step~\cite{low_denser_2026}. In both cases, the number of these operations grows only linearly with the number of qubits in the cold storage, although the Hamiltonian has many more terms, in agreement with Ref.~\cite{low_denser_2026}, which showed that a block-encoding can be implemented with a number of qubit transfers between a slow memory and the compute region that grows linearly with the number of qubits in the memory and does not depend on the number of terms (their Theorem~1). For the systems considered here, the number of these operations is 1.6 to 6.2 times smaller in first quantisation. This may allow a simpler layout of the first-quantised algorithm, with lower latency from accessing the cold storage. A rigorous comparison of the number of physical qubits and the run time requires developing a compilation of the first-quantised algorithm to lattice surgery, including its layout and magic-state factories, as Ref.~\cite{low_denser_2026} did for second quantisation, which we leave to future work.

Overall, our results show that first quantisation is a practical route for simulating chemistry in the low-filling regime when the number of logical qubits is limited.

\section{Data and code availability}\label{sec:data}
The data and code that support the findings of this article will be made available at a later date.

\section{Use of AI tools}\label{sec:ai}
The authors developed the sum-of-squares factorisation, including the layered unitaries, their truncation and most of their orderings. They also developed the objective function~\eqref{eq: objective_function}, the block-encodings and the code for the factorisations in first and second quantisation. We used large language models by Anthropic: Claude Opus~4 and Claude Opus~5.5 during the project, and Claude Fable~5.2 for the final checks. The language models implemented the layered unitaries in the factorisation code, the cost models and the analysis and plotting scripts. At the authors' request, the language models also proposed two ingredients: the triangle wave + CSD ordering (\cref{sec:us_orderings}), when asked for an ordering between triangle wave and CSD, and SELECT-XOR with SELECT-Z uncomputation (\cref{fig:window_select_xor}), when asked to replace swapping the ancilla qubits back in each layer by an uncomputation. They also ran some of the calculations on high-performance computing clusters, repeating runs set up by the authors from their scripts and examples. Finally, they processed the data and assisted in writing parts of the manuscript, starting from a draft written by the authors. The authors directed this work and checked its output. In particular, our cost model of the second-quantised algorithm reproduces the costs published in Ref.~\cite{low_denser_2026} (\cref{fig:sq_validation}), and the accuracy criterion for the factorisations was tested against FCI and DMRG energies (\cref{app:sq_validation_low}). The authors take full responsibility for the content of the paper.

\section{Acknowledgment} We thank Marius Bothe for helpful discussions on the manuscript and Anastasia Istratuca for the initial implementation of second-quantised factorisations. This work was part of the QUICHE project (QUantum Integrated CHEmistry, reference: 10150101) funded by Innovate UK.

\appendix
\crefalias{section}{appendix}
\section{Hartree--Fock energy of the factorised Hamiltonian}\label{app:hf_energy}
Let $\hat C^{\sigma}$, $\sigma\in\{\uparrow,\downarrow\}$, be the $D\times N_{\sigma}$ matrices of the occupied orbitals of the spin-unrestricted HF solution of the unfactorised Hamiltonian, $N_{\uparrow}+N_{\downarrow}=N$. We obtain them with a spin-unrestricted HF calculation followed by a stability analysis, so that spin symmetry can be broken, and $E^0_{\rm HF}$ is the HF energy of this solution. The term $E_{\rm HF}$ in~\cref{eq: objective_function} is the expectation value of the factorised Hamiltonian in the same Slater determinant. With $\hat P^{\sigma}=\hat C^{\sigma}\hat C^{\sigma\dagger}$ and $g_{pqrs}=\sum_{vl}\omega_{vl}W_{vlpq}W_{vlrs}$,
\begin{equation}
    E_{\rm HF} = \beta N + \sum_{\sigma}{\rm Tr}\big[(\hat T+\hat B)\hat P^{\sigma}\big]
    + \frac{1}{2}\sum_{vl}\omega_{vl}\left[\Big(\sum_{\sigma}{\rm Tr}\big[\hat W^{(v,l)}\hat P^{\sigma}\big]\Big)^2 - \sum_{\sigma}{\rm Tr}\big[\hat W^{(v,l)}\hat P^{\sigma}\hat W^{(v,l)}\hat P^{\sigma}\big]\right],
\end{equation}
where the first term in the square brackets is the Coulomb and the second the exchange contribution. We never form $g_{pqrs}$ or $\hat W^{(v,l)}$: with the factors~\eqref{eq:joint_diagonalization} transformed to the occupied orbitals,
\begin{equation}
    \bar R^{\sigma}_{vjm} = \sum_{p} C^{\sigma}_{pj} R_{vpm},\qquad
    X^{\sigma}_{vl} = \sum_{j=1}^{N_\sigma}\sum_{m} f_{vlm}\left(\bar R^{\sigma}_{vjm}\right)^2,\qquad
    Y^{\sigma}_{vl,ij} = \sum_{m}\bar R^{\sigma}_{vim} f_{vlm}\bar R^{\sigma}_{vjm},
\end{equation}
the HF energy is
\begin{equation}\label{eq:hf_energy_factors}
    E_{\rm HF} = \beta N + \sum_{\sigma}\sum_{j=1}^{N_\sigma}\big[\hat C^{\sigma\dagger}(\hat T+\hat B)\hat C^{\sigma}\big]_{jj}
    + \frac{1}{2}\sum_{vl}\omega_{vl}\left[\big(X^{\uparrow}_{vl}+X^{\downarrow}_{vl}\big)^2 - \sum_{\sigma}\sum_{i,j=1}^{N_\sigma}\big(Y^{\sigma}_{vl,ij}\big)^2\right],
\end{equation}
which costs $O(VMDN+VLMN^2)$ operations. The orbitals $\hat C^{\sigma}$ are those of the unfactorised Hamiltonian and are not re-optimised for the factorised one, so $E_{\rm HF}$ is an upper bound on the HF energy of the factorised Hamiltonian. For the rotated ansatz, $\hat R^{(v)}$ and $\hat B$ in~\cref{eq:hf_energy_factors} are those of~\cref{eq:R_rotated,eq:B_rotated} in the original frame, so $\hat C^{\sigma}$ are used without rotation.

\section{Operator block-encoded by the circuit of~\cref{fig: quantum_circuit_be_full}}\label{app:be_derivation}
Here we derive the operator that the circuit of~\cref{fig: quantum_circuit_be_full} block-encodes when the control qubit is in $\ket{1}$ state. We denote the block-encoding of $\sum_j(\hat W^{(v,l)})_j/(N\epsilon_{vl})$ of~\cref{fig: quantum_circuit_be_sum_over_electrons} by ${\rm BE}_{vl}$, and its block by $\bra{0}\cdot\ket{0}$, which is flagged by $\ket{0}_{\rm ne}\ket{0}_{\rm flag}$. This circuit is its own inverse, ${\rm BE}_{vl}^2=\hat I$ (\cref{sec: be_hamiltonian}). According to~\cref{eq: PREP}, PREP prepares the term $(v,l)$ with the probability
\begin{equation}\label{eq:be_probabilities}
    p_{vl}=\frac{N^2\epsilon_{vl}^2}{4\Lambda}\quad\text{for } v<V,\qquad p_{V0}=\frac{N(\delta_{D-1}-\delta_0)}{2\Lambda},
\end{equation}
and these probabilities add up to one by~\cref{eq: Lambda'}.

For a two-body term, the circuit applies ${\rm BE}_{vl}$, the reflection ${\rm REF}=2\ketbra{0}{0}-\hat I$ on the registers $\ket{\cdot}_{\rm ne}\ket{\cdot}_{\rm flag}$, the sign $(-1)^{\theta_{vl}}$, and ${\rm BE}_{vl}$ again. The block of this sequence is
\begin{align}\label{eq:be_two_body_block}
    (-1)^{\theta_{vl}}\bra{0}{\rm BE}_{vl}\,{\rm REF}\,{\rm BE}_{vl}\ket{0}
    &=(-1)^{\theta_{vl}}\left[2\left(\bra{0}{\rm BE}_{vl}\ket{0}\right)^2-\bra{0}{\rm BE}_{vl}^2\ket{0}\right]\nonumber\\
    &=(-1)^{\theta_{vl}}\left[\frac{2}{N^2\epsilon_{vl}^2}\Big(\sum_{j=0}^{N-1}(\hat W^{(v,l)})_j\Big)^2-\hat I\right],
\end{align}
which is the Chebyshev polynomial of second order~\cite{low_hamiltonian_simulation_qubitization_2019}. For the one-body term, the circuit applies ${\rm BE}_{V0}$ once. With~\cref{eq: factorizations_one_body} and $\epsilon_{V0}=1$, its block is
\begin{equation}\label{eq:be_one_body_block}
    \bra{0}{\rm BE}_{V0}\ket{0}=\frac1N\sum_{j=0}^{N-1}(\hat W^{(V,0)})_j=\frac{2}{N(\delta_{D-1}-\delta_0)}\sum_{j=0}^{N-1}\big(\hat W_0-\delta_0\hat I\big)_j-\hat I.
\end{equation}

All gates between PREP and UNPREP are diagonal in $(v,l)$. The block of the whole circuit is therefore the sum of~\cref{eq:be_two_body_block,eq:be_one_body_block} weighted with the probabilities~\eqref{eq:be_probabilities}:
\begin{align}\label{eq:be_total_block}
    &\frac1\Lambda\Bigg[\sum_{v=0}^{V-1}\sum_{l=0}^{L-1}\frac{(-1)^{\theta_{vl}}}{2}\Big(\sum_{j=0}^{N-1}(\hat W^{(v,l)})_j\Big)^2+\sum_{j=0}^{N-1}\big(\hat W_0\big)_j-\delta_0N\Bigg]-\Big(\sum_{v,l}(-1)^{\theta_{vl}}p_{vl}+p_{V0}\Big)\hat I\nonumber\\
    &\qquad=\frac{\hat H''-\delta_0N}{\Lambda}-\Big(1-2\sum_{vl:\theta_{vl}=1}p_{vl}\Big)\hat I
    =\frac{\hat H''+\Omega''}{\Lambda}-\hat I,\qquad
    \Omega''=-\delta_0N+\sum_{vl:\theta_{vl}=1}\frac{N^2\epsilon_{vl}^2}{2},
\end{align}
where we used~\cref{eq: hamiltonian_for_be}. A square with $\theta_{vl}=0$ contributes $-p_{vl}\hat I$ to the constant, and a square with $\theta_{vl}=1$ contributes $+p_{vl}\hat I$. This is the origin of the additional shift $N^2\epsilon_{vl}^2/2$ of each square with a negative weight. Finally, a block of a unitary has a norm of at most one. Therefore $(\hat H''+\Omega'')/\Lambda-\hat I\succeq-\hat I$, i.e.\ $\hat H''+\Omega''$ is positive semidefinite.

\section{Unitary synthesis when $D$ is not a power of 2}\label{app:nonpow2}
When $D$ is not a power of 2, we use a register of $d=\lceil\log_2D\rceil$ qubits. Of its $2^d$ basis states $\ket{p}$, the first $D$, $p<D$, are the orbitals, and the remaining $2^d-D$ states are padding.
In our ansatz for the unitaries, multiplexed rotation gates \eqref{eq:multiplexor} are executed. Some of the rotations touch basis states inside the padding. We set all of those rotation angles to zero.
For costing purposes, we load explicit 0 values in the usual way in the QROAM. Some savings are conceivable with a variable-spaced QROAM, but here we do not consider this.
In the remainder of the appendix, we make this method more precise.

A multiplexed rotation $\hat Q_s(\vec\theta)$~\eqref{eq:multiplexor} rotates the target qubit $s$ by an angle selected by the other $d-1$ qubits. Notice this corresponds to the product of $2^{d-1}$ commuting Givens rotations, each acting on a pair of basis states $(p,\,p+2^s)$, where bit $s$ of $p$ is zero, with one angle per pair. To see this, write the index of a basis state in binary, $p=\sum_{k}p_k2^k$, and let $c_s(p)$ be the $(d-1)$-bit integer obtained by deleting bit $p_s$. The projector $\ketbra{c}{c}_{\bar s}$ in~\eqref{eq:multiplexor} selects the single term $c=c_s(p)$, so only the target qubit is rotated. With $\hat R_y(\theta)\ket{0}=\cos\theta\ket{0}+\sin\theta\ket{1}$ and $\hat R_y(\theta)\ket{1}=-\sin\theta\ket{0}+\cos\theta\ket{1}$, and since flipping bit $s$ from 0 to 1 adds $2^s$ to the index, for every $p$ with $p_s=0$ we obtain
\begin{equation}\label{eq:layer_on_pair}
\begin{aligned}
    \hat Q_s(\vec\theta)\ket{p} &= \cos\theta_{c}\ket{p}+\sin\theta_{c}\ket{p+2^s},\\
    \hat Q_s(\vec\theta)\ket{p+2^s} &= -\sin\theta_{c}\ket{p}+\cos\theta_{c}\ket{p+2^s},
\end{aligned}
\qquad c=c_s(p)=c_s(p+2^s).
\end{equation}
Both states of the pair $(p,\,p+2^s)$ see the same angle, because they differ only in bit $s$. Thus, $\hat Q_s$ acts on the pair as the Givens rotation $\hat G_{p,p+2^s}(\theta_{c_s(p)})$, which rotates $\ket{p}$ and $\ket{p+2^s}$ as in~\eqref{eq:layer_on_pair} and acts as the identity on all other basis states. The $2^{d-1}$ pairs, one for each value of $c$, are disjoint, so their Givens rotations commute and
\begin{equation}\label{eq:layer_givens}
    \hat Q_s(\vec\theta)=\prod_{p\,:\,p_s=0}\hat G_{p,\,p+2^s}\big(\theta_{c_s(p)}\big).
\end{equation}
For example, for $d=2$ and $s=0$, the pairs are $(0,1)$ with the angle $\theta_0$ and $(2,3)$ with the angle $\theta_1$.

A Givens rotation on a pair with $p<D\le p+2^s$ would rotate an orbital into a padding state. Therefore, we set to zero the angle of every pair that contains a padding state:
\begin{equation}\label{eq:padding_rule}
    \theta\ \text{of the pair}\ (p,\,p+2^s)\ =\ 0\quad\text{if}\quad p+2^s\ge D .
\end{equation}
For example, for $D=3$ and $d=2$, the state $\ket{3}$ is padding: in~\eqref{eq:layer_givens}, the layer on qubit 0 keeps the angle of the pair $(0,1)$ and sets the angle of $(2,3)$ to zero, and the layer on qubit 1 keeps $(0,2)$ and sets $(1,3)$ to zero. The pairs with both states in the padding do not mix orbitals with padding, and we set their angles to zero only for convenience. A Givens rotation with zero angle is the identity, so, with~\eqref{eq:padding_rule}, every layer acts as $\hat Q_s=\hat M_s\oplus\hat I_{2^d-D}$, where $\hat M_s$ is a $D\times D$ orthogonal matrix. The product of such layers in any ordering and of any depth has the same block structure. Thus, the ansatz is an orthogonal $D\times D$ matrix for any choice of the remaining angles, and no amplitude leaks into the padding. The rule~\eqref{eq:padding_rule} depends only on $D$, so the angles set to zero are the same for all $V$ multiplexed unitaries, and in the variational optimisation we keep them fixed at zero. A unitary of dimension $D$ is represented exactly with $N_S=2^d-1$ layers rather than $D-1$; in this work, $N_S<2^d-1$ for every system with $D\neq2^d$, so the depth is a variational parameter as for $D=2^d$.

We cost a multiplexed unitary of dimension $D$ as one of the padded dimension $2^d$, with the same ordering and number of layers: the angles set to zero by~\eqref{eq:padding_rule} are loaded by \qroam{} and selected by \selxor{} as any other angle. This gives an upper bound on the cost. The zero angles occupy the last addresses of the lookup table of a window, or the last rows $j_{\rm t}$ of a layer, so part of this cost can be saved: a \qroam{} need not resolve the addresses of a block on which every loaded word is the same, as in the QROM with variable spacing of Ref.~\cite{sanders2020compilation}, and, if the reference row $j_{\rm t}=0$ of~\cref{eq:selxor_def} is replaced by a row whose angles are all zero, the zero rows need not be loaded or selected. We do not use these savings here. Of the systems in~\cref{tab:factorisations}, they affect H$_4$ in the def2-TZVPP basis ($D=56$, $d=6$) and [Fe$_2$S$_2$]$^{2-}$ ($D=122$, $d=7$).

\section{Validation of Our Implementation of Sum-of-Squares in Second Quantisation}\label{app:sq_implementation}
The second-quantised sum-of-squares representation described in this appendix was introduced in Ref.~\cite{low_fast_2025}, and our implementation mostly follows it. We include it only for reproducibility: since for large $D$ we found it difficult to obtain second-quantised factorisations that reach the target accuracy (\cref{app:factorisation_search}), we describe our implementation here in full, so that it can be reproduced and checked. At the end of this appendix, we validate it on the Fe$_2$S$_2$ Hamiltonian of Ref.~\cite{li_spin-projected_2017}, which is used in Ref.~\cite{low_fast_2025} and for which energies from DMRG and FCI can be computed.

The Hamiltonian in second quantisation can be written in terms of unitary group generators (not to be confused with $\hat E_{pq}$ from above),
\begin{equation}
    \hat E_{pq} = \sum_{\sigma} a^{\dagger}_{p\sigma} a_{q\sigma},
\end{equation}
as follows:
\begin{equation}\label{eq:second_quantization_ham}
    \hat H_0 = \alpha \hat I  + \sum_{pq}h_{pq}\hat E_{pq} + \frac{1}{2}\sum_{pqrs}h_{pqrs} \hat E_{pq} \hat E_{rs},
\end{equation}
with
\begin{equation}
    h_{pq} = t_{pq} - \frac{1}{2}\sum_{r}h_{rprq}.
\end{equation}
This Hamiltonian acts on Hilbert space with variable number of electrons. As we are interested in $N$-electron symmetry sector then we can apply BLISS approach
\begin{equation}\label{eq:second_quantization_ham_bliss}
    \hat H_0 = (\alpha + \beta N) \hat I + \sum_{pq}\left(h_{pq} + b_{pq} - \beta \delta_{pq} \right) \hat E_{pq}
    + \frac{1}{2}\sum_{pqrs}\left(h_{pqrs} - \frac{1}{N}\left(b_{pq}\delta_{rs} + \delta_{pq}b_{rs}\right)\right) \hat E_{pq} \hat E_{rs}
\end{equation}
Now, we can consider the following Hamiltonian ansatz,
\begin{equation}
    \hat H' = \sum_{pq}W_{0pq} \hat E_{pq} + \sum_{vl}\frac{1}{2}\left(\alpha_{vl} \hat I + \sum_{pq}W_{vlpq}\hat E_{pq} \right)^2 - \frac{1}{2}\sum_{vl}\alpha_{vl}^2.
\end{equation}
Expanding the square, we obtain
\begin{equation}
    \hat H' = \sum_{pq} \left(W_{0pq}  +  \sum_{vl}\alpha_{vl}W_{vlpq}\right) \hat E_{pq} + \frac{1}{2} \sum_{pqrs}g_{pqrs}\hat E_{pq}\hat E_{rs}
\end{equation}
with $g_{pqrs}=\sum_{vl}W_{vlpq}W_{vlrs}$.
We choose $W_{0pq}$ so that the one-body term of $\hat H'$ is that of~\cref{eq:second_quantization_ham_bliss},
\begin{equation}
    W_{0pq} = h_{pq} + b_{pq} - \beta \delta_{pq} - \sum_{vl}\alpha_{vl} W_{vlpq}.
\end{equation}
Therefore, $\hat H'$ differs from~\cref{eq:second_quantization_ham_bliss} only in the two-body term, where $g_{pqrs}$ approximates $h_{pqrs} - \frac{1}{N}\left(b_{pq}\delta_{rs} + \delta_{pq}b_{rs}\right)$. To calculate the energy error of the approximate Hamiltonian, we remove the BLISS shift, which does not change the Hamiltonian in the $N$-electron sector, and carry out calculations with the exact one-body coefficients $h_{pq}$ and the two-body coefficients
\begin{equation}
    h'_{pqrs} = g_{pqrs} + \frac{1}{N}\left(b_{pq}\delta_{rs} + \delta_{pq}b_{rs}\right).
\end{equation}
We now find the subnormalisation factor, $\Lambda$, and the shift $\Omega$ in the block-encoding of
\begin{equation}
    \frac{\hat H_0 + \Omega}{\Lambda} - \hat I,
\end{equation}
The subnormalisation factor is
\begin{equation}
    \Lambda = \|\hat W_0 \|_{\rm SC} + \frac{1}{4}\sum_{vl}  \lambda_{vl}^2, 
\end{equation}
where $\|\hat W_0\|_{\rm SC}=\sum_{j=0}^{D-1}|\delta_j|$ is the Schatten 1-norm of $\hat W_0$, i.e.\ the sum of the absolute values of its eigenvalues $\delta_0\le\dots\le\delta_{D-1}$, as in~\cref{sec:sos}, and $\lambda_{vl}$ comes from block-encoding of
\begin{equation}
    \alpha_{vl}\hat I + \sum_{pq} W_{vlpq} \hat E_{pq}. 
\end{equation}
We can use Majorana operators instead of $\hat E_{pq}$. The new constant shift is then 
\begin{equation}
    \left(\alpha_{vl} + {\rm Tr}[\hat W^{(v,l)}]\right) \hat I, 
\end{equation}
and the remaining term
\begin{equation}
    \frac{i}{2}\sum_{pq\sigma}W_{vlpq}\gamma_{p\sigma,0}\gamma_{q\sigma,1},
\end{equation}
which can be block-encoded with one-norm that depends on a factorisation. As in Ref.~\cite{low_fast_2025}, we use $W_{vlpq}=\sum_{m}f_{vlm}R_{vpm}R_{vqm}$, where the vectors $R_{v\cdot m}$ are normalised, $\sum_{p}R_{vpm}^2=1$, so that the one-norm is $\sum_m|f_{vlm}|$. Thus, 
\begin{equation}
    \lambda_{vl} = |\alpha_{vl} + {\rm Tr}[\hat W^{(v,l)}]| + \sum_{m}|f_{vlm}|.
\end{equation}
The constant shift to make the Hamiltonian semi-positive definite is
\begin{equation}
    \Omega = \sum_{j=0}^{D-1} (|\delta_j| - \delta_j) + \frac{1}{2}\sum_{vl}\alpha_{vl}^2 - (\alpha + \beta N) = \|\hat W_0\|_{\rm SC} - {\rm Tr}[\hat W_0] + \frac{1}{2} \sum_{vl} \alpha^2_{vl} - (\alpha + \beta N)
\end{equation}

\subsection{Validation on the Fe$_2$S$_2$ Hamiltonian used by Low \emph{et al.}}\label{app:sq_validation_low}
We test our implementation on the Fe$_2$S$_2$ Hamiltonian with 30 electrons in 20 orbitals. It was constructed in Ref.~\cite{li_spin-projected_2017} and is one of the systems studied in Ref.~\cite{low_fast_2025}. This system is small enough for its energies to be computed with methods that do not rely on a single reference determinant: the density-matrix renormalisation group (DMRG) and full configuration interaction (FCI). We can therefore check two things: that we reproduce the factorisations of Ref.~\cite{low_fast_2025}, and that the factorisations accepted by our accuracy criterion are accurate. We use the published integrals~\cite{rubin_fast_2025} and the four sets of ranks of Table~VI of Ref.~\cite{low_fast_2025}, which are $(V,L,M)=(12,10,20)$, $(14,10,10)$, $(14,5,15)$ and $(11,10,15)$ in our notation. The resource estimate of Ref.~\cite{low_fast_2025} uses the third set. All factorisations discussed below were obtained in this work.

\paragraph{Protocols.}
We compare two protocols. Each is run without and with the HF term of~\cref{eq: objective_function}, from ten initial guesses for every set of ranks. Each run gives one factorised Hamiltonian, so there are 40 factorised Hamiltonians in each of the four cases.

The first protocol is the objective function of Ref.~\cite{low_fast_2025}. It has the gap term of~\cref{eq:gap_term_low} with $E_{\rm reg}=0.977$~Ha, the gap of their solution~\cite{rubin_fast_2025}, and $\epsilon_{\rm fro}=10^{-2}$~Ha and $\epsilon_{\Lambda}=20$~Ha, as in their Table~IV. It is minimised in a single stage of $10^6$ steps, with a learning rate that decreases from $10^{-1}$ to $10^{-5}$. When the HF term is added, we choose $\epsilon_{\rm HF}=5\times10^{-3}$~Ha.

The second protocol is our production protocol (\cref{app:factorisation_search}), with $\epsilon_{\rm fro}=10^{-4}$~Ha, $\epsilon_{\Lambda}=25$~Ha and $\epsilon_{\rm gap}=0.5$~Ha. Its first stage has 500k steps and $\epsilon_{\rm HF}=10^{-4}$~Ha. Its second stage is the one described in~\cref{app:factorisation_search}.

In all runs, the initial parameters are drawn from a normal distribution with a standard deviation of 0.1.

\paragraph{Two measures of the error.}
Our accuracy criterion is the same as for all systems of this work,
\begin{equation}\label{eq:energy_criterion}
    \max\left(|\Delta E_{\rm tot}|,|\Delta E_{\rm corr}|\right)<0.3~{\rm mHa},
\end{equation}
where $\Delta E_{\rm tot}$ and $\Delta E_{\rm corr}$ are the errors of the total energy and of the correlation energy, computed with UCCSD(T). We say that a run that satisfies~\cref{eq:energy_criterion} is in the box.

For this small system, we can check the criterion against a more accurate method. The ground state of this model has total spin $S=0$~\cite{li_spin-projected_2017}, and it is the state targeted by QPE. We compute the total energy of this state for the unfactorised and for the factorised Hamiltonian with spin-adapted DMRG~\cite{zhai_block2_2023}, and call the difference $\Delta E_{\rm DMRG}$. We say that a run is accurate if $|\Delta E_{\rm DMRG}|<0.3$~mHa. DMRG is not exact: its accuracy is set by the bond dimension, for which we use 500. As a check, we repeated the calculation with a bond dimension of 1000 for 33 of the factorised Hamiltonians, including all runs in the box, and $\Delta E_{\rm DMRG}$ changed by less than 0.03~mHa. The DMRG energies of~\cref{tab:sq_validation_energies} are computed with a bond dimension of 1000. For two of its factorised Hamiltonians, a bond dimension of 2000 changes $\Delta E_{\rm DMRG}$ by 0.001~mHa.

The state with $S=0$ is not described by a single determinant. For the HF term and for UCCSD(T), we therefore use a broken-symmetry unrestricted HF determinant with $N_\uparrow=N_\downarrow$, as for the manganese dimers (\cref{app:chemical_systems}). Ref.~\cite{low_fast_2025} also uses the energy of the state with $S=0$ for their gap and for the errors in the last column of their Table~VI. They select factorisations, however, with the CCSD(T) correlation energy of a different state, the high-spin state with eight unpaired electrons. They give two reasons: CCSD(T) converges for this state, and earlier resource estimates used it. They also note that this test may not be a good measure of the error for the state with $S=0$.

\begin{figure*}[htb!]
\centering
\includegraphics[width=\textwidth]{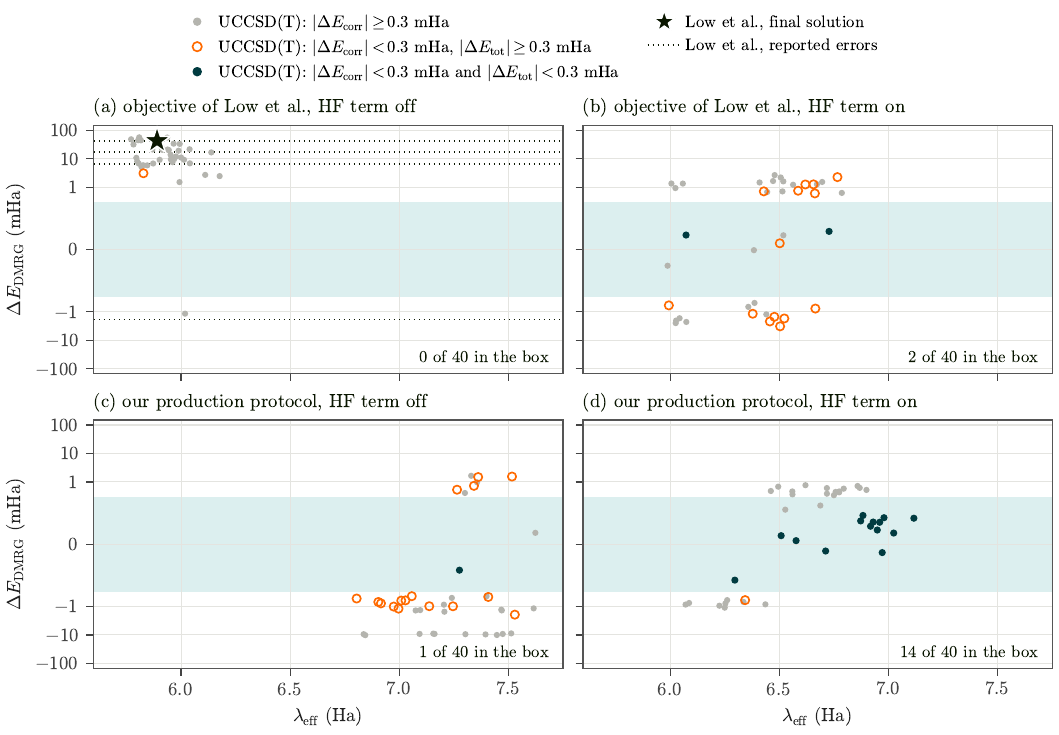}
\caption{Validation on the Fe$_2$S$_2$ Hamiltonian with 30 electrons in 20 orbitals of Ref.~\cite{li_spin-projected_2017}, which is used in Ref.~\cite{low_fast_2025}: error of the total energy of the ground state, which has $S=0$, computed with DMRG, $\Delta E_{\rm DMRG}$ (factorised minus unfactorised Hamiltonian), against $\lambda_{\rm eff}$. Each point is one factorisation obtained in this work, and each panel contains ten initial guesses for each of the four sets of ranks. (a, b) objective function of Ref.~\cite{low_fast_2025}, i.e.\ with the gap term of~\cref{eq:gap_term_low} and their parameters; (c, d) our production protocol; left: without the HF term, right: with it. The markers give the result of our accuracy criterion, which uses UCCSD(T), as in~\cref{fig:search_sq}. Filled circles: runs in the box, $\max(|\Delta E_{\rm tot}|,|\Delta E_{\rm corr}|)<0.3$~mHa; open circles: $|\Delta E_{\rm corr}|<0.3$~mHa but $|\Delta E_{\rm tot}|\geq0.3$~mHa; grey points: $|\Delta E_{\rm corr}|\geq0.3$~mHa. Shaded band: $|\Delta E_{\rm DMRG}|<0.3$~mHa; the vertical axis is linear inside it and logarithmic outside it. Star: solution of Ref.~\cite{low_fast_2025}; dotted lines: errors of their Table~VI for the four sets of ranks, with the signs that follow from the energies listed there. $\lambda_{\rm eff}$ is evaluated as in~\cref{tab:sq_validation_low}.}
\label{fig:sq_validation_low}
\end{figure*}

\paragraph{Results for the state with $S=0$.}
\Cref{tab:sq_validation_low} and~\cref{fig:sq_validation_low} show the results. With the objective function of Ref.~\cite{low_fast_2025} and no HF term, our $\Lambda$, $E_{\rm gap}$ and $\lambda_{\rm eff}$ are close to those of their solution. The total energy of the state with $S=0$ is too high, by more than 3~mHa in 36 of the 40 runs and by up to 57~mHa. Their Table~VI reports errors of $-1.9$, $6.5$, $17.1$ and $43.0$~mHa for the four sets of ranks. None of our runs is in the box.

Adding the HF term to this objective function brings all errors below 3.3~mHa in magnitude. It increases $\lambda_{\rm eff}$ by 3 to 14\%, depending on the ranks. With our production protocol and the HF term, all errors are below 1.1~mHa in magnitude, and 14 of the 40 runs are in the box. Without the HF term, one run is.

\Cref{fig:sq_validation_parity}(a) compares the two measures of the error for all 160 runs. $\Delta E_{\rm tot}$ from UCCSD(T) follows $\Delta E_{\rm DMRG}$: they differ by 0.05~mHa (median) and by at most 0.36~mHa. In total, 17 runs are in the box, and all of them are accurate, with $|\Delta E_{\rm DMRG}|\leq0.23$~mHa. Therefore, there is no run that UCCSD(T) accepts and DMRG rejects. The opposite case occurs: of the 24 accurate runs, seven are not in the box, because $\Delta E_{\rm tot}$ or $\Delta E_{\rm corr}$ from UCCSD(T) is between 0.3 and 0.72~mHa in magnitude. The correlation energy alone is not a sufficient test: 49 runs have $|\Delta E_{\rm corr}|<0.3$~mHa, and $|\Delta E_{\rm DMRG}|$ reaches 3.3~mHa among them. \Cref{tab:sq_validation_energies} lists the total energies of the unfactorised Hamiltonian and, for each set of ranks, of the best factorisation, i.e.\ the run with the lowest $\lambda_{\rm eff}$ among those that are both in the box and accurate. With our production protocol and the HF term, every set of ranks has such a run. For $(14,5,15)$, the ranks of the resource estimate of Ref.~\cite{low_fast_2025}, it has $\lambda_{\rm eff}=6.51$~Ha and $\Delta E_{\rm DMRG}=0.05$~mHa, against 5.89~Ha and 43.0~mHa for their solution.

\paragraph{State with $S=4$.}
For the state with $S=0$, the criterion is wrong in one direction only: it rejects some accurate factorisations, but it accepts no inaccurate one. This is not a general property of UCCSD(T). To show this, we repeat the test for the lowest state with $S=4$, which has eight unpaired electrons, as the state that Ref.~\cite{low_fast_2025} uses for CCSD(T). We run our production protocol with the UHF determinant with eight unpaired electrons as the reference of the HF term, of $E_0$ and of UCCSD(T). For this state, the total energies can be computed exactly with FCI. We call the error $\Delta E_{\rm FCI}$, and a run is accurate if $|\Delta E_{\rm FCI}|<0.3$~mHa.

\begin{figure*}[htb!]
\centering
\includegraphics[width=\textwidth]{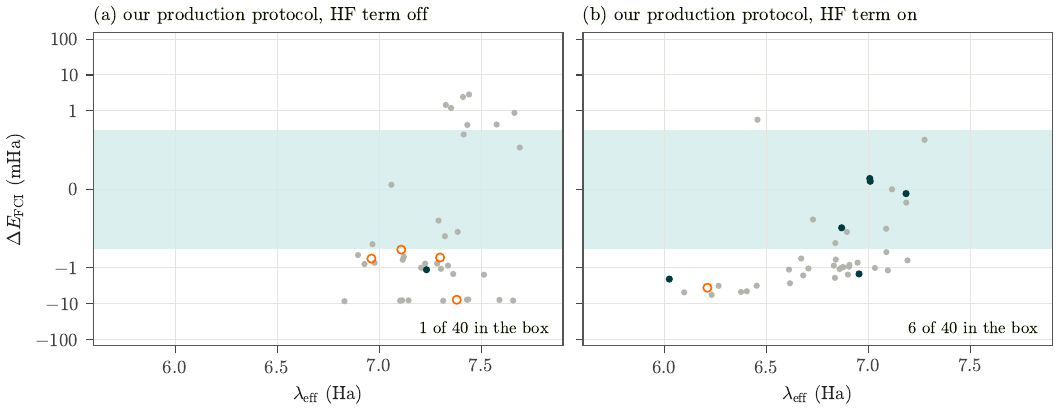}
\caption{As~\cref{fig:sq_validation_low}, for the lowest state with $S=4$ and our production protocol, (a) without and (b) with the HF term: error of the total energy of this state computed with FCI, $\Delta E_{\rm FCI}$, against $\lambda_{\rm eff}$. The reference of the HF term and of UCCSD(T) is the UHF determinant with eight unpaired electrons. Markers as in~\cref{fig:sq_validation_low}; shaded band: $|\Delta E_{\rm FCI}|<0.3$~mHa. Filled circles outside the band are runs that the criterion accepts although they are not accurate. Conversely, grey points and open circles inside the band are runs that are accurate according to FCI but are rejected by the UCCSD(T) criterion.}
\label{fig:sq_validation_s4}
\end{figure*}

\begin{figure}[htb!]
\centering
\includegraphics[width=0.75\textwidth]{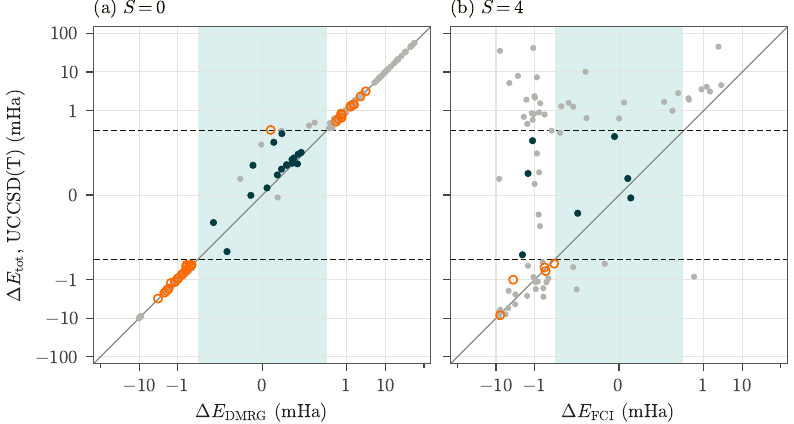}
\caption{Error of the total energy from UCCSD(T), $\Delta E_{\rm tot}$, against the error of the total energy from DMRG or FCI, with the markers of~\cref{fig:sq_validation_low}. (a) State with $S=0$: the 160 runs of~\cref{fig:sq_validation_low}, against $\Delta E_{\rm DMRG}$. (b) State with $S=4$: the 80 runs of~\cref{fig:sq_validation_s4}, against $\Delta E_{\rm FCI}$. Points on the diagonal have the same error in both. Shaded band: $|\Delta E_{\rm DMRG}|<0.3$~mHa in (a) and $|\Delta E_{\rm FCI}|<0.3$~mHa in (b); dashed lines: $|\Delta E_{\rm tot}|=0.3$~mHa. Both axes are linear between $-0.3$ and $0.3$~mHa and logarithmic outside.}
\label{fig:sq_validation_parity}
\end{figure}

\Cref{fig:sq_validation_s4} and~\cref{fig:sq_validation_parity}(b) show that, for this state, the error estimated with UCCSD(T) can be too optimistic for one factorisation and too pessimistic for another. $\Delta E_{\rm tot}$ from UCCSD(T) differs from $\Delta E_{\rm FCI}$ by 1.4~mHa (median) and by up to 43~mHa.

The estimate can be too optimistic. Of the seven factorisations in the box, i.e.\ those that satisfy our criterion based on UCCSD(T),~\cref{eq:energy_criterion}, three are not accurate, with $\Delta E_{\rm FCI}$ of $-1.1$, $-1.5$ and $-2.1$~mHa. For example, for the run in the box with the lowest $\lambda_{\rm eff}$, the error is $-0.28$~mHa according to UCCSD(T) and $-2.06$~mHa according to FCI.

The estimate can also be too pessimistic. Of the 18 accurate factorisations, i.e.\ those with $|\Delta E_{\rm FCI}|<0.3$~mHa, 14 are not in the box, i.e.\ they do not satisfy~\cref{eq:energy_criterion}. For example, for one run with $(V,L,M)=(12,10,20)$ and the HF term, the error is $-1.80$~mHa according to UCCSD(T) and $-0.20$~mHa according to FCI. For these reasons, the factorisation listed for this state in~\cref{tab:sq_validation_energies} is the best run that is both in the box and accurate.

For the unfactorised Hamiltonian, UCCSD(T) is 47~mHa above FCI for the state with $S=4$, against 9.6~mHa above DMRG for the state with $S=0$ (\cref{tab:sq_validation_energies}). In addition, UCCSD does not converge within 150 iterations in 13 of the 80 runs. Therefore, the agreement found for $S=0$ cannot be assumed for other states. The error of the CCSD(T) energy is, nevertheless, the metric that is commonly used in the literature to select factorisations~\cite{lee_even_2021,goings_reliably_2022,rocca_reducing_the_runtime_2024,oumarou_accelerating_quantum_2024,caesura_faster_quantum_chemistry_2025,low_fast_2025}, since DMRG and FCI are not affordable for large active spaces. Our goal is to compare first and second quantisation. We therefore apply the same criterion to both, so that the comparison is consistent.

\begin{table*}[htbp]
\centering
\begin{tabular}{lllccccccc}
\hline\hline
State, method & Protocol & HF term & $\Lambda$ (Ha) & $E_{\rm gap}$ (Ha) & $\lambda_{\rm eff}$ (Ha) & $\Delta E$ (mHa) & In the box & Accurate & Both \\
\hline
$S=0$, DMRG & Gap term of~\cref{eq:gap_term_low} & off & 16.8--19.9 & 0.977--1.035 & 5.8--6.2 & $-1.2$ to $+56.6$ & 0 & 0 & 0 \\
 & & on & 18.9--24.1 & 0.972--0.976 & 6.0--6.8 & $-3.3$ to $+2.7$ & 2 & 6 & 2 \\
 & Our production protocol & off & 27.8--34.7 & 0.836--0.850 & 6.8--7.6 & $-10.2$ to $+1.6$ & 1 & 2 & 1 \\
 & & on & 22.1--30.5 & 0.832--0.852 & 6.1--7.1 & $-1.1$ to $+0.8$ & 14 & 16 & 14 \\
 & Solution of Ref.~\cite{low_fast_2025} & off & 17.5 & 1.020 & 5.9 & $+43.0$ & & & \\
\hline
$S=4$, FCI & Our production protocol & off & 27.7--34.7 & 0.854--0.865 & 6.8--7.7 & $-8.5$ to $+2.9$ & 1 & 7 & 0 \\
 & & on & 21.3--31.2 & 0.849--0.870 & 6.0--7.3 & $-5.7$ to $+0.6$ & 6 & 11 & 4 \\
\hline\hline
\end{tabular}
\caption{Validation on the Fe$_2$S$_2$ Hamiltonian of Ref.~\cite{li_spin-projected_2017}, which is used in Ref.~\cite{low_fast_2025}. All factorisations were obtained in this work, with the objective function that has the gap term of~\cref{eq:gap_term_low} and the parameters of Ref.~\cite{low_fast_2025}, or with our production protocol; the solution of Ref.~\cite{low_fast_2025} is listed for comparison. Each line summarises 40 runs (four sets of ranks, ten initial guesses): the smallest and largest values of $\Lambda$, $E_{\rm gap}$, $\lambda_{\rm eff}$ and of the error of the total energy of the state, $\Delta E$, which is $\Delta E_{\rm DMRG}$ for $S=0$ and $\Delta E_{\rm FCI}$ for $S=4$. In the box: runs accepted by our UCCSD(T) criterion. Accurate: runs with $|\Delta E|<0.3$~mHa. Both: runs that are in the box and accurate. $E_{\rm gap}=E_0+\Omega$ and $\lambda_{\rm eff}$,~\cref{eq:effective_norm}, are evaluated with the total energy of the state from DMRG or FCI for the factorised Hamiltonian as $E_0$. The solution of Ref.~\cite{low_fast_2025} has $(V,L,M)=(14,5,15)$; its values are from the published data~\cite{rubin_fast_2025} and their Table~VI.}
\label{tab:sq_validation_low}
\end{table*}

\begin{table*}[htbp]
\centering
\begin{tabular}{lccccccc}
\hline\hline
 & & & & & \multicolumn{3}{c}{Total energy} \\
Hamiltonian & $(V,L,M)$ & $\Lambda$ (Ha) & $E_{\rm gap}$ (Ha) & $\lambda_{\rm eff}$ (Ha) & UHF & UCCSD(T) & DMRG or FCI \\
\hline
\multicolumn{8}{l}{\emph{State with $S=0$; last column: DMRG}} \\
Unfactorised: $E$ (Ha) & & & & & $-116.512692$ & $-116.596038$ & $-116.605602$ \\
Gap term of~\cref{eq:gap_term_low}: $\Delta E$ (mHa) & (14, 10, 10) & 19.42 & 0.974 & 6.07 & $+0.403$ & $+0.286$ & $+0.083$ \\
 & (12, 10, 20) & 23.70 & 0.975 & 6.73 & $-0.059$ & $+0.139$ & $+0.115$ \\
Our production protocol: $\Delta E$ (mHa) & (14, 10, 10) & 23.82 & 0.847 & 6.30 & $-0.140$ & $-0.128$ & $-0.233$ \\
 & (14, 5, 15) & 25.76 & 0.836 & 6.51 & $-0.023$ & $+0.246$ & $+0.053$ \\
 & (11, 10, 15) & 26.08 & 0.843 & 6.58 & $-0.027$ & $+0.033$ & $+0.020$ \\
 & (12, 10, 20) & 28.43 & 0.843 & 6.87 & $-0.005$ & $+0.173$ & $+0.148$ \\
\hline
\multicolumn{8}{l}{\emph{State with $S=4$; last column: FCI}} \\
Unfactorised: $E$ (Ha) & & & & & $-116.484910$ & $-116.541982$ & $-116.589000$ \\
Our production protocol: $\Delta E$ (mHa) & (12, 10, 20) & 27.87 & 0.859 & 6.87 & $-0.003$ & $-0.085$ & $-0.193$ \\
\hline\hline
\end{tabular}
\caption{Total energies of the unfactorised Hamiltonian and energy errors of factorisations obtained with the HF term. For the unfactorised Hamiltonian, we list the total energy $E$ in Ha. For each factorised Hamiltonian, which is named by the protocol that gave it, we list the error $\Delta E$ in mHa, i.e.\ its total energy minus that of the unfactorised Hamiltonian, computed with the method of the column. For each protocol and each set of ranks, we list the run with the lowest $\lambda_{\rm eff}$ among the runs that are both in the box and accurate in~\cref{fig:sq_validation_low}(b),~\cref{fig:sq_validation_low}(d) and~\cref{fig:sq_validation_s4}(b). A set of ranks is omitted if it has no such run: $(14,5,15)$ and $(11,10,15)$ for the gap term of~\cref{eq:gap_term_low}, and all sets except $(12,10,20)$ for $S=4$. The resource estimate of Ref.~\cite{low_fast_2025} uses $(14,5,15)$. $E_{\rm gap}$ and $\lambda_{\rm eff}$ are evaluated as in~\cref{tab:sq_validation_low}. UHF: the broken-symmetry determinant with $N_\uparrow=N_\downarrow$ for $S=0$ and the determinant with eight unpaired electrons for $S=4$, converged for each Hamiltonian. Last column: DMRG with a bond dimension of 1000 for $S=0$, and FCI for $S=4$.}
\label{tab:sq_validation_energies}
\end{table*}

\section{Molecular models and active spaces}\label{app:chemical_systems}
We compute the one- and two-electron integrals of all systems, and their Hartree--Fock and UCCSD(T) energies with PySCF~\cite{sun_python_2026}.

\paragraph{Hydrogen square.}
H$_4$ consists of four hydrogen atoms at the corners of a square with a side of 2~bohr (1.058~\AA), as in Refs.~\cite{lee_even_2021,georges_quantum_2025}. We use the def2-SVPD and def2-TZVPP basis sets and keep all their orbitals, $D=32$ and $D=56$, respectively.

\paragraph{Iron--sulfur cluster.}
For [Fe$_2$S$_2$]$^{2-}$, we use the active-space Hamiltonian of Ref.~\cite{georges_quantum_2025}. It has the same 14 active electrons as the CAS(14,12) model of the [2Fe-2S] cluster in Ref.~\cite{lee_evaluating_2023} (their Supplementary Section~S4.1), but also includes virtual orbitals. It is computed using the cc-pVTZ-DK basis set with the sf-X2C Hamiltonian, starting from high-spin restricted open-shell BP86 orbitals, which are localised separately in the occupied and virtual spaces. The active space consists of 12 occupied orbitals, the five 3d orbitals of each iron atom and one 3p orbital of each bridging sulfur atom, and of localised virtual orbitals, selected as in Ref.~\cite{georges_quantum_2025}. Of the 128 orbitals of Ref.~\cite{georges_quantum_2025}, we keep the first 122, which are all centred on the iron or bridging sulfur atoms. We consider the state with eight unpaired electrons, $N_\uparrow=11$ and $N_\downarrow=3$.

\paragraph{Permanganate.}
The geometry of [MnO$_4$]$^-$ was optimised in vacuum assuming $T_d$ symmetry with BLYP and the 6-31G basis set, as in Ref.~\cite{olsen_modeling_2017}, resulting in a Mn--O distance of 1.640~\AA. The Hamiltonian is computed using the cc-pVTZ-DK basis set from restricted B3LYP orbitals. The active space consists of the seven highest doubly occupied orbitals (14 electrons), the same number of active electrons as in the CAS(14,12) space of Ref.~\cite{olsen_modeling_2017}. Apart from the occupied orbitals, the active space also contains virtual orbitals selected from MP2 natural orbitals with the largest occupation numbers, up to 128 orbitals in total.

\paragraph{Geometries of the manganese dimers.}
Both manganese dimers are modelled by replacing the bulky terminal ligands of the experimental complexes with NH$_3$ around the same donor atoms, thus preserving the coordination number of each manganese atom but removing the ligand $\pi$ orbitals from the active space. [Mn$_2$($\mu$-O)$_2$(NH$_3$)$_8$]$^{3+}$ is a model of the mixed-valence Mn(III)Mn(IV) complex [(phen)$_2$Mn($\mu$-O)$_2$Mn(phen)$_2$]$^{3+}$~\cite{stebler_1986,larsen_2002}, in which each 1,10-phenanthroline (phen) is replaced by two NH$_3$. Starting from an idealised structure, the geometry was optimised with unrestricted B3LYP using the cc-pVDZ-DK basis set in the doublet state. In the optimised structure the Mn--Mn distance is 2.79~\AA\ and the two manganese atoms are inequivalent: the additional electron of Mn(III) remains localised on one atom which has longer Mn--O bonds (1.87 and 1.94~\AA) than the Mn(IV) atom (1.74 and 1.77~\AA). [Mn$_2$($\mu$-O)$_3$(NH$_3$)$_6$]$^{2+}$ is a model of the Mn(IV)Mn(IV) complex [Mn$_2$($\mu$-O)$_3$(Me$_3$tacn)$_2$]$^{2+}$ (Me$_3$tacn = 1,4,7-trimethyl-1,4,7-triazacyclononane)~\cite{wieghardt_1988}, in which each Me$_3$tacn is replaced by three NH$_3$, as in Ref.~\cite{delfs_stranger_2000}. Starting from a $D_{3h}$ structure with the experimental Mn--Mn distance of 2.296~\AA~\cite{wieghardt_1988}, the geometry was optimised with broken-symmetry unrestricted B3LYP, cc-pVDZ-DK and the spin-free exact two-component (sf-X2C) Hamiltonian, which gives Mn--Mn 2.359~\AA\ and Mn--O 1.826~\AA.

\paragraph{Active spaces of the manganese dimers.}
The Hamiltonians are computed using the cc-pVTZ-DK basis set with the sf-X2C Hamiltonian, starting from restricted open-shell B3LYP orbitals. The occupied active orbitals are chosen with the atomic valence active space (AVAS) projection~\cite{sayfutyarova_2017} onto the Mn 3d and O 2p orbitals of a minimal basis: the doubly occupied orbitals are rotated to diagonalise this projection and those with weights above 0.98 are kept, while the remaining weights are below 0.33 and belong to NH$_3$ lone pairs; singly occupied orbitals are always active. This gives nine doubly occupied and one singly occupied orbital (19 electrons) for [Mn$_2$O$_2$]$^{3+}$, and twelve doubly occupied orbitals (24 electrons) for [Mn$_2$O$_3$]$^{2+}$, i.e.\ the d electrons of both manganese atoms and the 2p electrons of the bridging oxygen atoms. The remaining active orbitals are the virtual MP2 natural orbitals with the largest occupation numbers, up to 128 orbitals in total; in this MP2 calculation, the occupied orbitals outside the active space are frozen. The 128 orbitals contain 93\% and 95\% of the Mn 3d and O 2p minimal-basis character, respectively; almost all of the rest is in frozen occupied orbitals, mainly the Mn--NH$_3$ bonding orbitals, and less than 0.2\% is in the discarded virtual orbitals. The smallest occupation numbers of the natural orbitals kept are $6.4\times10^{-5}$ and $1.3\times10^{-4}$. The remaining occupied orbitals (62 and 54) are frozen and enter the active-space Hamiltonian through their mean field. The energy errors of the factorised Hamiltonians are evaluated with UCCSD(T) using a broken-symmetry unrestricted Hartree--Fock reference. For [Mn$_2$O$_2$]$^{3+}$ we find two nearly degenerate broken-symmetry solutions and use the one with the lower UCCSD(T) energy: at $D=128$ it lies 5.7~mHa above the other at the Hartree--Fock level but 55~mHa below it at the UCCSD(T) level. For [Mn$_2$O$_3$]$^{2+}$ we use the lowest Hartree--Fock solution found.

\section{Cost model}\label{app:cost_model}
In this appendix, we summarise how we obtain the Toffoli and qubit counts of the first-quantised QPE in~\cref{fig:be_vs_electrons_D,fig:qpe_fronts_cold,fig:qpe_fronts_monolithic}. The cost model counts Toffoli gates and logical qubits; Clifford gates are free. All rotations are performed by addition into a phase-gradient state, which requires Toffoli gates only, except for the two rotations of UNIFORM, which we convert to Toffoli gates below.

\paragraph{Composition rules.} Every gadget $G$ of the circuit has a set of compilations, each with a Toffoli count $T_G$ and a peak number $A_G$ of clean ancillae, which are returned to $\ket{0}$ when the gadget ends. For a budget $A$ of clean ancillae, we take the compilation with the fewest Toffolis and $A_G\le A$. Gadgets applied one after another add their Toffolis, and take the maximum of their ancillae, $T=T_{G_1}+T_{G_2}$, $A=\max(A_{G_1},A_{G_2})$; a gadget that runs while $k$ qubits are held, e.g.\ the output of a lookup during an addition, needs $A_G+k$ ancillae.  Sweeping $A$ over all values at which one of the gadgets changes its compilation gives the Toffoli--qubit front of~\cref{fig:be_vs_electrons_D,fig:qpe_fronts_cold,fig:qpe_fronts_monolithic}. The number of logical qubits is the number of persistent registers (\cref{tab:fq_registers}) plus $A$.

\paragraph{Primitives.} In addition to the elements of~\cref{eq:element_costs}, the circuit uses the following gadgets:
\begin{itemize}
    \item lookup rotation $\mathrm{LR}(L,b)$, a rotation by one of $L$ angles of $b$ bits selected by an address register~\cite{gidney_halving_2018,sanders2020compilation}: $\qroam(L,b)$ loads the angle~\cite{berry_qubitization_2019,low_trading_2024}, $\addpg(b)$ adds it into the phase-gradient state while the $b$-bit output is held~\cite{gidney_halving_2018,low_denser_2026}, and $\qroam^\dagger(L)$ uncomputes it~\cite{berry_qubitization_2019}. With the costs of~\cref{eq:element_costs},
    \begin{equation}\label{eq:lookup_rotation}
    \begin{aligned}
        T_{\rm LR}(L,b) &= \lceil L/\lambda\rceil+b(\lambda-1)+T_{\addpg}(b)+\lceil L/\lambda'\rceil+\lambda',\\
        A_{\rm LR}(L,b) &= \max\big(b\lambda+\lceil\log_2(L/\lambda)\rceil,\; b+A_{\addpg}(b),\; \lambda'+\lceil\log_2(L/\lambda')\rceil\big),
    \end{aligned}
    \end{equation}
    where $(T_{\addpg},A_{\addpg})=(b-1,b-1)$ or $(2b-2,0)$, and $\lambda$, $\lambda'$ and the adder are chosen for the ancilla budget. The controlled rotation, $\mathrm{LR}_{\rm c}(L,b)$, costs two Toffolis more, one in \qroam{} and one in $\qroam^\dagger$~\cite{babbush_encoding_2018}. A table with a single entry is loaded with Clifford gates;
    \item uniform superposition over $N=2^kL'$ states with $L'$ odd, prepared with amplitude amplification~\cite{babbush_encoding_2018}. The gate counts of its implementation according to Qualtran~\cite{harrigan_qualtran_2024} are: $3\lceil\log_2L'\rceil-1$ Toffolis and two single-qubit rotations, with $\lceil\log_2 L'\rceil+2$ clean ancillae; nothing for $L'=1$. Each rotation is counted as 11 $T$ gates, i.e.\ 5.5 Toffolis, since one CCZ state, the magic state consumed by a Toffoli gate, can be converted into two $T$ states~\cite{gidney_efficient_2019}; these rotations contribute less than $0.1\%$ of the Toffoli count of a walk step for the systems considered here;
    \item SWAP network that moves electron register $j$ into position 0~\cite{su_yuan_fault-tolerant_2021}: $(N-1)\lceil\log_2D\rceil$ Toffolis (controlled swaps), no ancillae;
    \item comparator of two $n$-bit registers~\cite{gidney_halving_2018}: $n$ Toffolis, $n$ ancillae, uncomputed by measurement; it is used only by alias sampling in second quantisation;
    \item reflection about $\ket{0}$ on $k$ qubits, a ladder of logical-AND gates~\cite{gidney_halving_2018}: $k-2$ Toffolis, $k-2$ ancillae.
\end{itemize}

\paragraph{Pure-state preparation.}
PREP prepares the outer register in the state
\[
  \sum_{j=0}^{X-1} \sqrt{p_j}\,\ket{j}
\]
of Eq.~\eqref{eq: PREP}, where $p_j$ are the probabilities $p_{vl}$ and
$p_{V0}$ of Eq.~\eqref{eq:be_probabilities}. The index
$j = v\,2^{\lceil\log_2 L\rceil} + l$ labels the two-body term $(v,l)$,
with $p_j = 0$ for the padding values $l \ge L$, and
$j = V\,2^{\lceil\log_2 L\rceil}$ labels the one-body term. The state
therefore has $X = V\,2^{\lceil\log_2 L\rceil} + 1$ amplitudes on
$n = \lceil\log_2 X\rceil$ qubits.

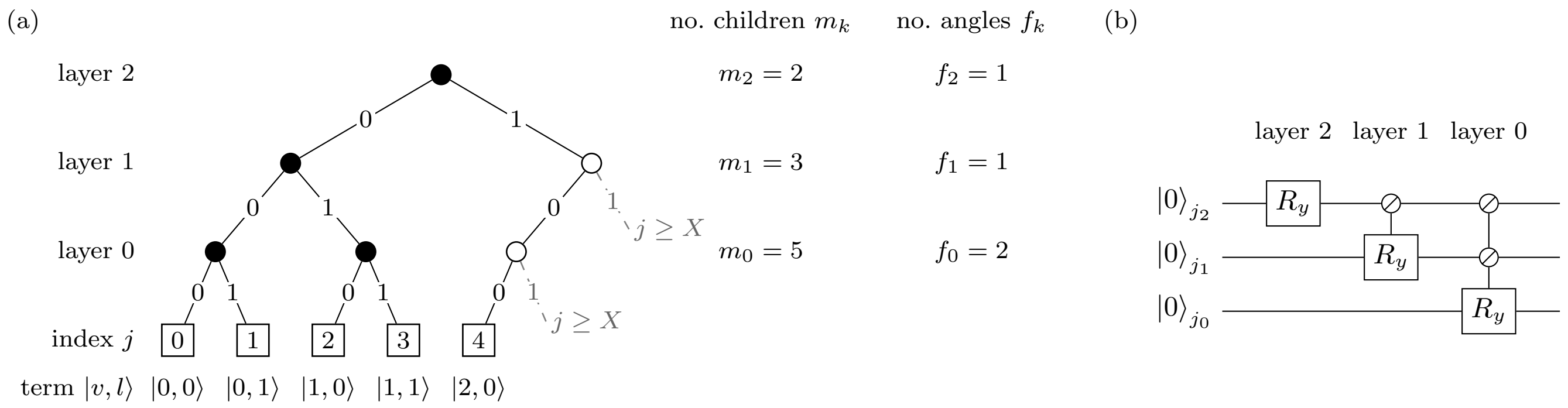
\begin{figure}[t]
\centering
\tikzset{
  spfull/.style={circle, draw, fill=black, line width=0.5pt, inner sep=0pt, minimum size=2.2mm},
  spopen/.style={spfull, fill=white},
  spleaf/.style={draw, line width=0.5pt, inner sep=0pt, minimum size=3.6mm, font=\footnotesize},
  spout/.style={draw=black!50, dashed, line width=0.5pt},
  splab/.style={font=\footnotesize, inner sep=1.5pt},
  spbit/.style={font=\footnotesize, inner sep=0.8pt, fill=white},
}%
\begin{tikzpicture}
\begin{scope}[x=0.85cm,y=1.0cm]
  \node[splab, anchor=west] at (-2.35,0.6) {(a)};
  \node[spfull] (r)  at (3.5,0) {};
  \node[spfull] (a0) at (1.5,-1) {};
  \node[spopen] (a1) at (5.5,-1) {};
  \node[spfull] (b0) at (0.5,-2) {};
  \node[spfull] (b1) at (2.5,-2) {};
  \node[spopen] (b2) at (4.5,-2) {};
  \foreach \j in {0,...,4} \node[spleaf] (l\j) at (\j,-3) {$\j$};
  \draw (r)  -- node[spbit] {0} (a0);  \draw (r)  -- node[spbit] {1} (a1);
  \draw (a0) -- node[spbit] {0} (b0);  \draw (a0) -- node[spbit] {1} (b1);
  \draw (a1) -- node[spbit] {0} (b2);
  \draw (b0) -- node[spbit] {0} (l0);  \draw (b0) -- node[spbit] {1} (l1);
  \draw (b1) -- node[spbit] {0} (l2);  \draw (b1) -- node[spbit] {1} (l3);
  \draw (b2) -- node[spbit] {0} (l4);
  \draw[spout] (a1) -- node[spbit, text=black!60] {1} (6.0,-1.75) node[splab, right, text=black!60] {$j\ge X$};
  \draw[spout] (b2) -- node[spbit, text=black!60] {1} (4.9,-2.8) node[splab, right, text=black!60] {$j\ge X$};
  \foreach \j/\v/\l in {0/0/0,1/0/1,2/1/0,3/1/1,4/2/0} \node[splab] at (\j,-3.55) {$\ket{\v,\l}$};
  \foreach \k in {0,1,2} \node[splab, anchor=east] at (-0.5,\k-2) {layer $\k$};
  \node[splab, anchor=east] at (-0.5,-3)    {index $j$};
  \node[splab, anchor=east] at (-0.5,-3.55) {term $\ket{v,l}$};
  \node[splab] at (7.75,0.6) {no.\ children $m_k$};
  \node[splab] at (10.55,0.6) {no.\ angles $f_k$};
  \foreach \k/\m/\f in {0/5/2,1/3/1,2/2/1} {
    \node[splab] at (7.75,\k-2) {$m_{\k}=\m$};
    \node[splab] at (10.55,\k-2) {$f_{\k}=\f$};
  }
\end{scope}
\node[splab, anchor=west] at (10.4,0.6) {(b)};
\begin{scope}[shift={(11.8,-1.2)}]
\begin{yquant}[operator/separation=5mm]
qubit {$\ket{0}_{j_2}$} q2;
qubit {$\ket{0}_{j_1}$} q1;
qubit {$\ket{0}_{j_0}$} q0;
[name=s2]
box {$R_y$} q2;
[multictrl, name=s1]
box {$R_y$} q1 ~ q2;
[multictrl, name=s0]
box {$R_y$} q0 ~ q2, q1;
\end{yquant}
\foreach \k in {0,1,2} \node[splab] at (s\k |- 0,0.55) {layer \k};
\end{scope}
\end{tikzpicture}
  \caption{Pure-state preparation~\cite{shende_synthesis_2006,low_trading_2024}
    of $X=5$ amplitudes on $n=3$ qubits, the state of PREP for $V=2$ and
    $L=2$. (a)~Binary tree of the index $j = 4j_2 + 2j_1 + j_0$. Layers
    $k$ are counted from the bottom; the nodes of layer $k$ rotate qubit
    $j_k$, and their branches 0 and 1 are the values of $j_k$. Squares:
    indices $j$, with the terms $\ket{v,l}$ that they label; $j=4$ is the
    one-body term. Black circle: qubit $j_k$ is rotated by an angle loaded
    by the lookup of layer $k$. White circle: branch 1 leads only to
    indices $j\ge X$ (dashed), so qubit $j_k$ stays in $\ket{0}$ and no
    angle is loaded. Right: layer $k$ has $m_k = \lceil X/2^k\rceil$
    children, the nodes just below it, and loads
    $f_k = \lfloor m_k/2\rfloor$ angles, in total
    $f_0+f_1+f_2 = 4 = X-1$. (b)~The circuit: layer $k$ rotates qubit
    $j_k$ by an angle that depends on the qubits above it (slashed
    circles: multiplexed control).}
  \label{fig:state-prep}
\end{figure}

Figure~\ref{fig:state-prep}(a) shows the binary tree of the state
preparation, for which we use the variant~\cite{low_trading_2024} of the
Shende--Bullock--Markov method~\cite{shende_synthesis_2006}. Each node of
the tree is a rotation, and each layer corresponds to the qubit that is
rotated. We count the layers $k = 0,\dots,n-1$ from the bottom. The
angle of a rotation depends on the qubits above it, so each layer is a
multiplexed rotation (Fig.~\ref{fig:state-prep}(b)), which we implement
as the lookup rotation $\mathrm{LR}$ of Eq.~\eqref{eq:lookup_rotation}.
Only the black nodes require a rotation to reach all $X$ leaves.

\textit{Number of angles.} Let $m_k$ be the number of children of layer $k$, i.e.\ of the nodes just
below it, for example $m_0 = X$. The nodes of layer $k$ take these children in consecutive pairs, so each node has two children, except that, if $m_k$ is odd, the last node has only one (white circles in~\cref{fig:state-prep}(a)). Therefore layer $k$ has
$m_{k+1} = \lceil m_k/2 \rceil$ nodes and, by induction,
$m_k = \lceil X/2^k \rceil$. Only the black nodes, which have two children each, need a rotation, so the number of angles of layer $k$ is the number of its black nodes, i.e.\ half the number of its children.
If $m_k$ is odd, the last child is the only child of a white node, which needs no rotation, so the half is rounded down, and layer $k$ loads
\begin{equation}
  f_k = \left\lfloor \frac{m_k}{2} \right\rfloor
      = \left\lfloor \frac{1}{2}\left\lceil \frac{X}{2^k}
        \right\rceil \right\rfloor
  \label{eq:state-prep-angles}
\end{equation}
angles. Since $\lfloor m/2 \rfloor = m - \lceil m/2 \rceil$ and
$\lceil m_k/2 \rceil = m_{k+1}$, the total telescopes,
\[
  \sum_{k=0}^{n-1} f_k
  = \sum_{k=0}^{n-1} \Bigl( m_k - \Bigl\lceil \frac{m_k}{2} \Bigr\rceil \Bigr)
  = \sum_{k=0}^{n-1} \bigl( m_k - m_{k+1} \bigr)
  = m_0 - m_n = X - 1,
\]
with $m_n = \lceil X/2^n \rceil = 1$, giving one angle for each free parameter
of a real normalised state of dimension $X$.\footnote{Eq.~(E27) of
Ref.~\cite{low_denser_2026} gives a smaller count because its last step
does not load all angles.}

\textit{Cost.} The layers are applied one after another, so PREP costs
\begin{equation}
  T_{\mathrm{PREP}} = \sum_{k=0}^{n-1} T_{\mathrm{LR}}(f_k, b),
  \qquad
  A_{\mathrm{PREP}} = \max_k A_{\mathrm{LR}}(f_k, b),
  \label{eq:state-prep-cost}
\end{equation}
and UNPREP, its inverse, costs the same.

\textit{Why not alias sampling.} As in Ref.~\cite{low_denser_2026}, we use
clean ancillae only and no coherent alias sampling, which keeps a large
number of junk qubits alive throughout the block-encoding. The
bottleneck is not PREP but the unitary synthesis inside the
block-encoding, so we leave these qubits to it as clean ancillae. With
dirty ancillae, the lookups of the block-encoding could borrow the junk
qubits. However, a borrowed qubit reduces the Toffoli count less than a
clean one, and PREP is a small part of the cost: for
$[\mathrm{MnO}_4]^-$, PREP, UNPREP and the other small gadgets of the
walk step together account for at most 3\% of its Toffoli count
(\cref{fig:be_breakdown}). We have not costed this option.

\paragraph{Walk step.} \Cref{tab:fq_walk} lists the gadgets of one step of the quantum walk (\cref{fig: quantum_circuit_be_full,fig: quantum_circuit_be_sum_over_electrons,fig: quantum_circuit_be_rectungular_matrices,fig: quantum_circuit_be_multiplex_be}) and their multiplicities. Only the two-body factors $\hat R^{(v)}$ (for $v<V$) are multiplexed: the one-body factor (for $v=V$) is the identity in the rotated frame~\eqref{eq:R_rotated}, so its angles are zero and are not loaded. A multiplexed unitary with $N_S$ layers in a given ordering is costed by the optimal windowing of~\cref{sec:us_cost}, with $L_{{\rm s},\omega}\to VL_{{\rm s},\omega}$ to account for the multiplexing on $v$. Unitaries of dimension $D$ that is not a power of 2 are costed at the padded dimension $2^{\lceil\log_2D\rceil}$ (\cref{app:nonpow2}). All angles have $b=16$ bits.

Three small gadgets of~\cref{tab:fq_walk} involve the qubit $\ket{\cdot}_{\rm b}$ of~\cref{fig: quantum_circuit_be_full}, which marks the one-body term ($v=V$): the computation of this qubit, the reflection REF, and the AND gate for the second block-encoding. If $V=2^k$, the lower $k$ qubits of $\ket{v}$ hold the two-body indices and its most significant qubit is set only for $v=V$, so it is the qubit $\ket{\cdot}_{\rm b}$ and costs nothing; otherwise, we compute $\ket{\cdot}_{\rm b}$ from $\ket{v}$ by testing $v=V$; \cref{fig: quantum_circuit_be_full} draws this as part of PREP. The reflection REF acts on $\ket{j}_{n_e}$ and the four flag qubits and is controlled by the control qubit and by $\ket{0}_{\rm b}$, i.e.\ it is a reflection on $\lceil\log_2N\rceil+6$ qubits. The second block-encoding is controlled by the same two qubits; we combine them into one qubit with an AND gate, which is held during the second block-encoding.

\begin{table}[htb!]
\centering
\renewcommand{\arraystretch}{1.3}
\begin{tabular}{l l}
\hline\hline
Register & Qubits \\
\hline
electrons $1,\dots,N-1$ (system, cold storage) & $(N-1)\lceil\log_2D\rceil$ \\
electron 0 (system), $\ket{p,m}$, padded to $\max(D,M)$ & $\lceil\log_2\max(D,M)\rceil$ \\
electron index $\ket{j}_{n_e}$ (cold storage) & $\lceil\log_2N\rceil$ \\
$\ket{v}$, $\ket{l}$ & $\lceil\log_2(V+1)\rceil$, $\lceil\log_2L\rceil$ \\
little-LCU qubit, ${\rm flag}_{0,1,2}$ & 4 \\
one-body qubit $\ket{\cdot}_{\rm b}$ (if $V\ne2^k$) & 1 \\
phase-gradient state & $b$ \\
QPE qubits~\cite{low_denser_2026} (cold storage) & 2 \\
\hline\hline
\end{tabular}
\caption{Persistent registers of the first-quantised QPE; one of the two QPE qubits controls the walk step.}
\label{tab:fq_registers}
\end{table}

\begin{table*}[htb!]
\centering
\renewcommand{\arraystretch}{1.3}
\begin{tabular}{l l l l c}
\hline\hline
Circuit & Gadget & Toffoli count & Clean ancillae & Per walk step \\
\hline
\Cref{fig: quantum_circuit_be_full} & PREP, UNPREP & \cref{eq:state-prep-cost} & \cref{eq:state-prep-cost} & 1 each \\
\Cref{fig: quantum_circuit_be_full} & one-body qubit $\ket{\cdot}_{\rm b}$ from $\ket{v}$ (if $V\ne2^k$) & $\lceil\log_2(V+1)\rceil-1$ & $\lceil\log_2(V+1)\rceil-1$ & 1 \\
\Cref{fig: quantum_circuit_be_sum_over_electrons}& UNIFORM, UNIFORM$^\dagger$ over electrons & $3\lceil\log_2L'\rceil+10$ & $\lceil\log_2 L'\rceil+2$ & 2 each \\
\Cref{fig: quantum_circuit_be_sum_over_electrons}& SWAP, SWAP$^\dagger$ & $(N-1)\lceil\log_2D\rceil$ & 0 & 2 each \\
\Cref{fig: quantum_circuit_be_rectungular_matrices} & $\hat U^{(v)\dagger}$, $\hat V^{(v)}$ and inverses & \cref{eq:total_toffoli} & \cref{eq:window_ancilla} & 2 each \\
\Cref{fig: quantum_circuit_be_rectungular_matrices}& singular values of $\hat R^{(v)}$ & $T_{\rm LR}(VD,b)$ & $A_{\rm LR}(VD,b)$ & 4 \\
\Cref{fig: quantum_circuit_be_multiplex_be} & little-LCU rotation and inverse & $T_{\rm LR}(VL+1,b)+2$ & $A_{\rm LR}(VL+1,b)$ & 2 each \\
\Cref{fig: quantum_circuit_be_multiplex_be} & ${\rm flag}_1$ & 1 & 0 & 2 \\
\Cref{fig: quantum_circuit_be_multiplex_be} & $\hat F^{(v,l)}$ & $T_{\rm LR}(VLM+D,b)+2$ & $A_{\rm LR}(VLM+D,b)$ & 2 \\
\Cref{fig: quantum_circuit_be_full} & sign $(-1)^{\theta_{vl}}$ & $\qroam(X,1)+\qroam^\dagger(X)$ & \cref{eq:element_costs} & 1 \\
\Cref{fig: quantum_circuit_be_full} & REF on $\lceil\log_2N\rceil+6$ qubits & $\lceil\log_2N\rceil+4$ & $\lceil\log_2N\rceil+4$ & 1 \\
\Cref{fig: quantum_circuit_be_full}& AND of control and $\ket{0}_{\rm b}$ & 1 & 1 & 1 \\
\hline\hline
\end{tabular}
\caption{Gadgets of one step of the first-quantised quantum walk, their Toffoli and clean-ancilla counts, and how often they appear per walk step. Where a row names several gadgets, the last column gives how often each of them appears; a gadget and its inverse have the same cost. The block-encoding of~\cref{fig: quantum_circuit_be_sum_over_electrons} appears twice per walk step, and each contains $\hat R^{(v)}$ and $\hat R^{(v)\dagger}$. The circuit of~\cref{fig: quantum_circuit_be_rectungular_matrices} therefore appears four times per walk step, twice as drawn and twice inverted: $\hat U^{(v)\dagger}$, $\hat V^{(v)}$ and their inverses appear twice each, and the lookup rotation of the singular values four times. UNIFORM includes two rotations of 5.5 Toffolis each, and costs nothing when $N$ is a power of 2 ($L'=1$), e.g.\ for H$_4$. The little-LCU rotation and $\hat F^{(v,l)}$ are controlled lookup rotations, controlled by the control qubit and by the little-LCU qubit, respectively (\cref{fig: quantum_circuit_be_multiplex_be}), hence the $+2$; the lookup rotation of the singular values is not controlled. The one-body qubit, REF and the AND gate are described in the text.}
\label{tab:fq_walk}
\end{table*}

\paragraph{QPE.} The QPE cost is $N_W$ times the Toffoli count of one walk step, with $N_W=\lceil\pi\lambda_{\rm eff}/(2\sigma_{\rm QPE})\rceil$,~\cref{eq: no qpe calls}, and $\sigma_{\rm QPE}=1.0$~mHa. We use two QPE qubits as in Ref.~\cite{low_denser_2026} and neglect the Toffolis of the QPE register and of the initial-state preparation.

\paragraph{Dirty ancillae.} A lookup of $L$ entries of $B$ bits may keep $\lambda-1$ of its $\lambda$ copies on borrowed (dirty) qubits~\cite{berry_qubitization_2019,low_denser_2026}: $2\lceil L/\lambda\rceil+4B(\lambda-1)$ Toffolis, $B(\lambda-1)$ dirty and $B+\lceil\log_2(L/\lambda)\rceil$ clean qubits; its measurement-based uncomputation costs $2\lceil L/\lambda'\rceil+4\lambda'$ Toffolis with $\lambda'-1$ dirty and $\lceil\log_2(L/\lambda')\rceil+1$ clean qubits, and the controlled versions cost two Toffolis more. For a lookup rotation, we also allow the two-pass scheme of Ref.~\cite{low_denser_2026}: one pass of $\lceil L/\lambda\rceil+2B(\lambda-1)$ Toffolis writes the angle and leaves the borrowed copies toggled, and the same pass after \addpg{} clears the output and restores them. In the windows of the multiplexed unitaries, only the copies of \qroam{} and of its phase fix-up may be dirty; the output register stays clean. A qubit may be borrowed only if no other gate acts on it while it holds a copy. The number of qubits that a gadget may borrow, its pool, is the number of qubits of the circuit that the gadget does not act on. The pools differ between the three stages of the block-encoding $\hat W^{(v,l)}$ (\cref{fig: quantum_circuit_be_multiplex_be}):
\begin{equation}\label{eq:fq_pools}
\begin{aligned}
    P_0 &= s\left[(N-1)\lceil\log_2D\rceil+\lceil\log_2N\rceil+1\right]+n_{\rm b},\\
    P_R &= P_0+\lceil\log_2L\rceil+3,\\
    P_{\rm LCU} &= P_0+3+\lceil\log_2\max(D,M)\rceil,\\
    P_F &= P_0+2.
\end{aligned}
\end{equation}
\begin{itemize}
    \item $P_0$ counts the qubits that are idle in all three stages: the $N-1$ electron registers that are not swapped into position 0, the electron index $\ket{j}_{n_e}$, the one-body qubit $\ket{\cdot}_{\rm b}$ ($n_{\rm b}=1$ if it is a separate qubit, i.e.\ if $V$ is not a power of 2, and $n_{\rm b}=0$ otherwise), and the idle QPE qubit. Here $s=1$ in a monolithic architecture and $s=0$ in the cold-storage architecture, where these $N-1$ electron registers, the electron index and the QPE qubits are in the cold storage and cannot be borrowed.
    \item $P_R$ is the pool of the multiplexed block-encodings $\hat R^{(v)}$ and $\hat R^{(v)\dagger}$, i.e.\ of the windows of $\hat U^{(v)\dagger}$ and $\hat V^{(v)}$ and of the lookup rotation of the singular values. They act on $\ket{v}$, $\ket{p,m}$, ${\rm flag}_0$ and the phase-gradient state; idle are $\ket{l}$, the little-LCU qubit, ${\rm flag}_1$ and ${\rm flag}_2$.
    \item $P_{\rm LCU}$ is the pool of the little-LCU rotation $R_y(\phi_{vl})$ and its inverse. They act on the control qubit, $\ket{v}$, $\ket{l}$, the little-LCU qubit and the phase-gradient state; idle are the three flags and the electron register $\ket{p,m}$ in position 0, which is not in the cold storage.
    \item $P_F$ is the pool of the lookup rotation $\hat F^{(v,l)}$. It acts on the little-LCU qubit, $\ket{v}$, $\ket{l}$, $\ket{p,m}$, ${\rm flag}_2$ and the phase-gradient state; idle are ${\rm flag}_0$ and ${\rm flag}_1$.
\end{itemize}
Only the lookups of the following three stages may borrow qubits: the unitary synthesis of $\hat U^{(v)\dagger}$ and $\hat V^{(v)}$, the lookup rotations of the singular values and of $\hat F^{(v,l)}$, and the little-LCU rotation. The other gadgets of~\cref{tab:fq_walk}, i.e.\ PREP and UNPREP, the one-body qubit, UNIFORM, SWAP, ${\rm flag}_1$, the sign $(-1)^{\theta_{vl}}$, REF and the AND gate, use clean ancillae only. The QPE qubit that controls the walk step is idle during $\hat R^{(v)}$ and $\hat F^{(v,l)}$, but, as in second quantisation, it is not borrowed, since both QPE qubits are in the cold storage, as in Ref.~\cite{low_denser_2026}. The register that the multiplexed unitaries act on, $\ket{p,m}$, is never borrowed by them.

\paragraph{Second quantisation.} We cost the block-encoding of Ref.~\cite{low_denser_2026} (App.~E, Fig.~45, stages E2--E4) with the same primitives and composition rules, with the following choices: the ranks $(R,B,C)=(V,M,L)$ of the factorisation; $b=16$ bits for all angles and coefficients; OUTER and INNER prepared either as pure states or by alias sampling, whichever is cheaper at each budget; the Givens angles streamed in batches of any size, or loaded once and held from $\hat R$ to $\hat R^\dagger$; and, with dirty ancillae, the idle system qubits, the two spin indices and the idle QPE qubit borrowed only in a monolithic architecture, since they are in the cold storage otherwise, and, during the Givens rotations, only the system qubits of the spin sector that is not rotated.

\begin{figure*}[htb!]
\centering
\includegraphics[width=0.75\textwidth]{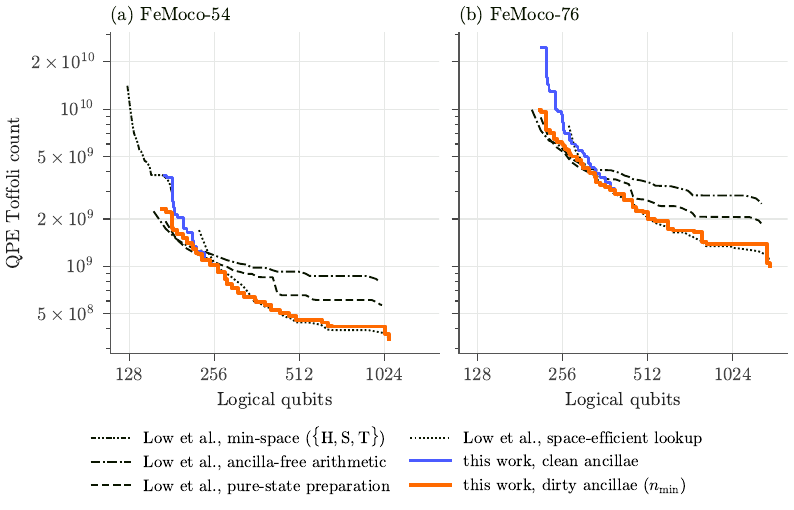}
\caption{Validation of our second-quantised cost model against Ref.~\cite{low_denser_2026}: Toffoli count of QPE with $\sigma_{\rm QPE}=1$~mHa against the number of logical qubits for FeMoco-54 and FeMoco-76, with the factorisations of Table~XII of Ref.~\cite{low_denser_2026} and two QPE qubits. Black: the curves of \cite[Fig.~46 (left)]{low_denser_2026}, digitised from the vector paths of the published figure. Orange: our cost model with dirty ancillae and the pool of dirty qubits of Ref.~\cite{low_denser_2026}, $n_{\rm min}$. Blue: our cost model with clean ancillae only.}
\label{fig:sq_validation}
\end{figure*}

\paragraph{Validation against Ref.~\cite{low_denser_2026}.} Ref.~\cite{low_denser_2026} reports the cost of the second-quantised QPE for FeMoco only graphically (their Fig.~46), so we digitised their curves from the vector paths of the published figure. With their factorisations (their Table~XII), their pool of dirty qubits and two QPE qubits, our cost model reproduces the lower envelope of their pure-state-preparation and space-efficient-lookup compilations (\cref{fig:sq_validation}). We compare the Toffoli counts at every qubit count at which our front changes, between 173 and 1\,031 logical qubits for FeMoco-54 and between 215 and 1\,352 for FeMoco-76. The median difference is $4\%$ and $5\%$, respectively, and the difference is below $20\%$ at almost all of these qubit counts. The largest differences, up to $30\%$, occur at small qubit counts, where our Toffoli counts are higher than theirs by about $10\%$ on average; they arise because some compilation choices change at slightly different qubit counts in the two models. Our fronts start at slightly fewer qubits (167 and 212 instead of 172 and 214), and at the Toffoli-optimal end our minimal Toffoli counts are $1\%$ and $8\%$ lower than theirs. We do not reproduce their ancilla-free and Clifford+T compilations, which we do not consider in this work (\cref{sec:resource_estimates}).

\paragraph{Code.} The implementation of this cost model will be made available with the data of this article (\cref{sec:data}).

\section{Total QPE cost in a monolithic architecture}\label{app:monolithic}
\begin{figure*}[htb!]
\centering
\includegraphics[width=\textwidth]{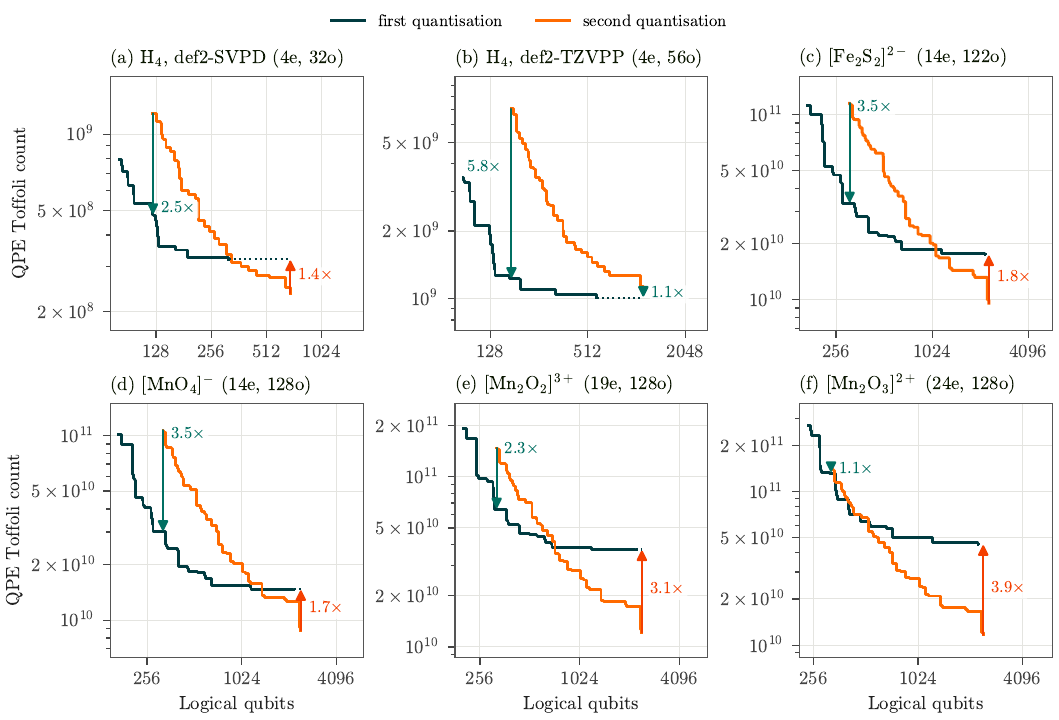}
\caption{As~\cref{fig:qpe_fronts_cold}, but in a monolithic architecture, where the qubits of the cold storage, i.e.\ the idle electron registers and the electron index (first quantisation), the spin-orbital qubits and spin indices (second quantisation) and the idle QPE qubit, can also be borrowed as dirty ancillae.}
\label{fig:qpe_fronts_monolithic}
\end{figure*}

In a monolithic architecture, all logical qubits are stored in the same way, and every register that is idle during a table lookup can be borrowed as dirty ancillae, including those kept in the cold storage in the architecture of Ref.~\cite{low_denser_2026} (\cref{sec:resource_estimates}): the idle QPE qubit and, in first quantisation, the $N-1$ electron registers that are not swapped into position 0 and the electron index, and in second quantisation, the $2D$ qubits of the spin orbitals and the two spin indices. As in the cold-storage architecture, the QPE qubit that controls the walk step is not borrowed (\cref{app:cost_model}). Otherwise, we use the same factorisations and compilation rules as in~\cref{fig:qpe_fronts_cold}. The fronts are shown in~\cref{fig:qpe_fronts_monolithic}. Since the second-quantised system register is larger, second quantisation gains more at low qubit counts: at its fewest qubits, first quantisation requires 2.5, 5.8, 3.5, 3.5, 2.3 and 1.1 times fewer Toffoli gates for H$_4$ (def2-SVPD and def2-TZVPP), [Fe$_2$S$_2$]$^{2-}$, [MnO$_4$]$^-$, [Mn$_2$O$_2$]$^{3+}$ and [Mn$_2$O$_3$]$^{2+}$, respectively, compared with 3.4 to 15.9 in the cold-storage architecture. The Toffoli-optimal points do not change, since there the clean ancillae are sufficient. First quantisation remains cheaper up to the same qubit counts as in~\cref{fig:qpe_fronts_cold}, except for [Mn$_2$O$_3$]$^{2+}$, where second quantisation is cheaper from 337 logical qubits.

\section{Rough estimate of the effect of the hot and cold storage}\label{app:cold_weight}
Here, we give an estimate of how much of the advantage of first quantisation is kept when the denser encoding of the cold storage is taken into account. It is a simplified estimate, which only aims to show that the advantage is robust; the precise number of physical qubits requires a compilation to the error-correcting code. We use a simple model of the number of physical qubits,
\begin{equation}
n^{\rm phys}=n^{\rm phys}_{\rm comp}+c_{\rm hot}\,n_{\rm hot}+c_{\rm cold}\,n_{\rm cold}.
\label{eq:physical_qubits}
\end{equation}
Here, $n^{\rm phys}_{\rm comp}$ is the number of physical qubits in the compute region, $n_{\rm hot}$ and $n_{\rm cold}$ are the numbers of logical qubits in the hot and cold storage, and $c_{\rm hot}$ and $c_{\rm cold}$ are the numbers of physical qubits per logical qubit in each of them. Dividing~\eqref{eq:physical_qubits} by $c_{\rm hot}$ gives
\begin{equation}
\frac{n^{\rm phys}}{c_{\rm hot}}=\frac{n^{\rm phys}_{\rm comp}}{c_{\rm hot}}+n_{\rm eff},
\label{eq:effective_qubits}
\end{equation}
where
\begin{equation}
n_{\rm eff}=n_{\rm hot}+\rho\,n_{\rm cold},\qquad\rho=\frac{c_{\rm cold}}{c_{\rm hot}},
\label{eq:n_eff}
\end{equation}
is the effective number of logical qubits and $\rho$ is the cost of a qubit in the cold storage relative to one in the hot storage. Thus, $c_{\rm hot}n_{\rm eff}$ is the number of physical qubits in the hot and cold storage, and $n_{\rm eff}$ is the number of logical qubits in the hot storage that would require the same number of physical qubits. For simplicity, we assume that both algorithms use the same layout, with the same compute region and the same code distance, so that $n^{\rm phys}_{\rm comp}$, $c_{\rm hot}$ and $c_{\rm cold}$ are the same for both. The first term in~\eqref{eq:effective_qubits} then shifts the fronts of both algorithms by the same number of qubits and does not change the ratio of their Toffoli counts at an equal number of physical qubits. We therefore compare the two algorithms at the same $n_{\rm eff}$. At small qubit counts, second quantisation requires several times more Toffoli gates, and thus a longer run, which may require a larger code distance. The assumption of equal code distances therefore favours second quantisation, such that our estimates of the advantage of first quantisation is conservative. We also do not consider the time per Toffoli gate.

\begin{figure}[htb!]
\centering
\includegraphics[width=\textwidth]{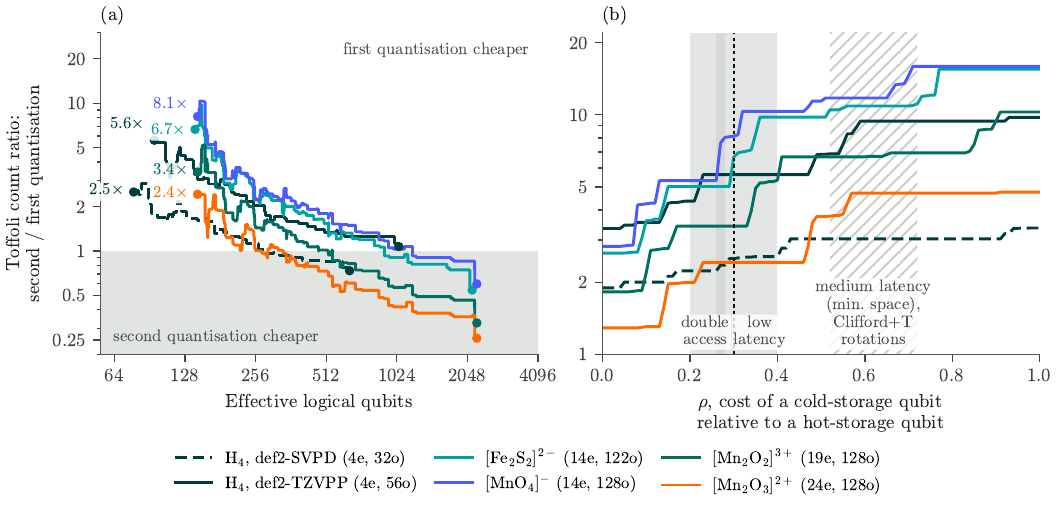}
\caption{Ratio of the Toffoli counts of quantum phase estimation in second and first quantisation for the six systems of~\cref{fig:summary}, (a) against the effective number of logical qubits, $n_{\rm eff}=n_{\rm hot}+0.3\,n_{\rm cold}$ (\cref{eq:n_eff}), and (b) against the cost $\rho$ of a qubit in the cold storage relative to one in the hot storage. In both panels, the two algorithms are compared at the same number of effective logical qubits. In (a), each curve starts at the fewest effective qubits on which the second-quantised algorithm can be implemented and ends where it reaches its minimal Toffoli count, as in~\cref{fig:summary}. In (b), the ratio is taken at the fewest effective qubits on which both algorithms can be implemented. For $\rho$ above 0.04 to 0.12, these are the fewest effective qubits of second quantisation, so that at $\rho=0.3$ (dashed line) panel (b) gives the left ends of the curves in (a). For smaller $\rho$, second quantisation can be implemented on fewer effective qubits than first quantisation, because it keeps more of its qubits in the cold storage, and the ratio is taken at the fewest effective qubits of first quantisation. Shaded: the values of $\rho$ in the layouts of Ref.~\cite{low_denser_2026} for the six systems of their Table~XXVII with double-access (minimum time) and low-latency hot storage, which overlap between 0.26 and 0.28; hatched: those with medium-latency (minimum space) hot storage, which use Clifford+T rotations.}
\label{fig:cold_weight}
\end{figure}

The ratios at the same $n_{\rm eff}$, \cref{fig:cold_weight}(a), depend on $\rho$, which is set by the layout and by the code distances of the hot and cold storage. In Ref.~\cite{low_denser_2026}, the cold storage packs logical qubits 4.5 times more densely than full-distance surface code patches, at the price of a high latency; in contrast, hot storage packs them the more densely, the slower its qubits are accessed: 2.5 times with medium and 1.4 times with low latency (their Table~VI). The value of $\rho$ is therefore set mainly by the hot storage: in the layouts of their Table~XXVII, a logical qubit in the cold storage requires 340 to 490 physical qubits, whereas one in the hot storage requires 590 to 840 with medium latency, 1\,000 to 1\,550 with low latency and 1\,440 to 1\,940 with double-access hot storage. In their layouts, the latency of the hot storage is matched to the rate at which the compute region consumes the magic states. The minimum-space layouts, with a single factory, therefore use medium-latency hot storage, with $\rho\approx2.5/4.5\approx0.56$, whereas the fastest layouts, with many factories and low-latency hot storage accessible from both sides, have $\rho\approx0.2$.

Ref.~\cite{low_denser_2026} reports layouts of the second-quantised algorithm for the six systems considered there, with 54 to 150 orbitals: two active spaces of FeMoco, one of cytochrome P450 and three of a ruthenium catalyst (their Table~XXVII). The minimum-space layouts use 132 to 386 logical qubits and have $\rho=0.52$--$0.72$. The low-latency layouts use 146 to 867 logical qubits and have $\rho=0.26$--$0.40$. The fastest layouts use 419 to 1\,951 logical qubits and have $\rho=0.20$--$0.28$. For the ruthenium catalyst with 100 and 150 orbitals, whose numbers of orbitals are closest to those of our systems, $\rho=0.52$--$0.65$, $0.26$--$0.34$ and $0.20$--$0.23$ in these three types of layouts, respectively. All minimum-space layouts synthesise the rotations from Clifford+T gates, which we do not consider (\cref{sec:resource_estimates}), whereas all layouts that, as in our compilation, perform rotations by addition into a phase-gradient state use low-latency or double-access hot storage. With up to 400 logical qubits, these layouts have $\rho=0.26$--$0.40$. We therefore take $\rho=0.3$, in the lower part of this range and within that of the ruthenium catalyst with 100 and 150 orbitals.

\Cref{fig:cold_weight}(a) shows the ratio of the Toffoli counts against the effective number of logical qubits for $\rho=0.3$. At the fewest effective qubits on which second quantisation can be implemented, first quantisation requires 2.5, 5.6, 6.7, 8.1, 3.4 and 2.4 times fewer Toffoli gates for H$_4$ in the def2-SVPD and def2-TZVPP basis sets, [Fe$_2$S$_2$]$^{2-}$, [MnO$_4$]$^-$, [Mn$_2$O$_2$]$^{3+}$ and [Mn$_2$O$_3$]$^{2+}$, respectively, and it remains cheaper up to between about 270 and 1\,200 effective qubits, and for H$_4$ in def2-TZVPP at every qubit count. \Cref{fig:cold_weight}(b) shows the ratio at the fewest effective qubits on which both algorithms can be implemented as a function of $\rho$, together with the values of $\rho$ in the layouts of Ref.~\cite{low_denser_2026}. The ratio increases with $\rho$, since second quantisation keeps more of its qubits in the cold storage. In the low-latency layouts, $\rho=0.26$--$0.40$, first quantisation requires 2.2 to 10.3 times fewer Toffoli gates, and in the fastest layouts, $\rho=0.20$--$0.28$, 2.0 to 8.1 times fewer. The fastest layouts, however, minimise the run time with many more qubits, whereas the advantage of first quantisation lies at small qubit counts. In the minimum-space layouts, $\rho=0.52$--$0.72$, first quantisation would require 3.0 to 15.9 times fewer Toffoli gates, but these layouts use Clifford+T rotations, which would increase the Toffoli counts of both algorithms. The ratios at the Toffoli-optimal end do not depend on $\rho$.

Finally, note that we have compared the two algorithms at the same $\rho$, i.e.\ in the same layout. In practice, each algorithm would be compiled to its own layout. Since first quantisation requires several times fewer Toffoli gates at small qubit counts, it can consume the magic states at a lower rate and still complete the phase estimation in the same time as second quantisation. It could therefore use a slower but denser hot storage, in which each logical qubit requires fewer physical qubits. Since first quantisation keeps more of its qubits in the hot storage than second quantisation, this would reduce its number of physical qubits relative to that of second quantisation. Such a comparison of two different layouts must be assessed by the number of physical qubits,~\eqref{eq:physical_qubits}, not by $n_{\rm eff}$, whose unit, a logical qubit in the hot storage, would then differ between the two algorithms. Therefore, the comparison at the same $\rho$ is likely to be pessimistic for first quantisation.

\section{Search for optimal factorisations}\label{app:factorisation_search}
For each system and representation, we minimise the objective function~\eqref{eq: objective_function} for several factorisation ranks $(V,L,M)$ and regularisation parameters $\epsilon_{\rm fro}$, $\epsilon_{\Lambda}$ and $\epsilon_{\rm gap}$ and, in first quantisation, for several orderings and numbers of layers, $N_S$, of the unitaries. We call one choice of these parameters a setting. For each setting, we perform 10 runs, each from a different initial guess, and we choose the factorisations used in~\cref{sec:resource_estimates} among the runs of all settings.

We optimise in two stages with the Adam optimiser~\cite{kingma_adam_2015}, and refer to this as our production protocol. In the first stage, we start from random initial guesses, with every parameter drawn uniformly from $[-0.1,0.1]$, and optimise for 125k steps (200k or 300k in a few settings of H$_4$ and [Fe$_2$S$_2$]$^{2-}$) with the HF term of~\cref{eq: objective_function} not smoothed, $\eta_{\rm HF}=0$, and with $\epsilon_{\rm HF}=10^{-3}$~Ha. In the second stage, we continue from converged solutions for 25k steps with the smoothed HF term, $\epsilon_{\rm HF}=5\times10^{-3}$~Ha and $\eta_{\rm HF}=10^{-4}$, and vary $\epsilon_{\rm fro}$, $\epsilon_{\Lambda}$ and $\epsilon_{\rm gap}$. In first quantisation, we then fit the angles of $N_S<2^{\lceil\log_2D\rceil}-1$ layers to the two unitaries of the SVD of $\hat U_{\rm ref}^{\dagger}\hat R^{(v)}$ and re-optimise the whole factorisation in the rotated frame for 50k steps with the same HF term, as in~\cref{sec:us_trainability}. Each run starts from a different full-depth solution. In every stage, the learning rate grows linearly from $10^{-3}$ to $10^{-2}$ during the first 0.7 to 1.6\% of the steps and then decreases to $10^{-10}$ with a cosine schedule~\cite{loshchilov_sgdr_2017} ($10^{-6}$ in the longer runs); the gradient is rescaled when its norm exceeds 1. A few settings deviate from this protocol: in second quantisation, the first-stage runs of [Fe$_2$S$_2$]$^{2-}$ use $\epsilon_{\rm HF}=5\times10^{-5}$~Ha and 150k steps, and one second-stage setting of [MnO$_4$]$^-$ uses $\epsilon_{\rm HF}=10^{-3}$~Ha; for [Mn$_2$O$_3$]$^{2+}$, one first-stage setting in each representation uses the smoothed HF term.

\begin{figure*}[htb!]
\centering
\includegraphics[width=\textwidth]{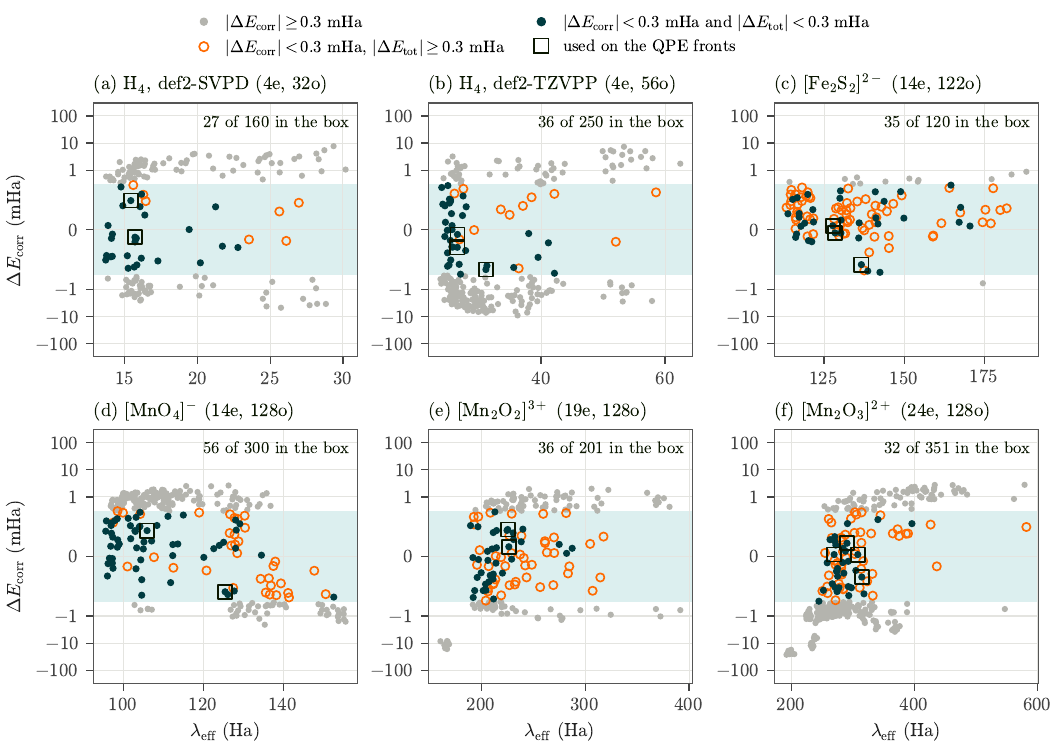}
\caption{Error of the correlation energy, $\Delta E_{\rm corr}$, against the effective subnormalisation factor, $\lambda_{\rm eff}$, of every first-quantised factorisation run (\cref{tab:search}). Filled circles: runs in the box, $\max(|\Delta E_{\rm tot}|,|\Delta E_{\rm corr}|)<0.3$~mHa; open circles: $|\Delta E_{\rm corr}|<0.3$~mHa (shaded band) but $|\Delta E_{\rm tot}|\geq0.3$~mHa; grey points: $|\Delta E_{\rm corr}|\geq0.3$~mHa. Squares: factorisations used on the QPE fronts of~\cref{fig:qpe_fronts_cold} (\cref{tab:factorisations}). The vertical axis is linear inside the shaded band and logarithmic outside it.}
\label{fig:search_fq}
\end{figure*}

\begin{figure*}[htb!]
\centering
\includegraphics[width=\textwidth]{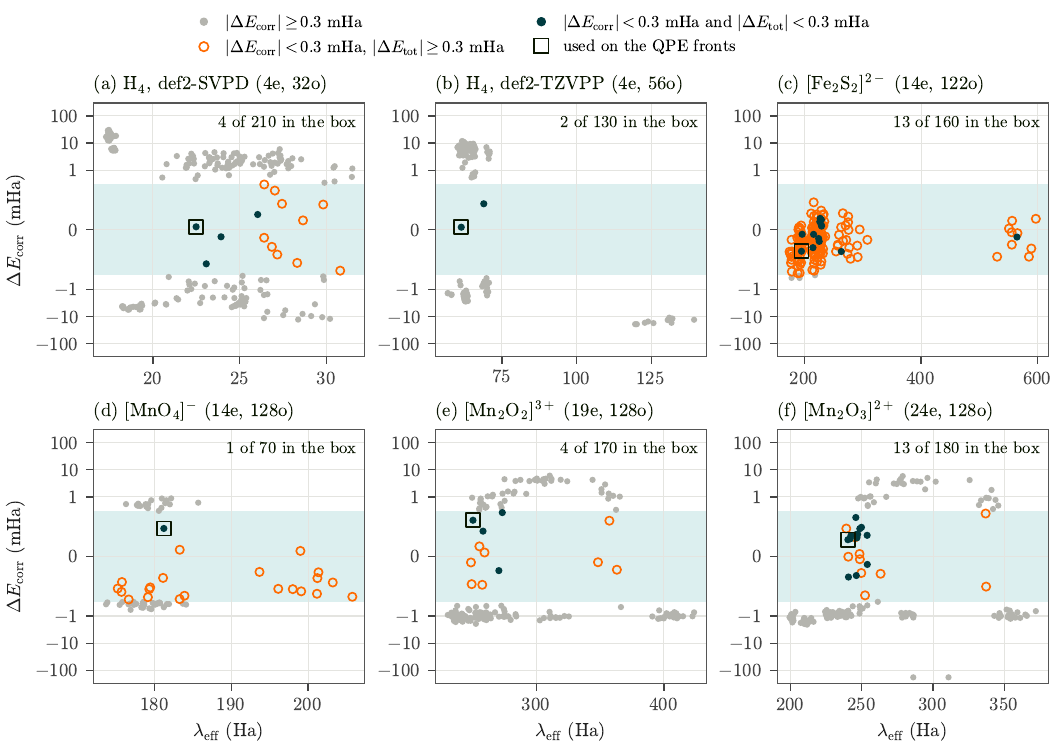}
\caption{As~\cref{fig:search_fq}, for second quantisation.}
\label{fig:search_sq}
\end{figure*}

\Cref{fig:search_fq,fig:search_sq} show the error of the correlation energy against $\lambda_{\rm eff}$ for every run, and~\cref{tab:search,tab:search_unitaries} list the settings. We distinguish the runs in the box, $\max(|\Delta E_{\rm tot}|,|\Delta E_{\rm corr}|)<0.3$~mHa, from the runs with an accurate correlation energy but an inaccurate total energy, which would be accepted if only the correlation energy were tested. For [Mn$_2$O$_2$]$^{3+}$, we include only the runs whose HF term uses the broken-symmetry solution with the lower UCCSD(T) energy (\cref{app:chemical_systems}). For H$_4$, [Fe$_2$S$_2$]$^{2-}$ and [MnO$_4$]$^-$, the full-depth first-quantised solutions from which the runs in the rotated frame start (20 to 70 per system) are not shown, since only the factorisations in the rotated frame enter~\cref{fig:qpe_fronts_cold} for these systems. They were obtained in the first stage, with $\epsilon_{\rm HF}=5\times10^{-3}$~Ha for H$_4$ and [Fe$_2$S$_2$]$^{2-}$, and for [MnO$_4$]$^-$ most of them were continued in the second stage for 50k steps. For the manganese dimers, the full-depth factorisations are also candidates for the fronts of~\cref{fig:qpe_fronts_cold} and are shown.

In first quantisation, every system has many runs in the box, obtained with many different settings, i.e.\ with different ranks, orderings and numbers of layers. In second quantisation, we performed a comparable number of runs and, for most systems, scanned a wider range of ranks and regularisation parameters, but only about $4\%$ of the runs reach the box, against $16\%$ in first quantisation, and for the three manganese systems they come from one or two settings, all with $L=D/2$. Most second-quantised runs miss the box because of the correlation energy (grey points). For [Fe$_2$S$_2$]$^{2-}$ in both representations, and for [MnO$_4$]$^-$ in second quantisation, most runs with an accurate correlation energy miss the box because of the total energy (open circles).

In second quantisation, the factorisation used on the QPE front of each system is the run in the box with the lowest $\lambda_{\rm eff}$. In first quantisation, it is not: the runs in the box with the lowest $\lambda_{\rm eff}$ have more layers of the unitaries or larger ranks, which makes their block-encodings more expensive, so that the fewest QPE Toffoli gates are obtained with a slightly larger $\lambda_{\rm eff}$. Therefore, in first quantisation, exploring several orderings and numbers of layers lowers the cost of QPE. Neither search is exhaustive: other ranks, regularisation parameters or longer optimisations could give factorisations with a lower $\lambda_{\rm eff}$ in either representation.

\begin{table*}[htbp]
\centering
\footnotesize
\renewcommand{\arraystretch}{1.25}
\setlength{\tabcolsep}{4pt}
\begin{tabular}{l c c c l l l l}
\hline\hline
System & & Runs & Settings & $(V, L, M)$ & $\epsilon_{\rm fro}$ & $\epsilon_{\Lambda}$ & $\epsilon_{\rm gap}$ \\
 & & & & & (Ha) & (Ha) & (Ha) \\
\hline
H$_4$, def2-SVPD & FQ & 27/160 & 10/16 & \parbox[t]{3.7cm}{\raggedright\strut $\mathbf{(4, 32, 32)}$, $\mathbf{(7, 32, 32)}$\strut} & \parbox[t]{3.45cm}{\raggedright\strut $\mathbf{10^{-3}}$\strut} & \parbox[t]{1.9cm}{\raggedright\strut $\mathbf{1}$\strut} & \parbox[t]{1.65cm}{\raggedright\strut $\mathbf{1}$\strut} \\
 & SQ & 4/210 & 4/21 & \parbox[t]{3.7cm}{\raggedright\strut $(1, 64, 64)$, $(1, 80, 80)$, $\mathbf{(4, 32, 32)}$, $(7, 21, 21)$, $(7, 32, 32)$\strut} & \parbox[t]{3.45cm}{\raggedright\strut \mbox{$\mathbf{7.7\times10^{-6}}$\textbf{--}$\mathbf{3.1\times10^{-5}}$\textsuperscript{\dag}}, $\mathbf{10^{-3}}$\strut} & \parbox[t]{1.9cm}{\raggedright\strut $0.1$, $\mathbf{1}$, $4$, $25$, $\mathbf{100}$, $200$, $\mathbf{400}$\strut} & \parbox[t]{1.65cm}{\raggedright\strut $\mathbf{0.1}$\strut} \\
\hline
H$_4$, def2-TZVPP & FQ & 36/250 & 12/25 & \parbox[t]{3.7cm}{\raggedright\strut $\mathbf{(4, 56, 56)}$, $\mathbf{(7, 56, 64)}$\strut} & \parbox[t]{3.45cm}{\raggedright\strut $\mathbf{10^{-3}}$\strut} & \parbox[t]{1.9cm}{\raggedright\strut $\mathbf{0.175}$, $\mathbf{1.75}$\strut} & \parbox[t]{1.65cm}{\raggedright\strut $\mathbf{0.1}$, $\mathbf{1}$\strut} \\
 & SQ & 2/130 & 2/13 & \parbox[t]{3.7cm}{\raggedright\strut $(4, 56, 56)$, $\mathbf{(7, 37, 37)}$, $\mathbf{(7, 56, 56)}$\strut} & \parbox[t]{3.45cm}{\raggedright\strut \mbox{$\mathbf{1.8\times10^{-5}}$\textbf{--}$\mathbf{5.0\times10^{-5}}$\textsuperscript{\dag}}, $\mathbf{10^{-3}}$\strut} & \parbox[t]{1.9cm}{\raggedright\strut $0.175$, $1.75$, $\mathbf{5}$, $200$, $400$\strut} & \parbox[t]{1.65cm}{\raggedright\strut $\mathbf{0.1}$\strut} \\
\hline
[Fe$_2$S$_2$]$^{2-}$ & FQ & 35/120 & 11/12 & \parbox[t]{3.7cm}{\raggedright\strut $\mathbf{(15, 91, 128)}$\strut} & \parbox[t]{3.45cm}{\raggedright\strut $\mathbf{10^{-3}}$\strut} & \parbox[t]{1.9cm}{\raggedright\strut $\mathbf{3.81}$\strut} & \parbox[t]{1.65cm}{\raggedright\strut $\mathbf{1}$\strut} \\
 & SQ & 13/160 & 8/15 & \parbox[t]{3.7cm}{\raggedright\strut $\mathbf{(15, 91, 122)}$\strut} & \parbox[t]{3.45cm}{\raggedright\strut $\mathbf{5\times10^{-5}}$\strut} & \parbox[t]{1.9cm}{\raggedright\strut $\mathbf{1}$, $\mathbf{3.81}$, $\mathbf{25}$, $\mathbf{100}$, $\mathbf{200}$, $400$\strut} & \parbox[t]{1.65cm}{\raggedright\strut $10^{-2}$, $2\times10^{-2}$, $\mathbf{0.1}$, $\mathbf{0.5}$, $\mathbf{2}$\strut} \\
\hline
[MnO$_4$]$^{-}$ & FQ & 56/300 & 21/30 & \parbox[t]{3.7cm}{\raggedright\strut $\mathbf{(16, 64, 128)}$, $\mathbf{(16, 128, 128)}$\strut} & \parbox[t]{3.45cm}{\raggedright\strut $\mathbf{10^{-3}}$\strut} & \parbox[t]{1.9cm}{\raggedright\strut $\mathbf{1.6}$, $\mathbf{3.2}$, $\mathbf{16}$\strut} & \parbox[t]{1.65cm}{\raggedright\strut $\mathbf{0.1}$\strut} \\
 & SQ & 1/70 & 1/7 & \parbox[t]{3.7cm}{\raggedright\strut $\mathbf{(16, 64, 128)}$, $(16, 128, 128)$\strut} & \parbox[t]{3.45cm}{\raggedright\strut $2\times10^{-4}$, $\mathbf{10^{-3}}$\strut} & \parbox[t]{1.9cm}{\raggedright\strut $\mathbf{16}$, $28$\strut} & \parbox[t]{1.65cm}{\raggedright\strut $\mathbf{1}$\strut} \\
\hline
[Mn$_2$O$_2$]$^{3+}$ & FQ & 36/201 & 14/21 & \parbox[t]{3.7cm}{\raggedright\strut $\mathbf{(16, 64, 128)}$, $\mathbf{(16, 128, 128)}$\strut} & \parbox[t]{3.45cm}{\raggedright\strut $\mathbf{3\times10^{-4}}$, $10^{-3}$\strut} & \parbox[t]{1.9cm}{\raggedright\strut $\mathbf{2}$, $\mathbf{4}$\strut} & \parbox[t]{1.65cm}{\raggedright\strut $0.1$, $\mathbf{1}$\strut} \\
 & SQ & 4/170 & 2/17 & \parbox[t]{3.7cm}{\raggedright\strut $(8, 128, 128)$, $(16, 32, 128)$, $(16, 48, 128)$, $\mathbf{(16, 64, 128)}$, $(16, 85, 128)$, $(16, 128, 85)$, $(16, 128, 128)$, $(24, 128, 128)$\strut} & \parbox[t]{3.45cm}{\raggedright\strut $10^{-4}$, $\mathbf{2\times10^{-4}}$\strut} & \parbox[t]{1.9cm}{\raggedright\strut $2$, $4$, $\mathbf{8}$, $16$, $32$\strut} & \parbox[t]{1.65cm}{\raggedright\strut $\mathbf{1}$, $3$\strut} \\
\hline
[Mn$_2$O$_3$]$^{2+}$ & FQ & 32/351 & 15/36 & \parbox[t]{3.7cm}{\raggedright\strut $\mathbf{(16, 64, 128)}$, $\mathbf{(16, 128, 128)}$\strut} & \parbox[t]{3.45cm}{\raggedright\strut $\mathbf{10^{-4}}$, $\mathbf{2\times10^{-4}}$, $10^{-3}$\strut} & \parbox[t]{1.9cm}{\raggedright\strut $\mathbf{2}$, $4$, $8$, $16$\strut} & \parbox[t]{1.65cm}{\raggedright\strut $\mathbf{0.1}$, $\mathbf{0.5}$, $\mathbf{1}$\strut} \\
 & SQ & 13/180 & 2/18 & \parbox[t]{3.7cm}{\raggedright\strut $(8, 128, 128)$, $(16, 32, 128)$, $(16, 48, 128)$, $\mathbf{(16, 64, 128)}$, $(16, 85, 128)$, $(16, 128, 85)$, $(16, 128, 128)$, $(24, 128, 128)$\strut} & \parbox[t]{3.45cm}{\raggedright\strut $\mathbf{2\times10^{-4}}$, $10^{-3}$\strut} & \parbox[t]{1.9cm}{\raggedright\strut $2$, $4$, $\mathbf{8}$, $16$, $32$\strut} & \parbox[t]{1.65cm}{\raggedright\strut $\mathbf{1}$, $3$\strut} \\
\hline\hline
\end{tabular}
\caption{Settings of the factorisation runs in~\cref{fig:search_fq,fig:search_sq} for first (FQ) and second (SQ) quantisation. Runs: the number of runs in the box, $\max(|\Delta E_{\rm tot}|,|\Delta E_{\rm corr}|)<0.3$~mHa, out of all runs; Settings: the number of settings with at least one run in the box, out of all settings. The remaining columns list the factorisation ranks and the regularisation parameters of~\cref{eq: objective_function} that we tried; values used by at least one run in the box are in bold. The orderings and numbers of layers of the first-quantised unitaries are listed in~\cref{tab:search_unitaries}. $^\dagger$One value for each $(V,L,M)$.}
\label{tab:search}
\end{table*}

\begin{table*}[htbp]
\centering
\small
\renewcommand{\arraystretch}{1.25}
\setlength{\tabcolsep}{4pt}
\begin{tabular}{l l l l l l}
\hline\hline
System & Cyclic & Triangle wave & Cyclic + CSD & Triangle wave + CSD & Full depth \\
\hline
H$_4$, def2-SVPD & \parbox[t]{1.8cm}{\raggedright\strut $\mathbf{17}$, $\mathbf{25}$\strut} & \parbox[t]{2.3cm}{\raggedright\strut $17$, $\mathbf{25}$\strut} & \parbox[t]{3.6cm}{\raggedright\strut 1-1-1-1-1: $25$\\ 1-2-2: $\mathbf{21}$\\ 2-1-2: $\mathbf{17}$, $\mathbf{25}$\\ 2-3: $\mathbf{25}$\strut} & \parbox[t]{3.6cm}{\raggedright\strut 1-1-1-1-1: $25$\\ 1-1-2-1: $17$, $\mathbf{25}$\strut} & \parbox[t]{1.5cm}{\raggedright\strut --\strut} \\
\hline
H$_4$, def2-TZVPP & \parbox[t]{1.8cm}{\raggedright\strut $\mathbf{29}$, $\mathbf{43}$\strut} & \parbox[t]{2.3cm}{\raggedright\strut $\mathbf{29}$, $\mathbf{43}$\strut} & \parbox[t]{3.6cm}{\raggedright\strut 1-5: $31$\\ 2-2-2: $29$\strut} & \parbox[t]{3.6cm}{\raggedright\strut 1-1-2-1-1: $\mathbf{29}$, $\mathbf{43}$\\ 1-1-4: $31$\strut} & \parbox[t]{1.5cm}{\raggedright\strut --\strut} \\
\hline
[Fe$_2$S$_2$]$^{2-}$ & \parbox[t]{1.8cm}{\raggedright\strut $\mathbf{61}$, $\mathbf{93}$\strut} & \parbox[t]{2.3cm}{\raggedright\strut $\mathbf{61}$, $\mathbf{93}$\strut} & \parbox[t]{3.6cm}{\raggedright\strut 3-2-2: $\mathbf{77}$\\ 3-4: $\mathbf{61}$, $77$, $\mathbf{93}$\strut} & \parbox[t]{3.6cm}{\raggedright\strut 1-2-2-1-1: $\mathbf{77}$\\ 1-2-3-1: $\mathbf{61}$, $\mathbf{77}$, $\mathbf{93}$\strut} & \parbox[t]{1.5cm}{\raggedright\strut --\strut} \\
\hline
[MnO$_4$]$^{-}$ & \parbox[t]{1.8cm}{\raggedright\strut $\mathbf{65}$, $\mathbf{81}$, $\mathbf{97}$\strut} & \parbox[t]{2.3cm}{\raggedright\strut $65$, $\mathbf{97}$\strut} & \parbox[t]{3.6cm}{\raggedright\strut 2-2-3: $\mathbf{65}$, $\mathbf{87}$\\ 3-4: $\mathbf{65}$, $\mathbf{77}$, $\mathbf{81}$, $\mathbf{87}$, $\mathbf{97}$\strut} & \parbox[t]{3.6cm}{\raggedright\strut 1-1-2-2-1: $\mathbf{87}$\\ 1-2-2-1-1: $\mathbf{65}$\\ 1-2-3-1: $\mathbf{65}$, $77$\\ 1-3-2-1: $87$, $\mathbf{97}$\strut} & \parbox[t]{1.5cm}{\raggedright\strut --\strut} \\
\hline
[Mn$_2$O$_2$]$^{3+}$ & \parbox[t]{1.8cm}{\raggedright\strut $\mathbf{65}$, $\mathbf{81}$, $\mathbf{97}$\strut} & \parbox[t]{2.3cm}{\raggedright\strut --\strut} & \parbox[t]{3.6cm}{\raggedright\strut 2-2-3: $65$, $81$\\ 3-4: $65$, $81$, $\mathbf{97}$\strut} & \parbox[t]{3.6cm}{\raggedright\strut 1-3-2-1: $\mathbf{97}$\strut} & \parbox[t]{1.5cm}{\raggedright\strut \textbf{yes}\strut} \\
\hline
[Mn$_2$O$_3$]$^{2+}$ & \parbox[t]{1.8cm}{\raggedright\strut $65$, $\mathbf{81}$, $\mathbf{97}$\strut} & \parbox[t]{2.3cm}{\raggedright\strut --\strut} & \parbox[t]{3.6cm}{\raggedright\strut 2-2-3: $65$, $\mathbf{81}$\\ 3-4: $65$, $\mathbf{81}$, $97$\strut} & \parbox[t]{3.6cm}{\raggedright\strut 1-3-2-1: $\mathbf{97}$\strut} & \parbox[t]{1.5cm}{\raggedright\strut \textbf{yes}\strut} \\
\hline\hline
\end{tabular}
\caption{Orderings and numbers of layers, $N_S$, of the unitaries tried in the first-quantised runs of~\cref{fig:search_fq} (\cref{tab:search}). For the nested orderings, each line gives one partition (\cref{sec:us_orderings}) and the values of $N_S$ tried with it. Values of $N_S$ used by at least one run in the box, $\max(|\Delta E_{\rm tot}|,|\Delta E_{\rm corr}|)<0.3$~mHa, are in bold. Full depth: factorisations with dense unitaries, costed as unitaries in the cyclic + CSD ordering with all $2^{\lceil\log_2D\rceil}-1=127$ layers; none of them is used in~\cref{tab:factorisations}; --: not tried.}
\label{tab:search_unitaries}
\end{table*}

\clearpage
\bibliography{main.bib}
\end{document}